\documentclass[12pt, letterpaper]{article}
\pdfoutput=1

\usepackage{amsmath}
\usepackage{amsfonts}
\usepackage{amssymb}
\usepackage{graphicx}
\usepackage{caption}
\usepackage{subcaption}
\usepackage{graphics,epsfig}
\usepackage{epsf}
\usepackage{epstopdf}
\usepackage{color}
\usepackage{float}
\usepackage[utf8]{inputenc}

\newcommand{\nc}{\newcommand}
\nc{\ba}{\begin{eqnarray}}
\nc{\ea}{\end{eqnarray}}

\usepackage{tikz}
\usetikzlibrary{arrows,shapes}
\usetikzlibrary{trees}
\usetikzlibrary{matrix,arrows} 				
\usetikzlibrary{positioning}				
\usetikzlibrary{calc,through}				
\usetikzlibrary{decorations.pathreplacing}  
\usepackage{pgffor}							

\usetikzlibrary{decorations.pathmorphing}	
\usetikzlibrary{decorations.markings}
\tikzset{
	vector/.style={decorate, decoration={snake}, draw},
	provector/.style={decorate, decoration={snake,amplitude=2.5pt}, draw},
	antivector/.style={decorate, decoration={snake,amplitude=-2.5pt}, draw},
	fermion/.style={draw=black, postaction={decorate},
		decoration={markings,mark=at position .55 with {\arrow[draw=black]{>}}}},
	fermionbar/.style={draw=black, postaction={decorate},
		decoration={markings,mark=at position .55 with {\arrow[draw=black]{<}}}},
	fermionnoarrow/.style={draw=black},
	gluon/.style={decorate, draw=black,
		decoration={coil,amplitude=4pt, segment length=5pt}},
	scalar/.style={dashed,draw=black, postaction={decorate},
		decoration={markings,mark=at position .55 with {\arrow[draw=black]{>}}}},
	scalarbar/.style={dashed,draw=black, postaction={decorate},
		decoration={markings,mark=at position .55 with {\arrow[draw=black]{<}}}},
	scalarnoarrow/.style={dashed,draw=black},
	electron/.style={draw=black, postaction={decorate},
		decoration={markings,mark=at position .55 with {\arrow[draw=black]{>}}}},
	bigvector/.style={decorate, decoration={snake,amplitude=4pt}, draw},
}

\tikzstyle{block} = [draw, rectangle, 
minimum height=3em, minimum width=6em]

\usepackage{hyperref}			

\begin{document}
	
\begin{flushright} {\footnotesize YITP-20-100}  \end{flushright}
\vspace{0.5cm}
\begin{center}

\def\thefootnote{\fnsymbol{footnote}}

{\large {\bf 
Inflation with multiple vector fields and  non-Gaussianities
}}
\\[1cm]
{Mohammad Ali Gorji$^{1}$ \footnote{gorji@yukawa.kyoto-u.ac.jp}, 
Seyed Ali Hosseini Mansoori$^{2}$\footnote{shosseini@shahroodut.ac.ir}, 
Hassan Firouzjahi$^{3}$ \footnote{firouz@ipm.ir}
}
\\[0.5cm]
 
 {\small \textit{$^1$Center for Gravitational Physics, Yukawa Institute for Theoretical Physics, Kyoto University, \\ Kyoto 606-8502, Japan
}}\\

{\small \textit{$^2$Faculty of Physics, Shahrood University of Technology, \\ 
P.~O.~Box 3619995161 Shahrood, Iran }}
 
 {\small \textit{$^3$School of Astronomy, Institute for Research in Fundamental Sciences (IPM), \\ P.~O.~Box 19395-5531, Tehran, Iran
}}\\

\end{center}

\vspace{.8cm} \hrule
\vspace{0.3cm}

\begin{abstract}

We consider a model of inflation consisting a triplet of $U(1)$ vector fields with the parity violating interaction which is non-minimally coupled to inflaton. The vector field sector enjoys global $O(3)$ symmetry which admits isotropic configuration and provides not only vector modes but also scalar and tensor modes. We decompose the scalar perturbations into the adiabatic, entropy and isocurvature perturbations and  compute all power spectra and cross correlations of the scalar and the tensor sectors. The tensor modes associated with the vector fields contribute to the power spectrum of gravitational waves while the parity violating term generates chirality in  gravitational power spectra and bispectra. We study nonlinear scalar and tensor perturbations  and compute all bispectra and three-point cross-correlations.  In particular, it is shown that  the non-Gaussianity of curvature perturbations and gravitational waves are enhanced by the vector field perturbations.  We show that non-Gaussianities put strong constraints on the model parameters such as the parity violating coupling and the fractional energy of the vector fields. 
 
 \end{abstract}
\vspace{0.5cm} \hrule
\def\thefootnote{\arabic{footnote}}
\setcounter{footnote}{0}

\newpage

\section{Introduction}


The simplest models of inflation are based on a scalar field which slowly rolls on top of a nearly flat potential. Among the basic predictions of the models of inflation are that the primordial fluctuations are nearly scale invariant, nearly Gaussian and nearly adiabatic which are well consistent with cosmological observations \cite{Ade:2015lrj,Akrami:2018odb,Akrami:2019izv}. While inflation is the leading paradigm for early universe cosmology and a working setup for generating the observed large scale structures, yet it is still a phenomenological scenario looking for a deeper theoretical understanding. It is an important open question as how inflation may be embedded in a more fundamental theory of high energy physics, perhaps with some links to physics beyond the Standard Model (SM) or quantum gravity. In particular one may expect more fields or fields different than scalar fields during inflation. Indeed, vector/gauge fields are ubiquitous in theories of high energy physics and in SM. In this regard, it is reasonable to look for the imprints of beyond SM vector/gauge fields during inflation. 

In recent years, the roles of vector/gauge fields in the context of inflationary models are widely studied. One may generally classify these models in two categories: i) inflation is mainly driven by a scalar field while there are some vector fields which either interact with inflaton or are pure spectator fields so that they do not contribute to the inflationary background, ii) vector fields play the role of inflaton and drive inflation.  In the first type of models, the vector field may not necessarily be isotropic. One possibility is that vector fields do not have a vacuum expectation value (vev) and show up only at the level of perturbations. For instance, in inflationary scenarios with a pseudoscalar  \cite{Anber:2006xt,Barnaby:2010vf, Barnaby:2011vw, Barnaby:2011qe, Sorbo:2011rz}, a $U(1)$ vector field is coupled to an axion-like inflaton field and produces chiral gravitational waves (GWs). It is also possible that a $U(1)$ vector field acquires a vev during inflation. The most well-known example of this type is the so-called anisotropic inflation where a $U(1)$ gauge field is non-minimally coupled to the inflaton field  \cite{ Watanabe:2009ct, Kanno:2010ab, Bartolo:2012sd, Emami:2013bk, Abolhasani:2013zya, Emami:2015qjl}.  Among the second types of models are isotropic vector inflation scenario where inflation can be realized from a bunch of massive vector fields \cite{Golovnev:2008cf} or models with non-Abelian gauge fields with $SU(2)$ gauge symmetry \cite{Maleknejad:2011jw, Maleknejad:2011sq}. The interesting point of these models is that they admit isotropic background and, therefore, can be responsible for both the background evolution and the generation of perturbations in an inflationary setup. Note, however, that  models of massive vector inflation usually suffer from some pathologies \cite{Himmetoglu:2008zp}. 
  
Our setup in this work is mostly related to the first type which is an isotropic extension of the combination of the pseudoscalar inflation and anisotropic inflation setups. In both of these models, the inflaton field is coupled to a $U(1)$ vector field but through different coupling functions. In pseudoscalar inflation, the coupling is axionic-like ${\phi} F_{\mu\nu}{\tilde F}^{\mu\nu}$ where $\phi$ is the inflaton field with a slow-roll potential, $F_{\mu\nu}$ is the strength tensor of the $U(1)$ vector field and ${\tilde F}^{\mu\nu}$ is the dual of $F_{\mu\nu}$. The vector field usually does not have a vev in this scenario. Even if one considers some initially nonzero vev for the vector field, it is diluted in an expanding background. In anisotropic inflation, however, inflaton is non-minimally coupled to the vector field with interaction of the form $f(\phi)^2 F_{\mu \nu} F^{\mu \nu}$. It is shown in  \cite{Watanabe:2009ct} that by choosing an appropriate form of coupling $f(\phi)$, the electric field energy density survives the inflationary expansion and will not be diluted. This is shown to be an attractor solution in which the background electric field energy density furnishes a small but a nearly constant fraction of the total energy density.  Consequently, both curvature perturbations and GW power spectra are affected by the vector field perturbations  \cite{Chen:2014eua}.

In the presence of a single $U(1)$ vector field with non-vanishing vev, the spacetime is anisotropic in the form of Bianchi I Universe. With strong observational constraints on background anisotropy \cite{Kim:2013gka, Ade:2015hxq}, one may wish to extend the models of anisotropic inflation to obtain isotropic inflationary background. One suggestion is to extend the setup to multiple $U(1)$ vector fields. It is well known that having $N$ vector fields, anisotropy scales as $N^{-1}$ which leads to the isotropic vector field configuration for large $N$ \cite{Golovnev:2008cf}. The isotropic background  can also be realized from three orthogonal $U(1)$ vector fields. The setup is locally like three $U(1)$ vector fields while it enjoys internal (global) $O(3)$ symmetry which admits isotropic 
Friedmann-Lema\^{i}tre-Robertson-Walker (FLRW) solution \cite{Yamamoto:2012sq, Funakoshi:2012ym,Emami:2016ldl}. Another possibility is to consider non-Abelian gauge symmetry with a gauge symmetry homomorphic to the $O(3)$ symmetry \cite{Galtsov:1991un}. For the non-Abelian case the simplest choice is $SU(2)$ gauge symmetry \cite{Maleknejad:2011jw}. Moreover, the isotropic extension of pseudoscalar, known as chromo-natural inflation, is also investigated \cite{Adshead:2012kp}. 

In this paper, we consider the  isotropic extension of the anisotropic  inflation model in which inflaton is a scalar field that non-minimally couples to a triplet of vector fields which admits isotropic background. In the presence of a non-minimal coupling between the inflaton and the vector field we should consider not only the standard kinetic term $F_{\mu \nu} F^{\mu \nu}$ but also the parity violating interaction $F_{\mu\nu}{\tilde F}^{\mu\nu}$ since the latter is no longer a total derivative. One interesting feature of the models with isotropic configuration of vector/gauge fields is the presence of two extra tensor modes on top of the gravitational tensor modes. These tensor modes provide nonzero shear anisotropy for the GWs at the level of linear perturbations \cite{Maleknejad:2014wsa,Dimastrogiovanni:2016fuu}. Even if we consider gauge fields as spectator fields which do not contribute to the inflationary background dynamics, they can still significantly change the spectrum of GWs. Recently, non-Gaussianity (NG) of primordial GWs in the presence of pure spectator gauge fields in slow-roll scalar field inflation are studied \cite{Agrawal:2017awz,Thorne:2017jft,Agrawal:2018mrg,Dimastrogiovanni:2018xnn,Dimastrogiovanni:2018gkl,Bordin:2018pca,Fujita:2018vmv,Iacconi:2020yxn}. In our model, vector fields are not spectator fields as they contribute to the background energy density $I\neq0$. Therefore, they change not only the GWs NG but also curvature perturbations NG.

The rest of the paper is organized as follows. In Section \ref{sec-model} we present our setup with a brief review of the background dynamics and the decomposition of perturbations in gravitational and matter sectors. In Section \ref{sec-scalar} we decompose the scalar perturbations into the adiabatic and entropy perturbations and calculate the curvature perturbations and the entropy perturbations power spectra and their cross correlations. This analysis is extended to tensor perturbations in Section \ref{tensor-linear}. The bispectra of the curvature perturbations and GWs and mixed NG between them are presented in Section \ref{NG}. The three-point cross-correlations between vector fields modes and curvature perturbations and GWs are found in Section \ref{NG-GF}. In Section \ref{compare} we compare our setup with various models of inflation containing non-Abelian gauge fields followed by summary and discussions in Section \ref{summary}. Many technical details such as the quadratic and cubic actions and the 
analysis of the in-in integrals are relegated to the appendices.

\vspace{1cm}


\section{The model}\label{sec-model}

In this section we present our model and briefly study the background inflationary dynamics. We also present the Scalar-Vector-Tensor (SVT) decompositions of perturbations in gravitational and matter sectors. 

The model contains the inflaton field $\phi$ with a nearly flat potential $V(\phi)$ and an orthogonal triplet of $U(1)$ vector fields $A^{a}{}_\mu$ where $a=1,2,3$ are the internal field space indices. The three identical copies of $U(1)$ fields have the local $U(1) \times U(1) \times U(1)$ symmetry and internal (global) $O(3)$ symmetry which admits isotropic cosmological background solution \cite{Emami:2016ldl,Gorji:2018okn}. In our setup, the vector fields $A^{a}{}_\mu$ with field strength tensor $F^{a}{}_{\mu\nu} = \partial_{\mu} A^{a}{}_{\nu} - \partial_{\nu} A^{a}{}_{\mu}$ are non-minimally coupled to the inflaton field through the coupling function $f(\phi)$ as follows (with the Planck mass $M_{\rm P}$ set to unity)
\begin{equation}\label{action}
S =\frac{1}{2} \int d^{4}x \sqrt{-g} \Big[ R - \partial_{\mu} \phi \partial^{\mu} \phi 
-2 V(\phi) - \frac{1}{2} f^{2}(\phi) \, \sum_{a=1}^3 \big(F^{a}{}_{\mu \nu} F_{a}{}^{\mu \nu}
+ \theta F^{a}{}_{\mu \nu} {\tilde F}_{a}{}^{\mu \nu} \big) \Big] \,,
\end{equation}
where $\mu,\nu = 0,..,3$ denote the spacetime indices, $R$ is the Ricci scalar, and ${\tilde F}^{a \mu \nu} = \frac{1}{2} \sqrt{-g} \varepsilon^{\mu\nu\eta\sigma} F^{a}{}_{\eta\sigma}$ is the dual field of $F^{a}{}_{\mu\nu}$ with the totally antisymmetric tensor density $\varepsilon^{\mu \nu \eta\sigma}$. The field indices $a,b, ...$ are raised and lowered by the flat Cartesian metric $\delta_{ab}$ while the spatial indices $i,j$ are raised and lowered by the spatial metric $g_{ij}$ which is different than the Cartesian metric $\delta_{ab}$. Finally the coupling constant $\theta$ represents the parity violating term. 

In Maxwell theory with no conformal coupling, the background vector field energy density dilutes exponentially. In order to break the conformal invariance and to prevent the dilution of background electric field energy density, the non-minimal coupling $f(\phi)$ is inserted to drag energy from the inflaton sector to the vector fields sector. As shown in \cite{Watanabe:2009ct}, with an appropriate form of the conformal coupling $f(\phi)$, the system reaches an attractor regime in which the vector field energy density remains a constant fraction of the total energy density. At the level of perturbations, the vector field perturbations acquire a nearly scale invariant perturbations which can affect the large scale curvature perturbations. In models of anisotropic inflation with one copy of vector field, quadrupolar statistical anisotropy are generated \cite{ Kanno:2010ab, Bartolo:2012sd, Emami:2013bk, Abolhasani:2013zya, Emami:2015qjl}
which are constrained by the CMB observations \cite{ Kim:2013gka, Ade:2015hxq}. Our isotropic setup, however, would not produce any statistical anisotropy by construction. 

In the presence of a non-minimal coupling between the inflaton and vector fields we also allowed the parity violating term $F^{a}{}_{\mu \nu} {\tilde F}_{a}{}^{\mu \nu}$ as well as the Maxwell term.  In the absence of non-minimal coupling, the latter is a topological term which does not contribute to the equations of motion. However, it is no longer a boundary term in the presence of the  non-minimal coupling.\footnote{We comment that the setup in the form of action (\ref{action}) but with a single copy of a pure spectator $U(1)$ gauge field (which does not contribute to the background dynamics) was studied in \cite{Caprini:2014mja, Caprini:2017vnn, Almeida:2017lrq, Almeida:2019hhx}. For example, in  \cite{Caprini:2014mja} it is shown that the problems associated with  large scale primordial magnetic fields may be alleviated in that scenario.}

The setup with the action (\ref{action}) is the isotropic extension of anisotropic inflation with the parity violating interaction; for various works on anisotropic inflation see \cite{anisotropic-inflation}. The internal $O(3)$ symmetry for the triplet of $U(1)$ vector fields  allows one to obtain an isotropic FLRW background. This proposal was first put forward in  \cite{Yamamoto:2012sq, Funakoshi:2012ym,Yamamoto:2012tq} and  in this work we extend this idea in various directions. First, we allow for the violation of parity by adding the interaction with the coupling $\theta$. Second, since the vector fields contribute to the background energy density, the curvature perturbations are not solely determined by the perturbations of the inflaton but by a combination of the perturbations of the inflaton field and the longitudinal scalar mode of the vector fields. We therefore decompose our scalar modes into the adiabatic, entropy and isocurvature modes and calculate their power spectra, bispectra and cross-correlations. Finally, we calculate the tensor non-Gaussianities which were not studied in previous works. As we already mentioned in the introduction, the extension of $U(1)$ vector field to a triple of $U(1)$ vector fields with global $O(3)$ symmetry provides extra tensor modes. These tensor modes, as we will see, drastically affect tensor non-Gaussianities.

\subsection{Cosmological background}

In order to have isotropic and homogeneous background configuration, we consider the following ansatz for the background vector fields \cite{Golovnev:2008cf,ArmendarizPicon:2004pm}

\begin{equation}
\label{bac-A}
A^{a}{}_{\mu}= A(t)\, \delta^{a}_{\mu} \,,
\end{equation}
which is consistent with the spatially flat FLRW metric 
\begin{equation}
\label{FRW}
ds^2 = - d t^2+ a(t)^2 \delta_{ij} dx^i dx^j\, ,
\end{equation}
where $t$ is the  cosmic time, $a(t)$ is the scale factor, and $A(t)$ is the background value of the vector field. The ansatz (\ref{bac-A}) is isotropic and therefore it is not the most general configuration. In general, one expects that initially the vector fields can have different background values and the setup be anisotropic. However, it is 
shown in \cite{Yamamoto:2012tq} that the configuration (\ref{bac-A}) is the attractor solution of the  slow-roll inflationary background even if one starts with homogeneous anisotropic initial configuration. Our setup can also be realized from the global limit of non-Abelian gauge fields when the gauge coupling vanishes \cite{Emami:2016ldl}. 
If the gauge fields do not contribute to the background dynamics then  they provide isocurvature scalar modes and  tensor modes. In our model, however, 
the vector fields contribute to the background, so  they contribute to the curvature perturbations as well.  In a recent paper \cite{Wolfson:2020fqz} it is shown that models which include scalar field as inflaton and gauge fields as spectator are phenomenologically more viable than models that only includes non-Abelian gauge fields.


Varying the action (\ref{action}) with respect to the vector fields $A^{a}{}_{\mu}$ and then solving the resultant equation with the background metric (\ref{FRW}), we find
\begin{equation}\label{vector}
\dot{A} = \frac{q_{0}}{a f^{2}} \, ,
\end{equation}
where a dot denotes derivative with respect to the cosmic time and $q_{0}$ is an integration constant. 

Varying the action with respect to the metric gives the Einstein fields equations, which after substitutions from Eq. (\ref{vector}), yield the following equations
\begin{eqnarray}\label{EE}
3{ H}^2 & = & \frac{1}{2} {\dot \phi}^2 + V 
+ \frac{3}{2}\frac{q_{0}^{2}}{a^4 f^2} \,, \\
 2{\dot  H} + 3{ H}^2 & = & - \Big( \frac{1}{2} {\dot \phi}^2 - V 
+ \frac{1}{2}\frac{q_{0}^{2}}{a^4 f^2} \Big) \,,
\end{eqnarray}
where ${ H} \equiv \dot a/a$ is the Hubble parameter. Note that the parity violating term does not contribute to the background equations in isotropic configuration. In \cite{Firouzjahi:2018wlp} a setup similar to this model but containing three complex scalar fields charged under the three copies of the gauge fields are studied. Since the parity violating term does not contribute to the background equations, the background solutions here are exactly the same as those of \cite{Firouzjahi:2018wlp} by setting the charge coupling ${\bf e}=0$ in \cite{Firouzjahi:2018wlp}. So, here we only briefly review the background equations and refer the interested reader  to \cite{Firouzjahi:2018wlp} for more details of background dynamics. 

The energy density of the vector field $\rho_A=3q_{0}^{2}/2 a^4 f^2$ cannot be large in comparison with the total energy density $\rho_A+\rho_{\phi} \simeq \rho_{\phi} = {\dot \phi}^2/2 + V$. Otherwise, it destroys the slow-roll inflation. Let us define the parameter $I$ as the fraction of the background electric field energy density to the inflaton energy density via 
\begin{equation}\label{I}
\frac{\rho_{A}}{\rho_{\phi}} \equiv \frac{I}{2}\epsilon\, , \hspace{1cm} 
\epsilon \equiv -\frac{\dot H}{H^2} \, ,
\end{equation}
where $\epsilon$ is the slow-roll parameter. Then it is shown in \cite{Watanabe:2009ct,Firouzjahi:2018wlp} that by choosing 
the conformal coupling  $f(\phi)$ in the form 
\begin{equation}\label{a1}
f(\phi)= \exp\Big(\frac{2}{1-I} \int \frac{V}{V_{,\phi}} d\phi\Big) \,,
\end{equation} 
the system reaches the attractor solution in which the vector field energy density remains a small but a constant fraction of the total energy density. Note that the above choice is one working example for any inflationary model based on slow-roll potentials \cite{Watanabe:2009ct} while for non-slow-roll models one may look for other possibilities. The parameter $I$ is expected to be very small to allow for an attractor solution at the background \cite{Firouzjahi:2018wlp}. We will also see this fact at the level of perturbations. One can show that  in the slow-roll limit with small $I$,  the potential is related to the Hubble expansion rate via \cite{Firouzjahi:2018wlp}
\begin{equation}\label{V38}
V(\phi)\simeq 3 H^2 \Big( 1 - \frac{\epsilon}{6} (2+I) \Big) \,.
\end{equation}

\subsection{Cosmological perturbations}

Due to internal $O(3)$ symmetry, the so-called SVT theorem is applicable here and we can decompose the perturbations into the scalar, vector, and tensor types in both gravity and matter sectors. Since there are vector modes in the matter sector, the vector perturbations are dynamical in our scenario. They, however, completely decouple from the scalar and tensor modes thanks to the internal $O(3)$ symmetry and they will decay after inflation. Here we then only consider the scalar and tensor types perturbations which are enough to look at the spectra of curvature perturbations and primordial GWs. 

Scalar and tensor perturbations around the background configuration (\ref{bac-A}) and (\ref{FRW}) are given by \cite{Firouzjahi:2018wlp}
\begin{eqnarray}\label{perturbations}
&& \delta{A}^{b}{}_0 \delta_{ab} = \delta{A}^{a}{}_0 = \partial_a Y \,, \hspace{1cm}
\delta{A}^{b}{}_i \delta_{ab} = \delta{A_{ia}} = \delta{Q}\, \delta_{ia}+\partial_i \partial_a M + \epsilon_{iab}\partial_b{U} + t_{ia} \,,
\\ \nonumber
&& \delta{g_{00}} = - 2  \alpha \,, \hspace{1cm} \delta{g_{0i}} = a \partial_i \beta \,,
\hspace{1cm} \delta{g_{ij}} = a^2 ( 2\psi \delta_{ij} + 2\partial_i\partial_j E + \gamma_{ij} ) \,,
\end{eqnarray}
where $(Y,\delta{Q},M,U,\alpha, \beta, \psi,E)$ are scalar modes and $(\gamma_{ij},t_{ia})$ are tensor modes which satisfy the transverse and traceless conditions
\begin{eqnarray}\label{traceless-transverse}
\partial_i t^i{}_j = 0 = t^i{}_i \,, \hspace{1cm}
\partial_i \gamma^i{}_j = 0 = \gamma^i{}_i \,.
\end{eqnarray}
In addition, there is the inflaton field scalar perturbation $\delta \phi$.

Not all of these scalar modes are real physical degrees of freedom. The diffeomorphism invariance of the action (\ref{action}) allows us to work in the spatially flat gauge
\begin{equation}\label{SF-G}
\psi =0\,, \hspace{1cm} \mbox{and} \hspace{1cm} E=0\,.
\end{equation}

Moreover, the model (\ref{action}) enjoys the local symmetry $A^{a}{}_{\mu} \rightarrow{A}^{a}{}_{\mu}-\partial_{\mu}\Lambda^a$. Decomposing $\Lambda^a$ into $\Lambda^a = \partial^a \Lambda + \Lambda^{a}_{\perp}$ with 
$\partial_a\Lambda^a_{\perp}=0$, this local symmetry implies
\begin{equation}\label{LGS}
\delta{A}^{a}{}_{\mu}\rightarrow\delta{A}^{a}{}_{\mu} - 
\partial_{\mu}\partial^a \Lambda-\partial_{\mu}\Lambda^a_{\perp} \,,
\end{equation}
which after substituting from Eq. (\ref{perturbations}) yields
\begin{eqnarray}\label{LGF-perturbations}
Y \rightarrow Y -  \dot {\Lambda}\,, \hspace{.5cm} M \rightarrow M - \Lambda \,.
\end{eqnarray}
All other perturbations in decomposition (\ref{perturbations}) are invariant under the local symmetry (\ref{LGS}). The above transformations show that still one scalar mode is not an independent physical degree of freedom and we fix the gauge by
\begin{equation}\label{GF-G}
M=0 \,.
\end{equation}
In conclusion, after fixing the gauges, we are left with six scalar modes 
$(Y,\delta{Q},U,\alpha,\beta,\delta\phi)$. 

The tensor modes are all gauge invariant and there is no need for gauge fixing.

Note that since the setup enjoys global $O(3)$ symmetry then the scalar and tensor modes evolve separately at the linear order of perturbations.  However, they will mix at higher orders, for example when calculating non-Gaussianities. 


\section{Linear scalar perturbations}\label{sec-scalar}

In this section, we study linear scalar perturbations and obtain all two-point correlation functions including power spectrum of curvature perturbations, power spectrum of entropy perturbations, and cross-correlations between curvature perturbations and entropy perturbations.

After fixing all gauges in (\ref{SF-G}) and (\ref{GF-G}), we have six scalar modes $(Y,\delta{Q},U,\alpha,\beta,\delta\phi)$ among which $(Y, \alpha, \beta)$ are non-dynamical, i.e. they appear with no time derivatives in quadratic action. As a result, from their algebraic equations of motion, we can express them in terms of dynamical modes $(\delta{Q},U,\delta\phi)$. Plugging these solutions into the quadratic action, we can integrate out these non-dynamical modes and obtain the quadratic action in terms of the dynamical modes.  Moreover, since the vector fields contribute to the background dynamics, curvature perturbation receives contributions not only from inflaton perturbations $\delta\phi$ but also from vector field perturbation $\delta{Q}$ as we will explicitly show in the next subsection.

\subsection{Adiabatic/entropy decomposition}

The details of the calculations of the quadratic actions are presented in appendix \ref{quad-action}. For the scalar modes to leading order in small parameters $I$ and $\epsilon$, the quadratic action is given by
\begin{eqnarray}\label{S2-SS}
S^{\rm SS} &=& \frac{1}{2} \int d\tau d^3x 
\Big[\overline{\delta \phi}'^2-(\partial \overline{\delta \phi})^2 + \frac{2}{\tau^2} (1+2I) \overline{\delta \phi}^2 + \overline{\delta Q}'^2 - (\partial \overline{\delta Q})^2+ \frac{2}{\tau^2} \overline{\delta Q}^2\\
\nonumber &+& \partial \tilde{U}'^2-(\partial \partial \tilde{U})^2+\frac{2}{\tau^2} \partial \tilde{U}^2 
+ \frac{4}{\tau^2} \sqrt{I} (\overline{\delta Q} -2 \tau \overline{\delta Q}' ) \overline{\delta \phi}
+ \frac{4}{\tau} \theta \Big(\sqrt{I} \partial \overline{\delta \phi} + \partial \overline{\delta Q}\Big) \partial \tilde{U} \Big]\nonumber ,
\end{eqnarray} where we have defined
\begin{equation}\label{canonical-S0} 
\overline{\delta \phi} \equiv a \delta\phi \,, \hspace{1cm} \overline{\delta Q} \equiv \sqrt{2} f \delta Q \,, \hspace{1cm}  \tilde{U} \equiv \sqrt{2} f U \,.
\end{equation}
The new fields $\overline{\delta \phi}$ and $\overline{\delta Q}$ are canonically normalized while $\tilde{U}$ is not canonically normalized but $\partial \tilde{U}$ is as it is clear from Eq. (\ref{S2-SS}). This is the reason why we show the latter with a tilde and not with a bar. After going to the Fourier space, we define the canonically normalized field associated to  the mode $U$.

From Eq. (\ref{bac-A}) we see that the vector field $A^{a}{}_i$ has a background value and therefore it contributes to the background energy density. As a result the vector field cannot be treated as a test field and it contributes to the curvature perturbations. To see this fact explicitly, let us look at the comoving curvature perturbations $\mathcal{R}$ given by 
\begin{equation}\label{curvature-perturbation0}
{\cal R} \equiv \psi + H \delta u \,,
\end{equation}
where $\psi$ represents  the spatial curvature defined in (\ref{perturbations}) and $\delta{u}$ is the velocity potential defined as $\delta T^{t}_{i} \equiv  (\rho+p)\partial_{i} \delta u$ with $\rho$ and $p$ being the total background energy density and pressure of the system.  By expanding the energy-momentum tensor  with scalar perturbations defined in Eq. (\ref{perturbations}) and going to the spatially flat gauge defined in (\ref{SF-G}) with $\psi=0$, we obtain 
\begin{equation}\label{curvature-perturbation01}
{\cal R} = -aH\frac{\sqrt{2} f A' \overline{\delta Q} 
+ a\phi' \overline{\delta \phi}}{2 f^2 A'^2+a^2 \phi'^2} \, .
\end{equation}

Rewriting the background quantities in terms of the parameter $I$ defined in (\ref{I}) and then expanding for small values of $I$, the leading contributions are obtained to be 
\begin{eqnarray}\label{curvature-pertubation1}
\mathcal{R} & = & - \frac{{H}}{ \phi'} 
\Big[
(1-I) \overline{\delta \phi} - \sqrt{I}\, \overline{\delta Q}
\Big] \,.
\end{eqnarray}
This result is consistent with our intuition about the curvature perturbations since from Eq. (\ref{I}) we see that the ratio of the energy density of vector fields to the total energy density is proportional to $I$ and the contribution from the vector field to the curvature perturbations is proportional to $\sqrt{I}$ in the above relation. 

Our model then can be interpreted as a multiple field model of inflation and, in analogy to the logic of \cite{Gordon:2000hv}, we can decompose the scalar modes into the adiabatic and the entropy modes. The adiabatic mode is proportional to the curvature perturbations, and inspired by Eq. (\ref{curvature-pertubation1}), we define it as follows
\begin{equation}\label{adiabtic}
\overline{\delta \sigma} \equiv  \cos\vartheta\, \overline{\delta \phi} + \sin\vartheta\, \overline{\delta Q} \,,
\end{equation}
where we have defined the angular variable $\vartheta$ via
\begin{eqnarray}\label{sin-cos}
\cos\vartheta \equiv  \sqrt{1-I} \,, \hspace{1cm} \sin\vartheta \equiv - \sqrt{I} \,.
\end{eqnarray}
We then define the entropy mode to be orthogonal to the adiabatic mode (\ref{adiabtic}) in the field space as follows 
\begin{equation}\label{entropic}
\overline{\delta s} \equiv - \sin\vartheta\, \overline{\delta \phi} + \cos\vartheta\, \overline{\delta Q} \,.
\end{equation}

From Eqs. (\ref{adiabtic}), (\ref{sin-cos}) and (\ref{entropic}), we first see that the above decomposition corresponds to a rotation in the field space with the constant angle  $\vartheta=-\tan^{-1}(\sqrt{I}/\sqrt{1-I})$. Second, it shows that the adiabatic mode is mostly dominated by $\overline{\delta\phi}$ while the entropy mode obtains most of its contributions from $\overline{\delta{Q}}$. The comoving curvature perturbations for single field inflation can be recovered if we set $I=0$.  

Substituting from Eqs. (\ref{adiabtic}) and (\ref{sin-cos}) into  Eq. (\ref{curvature-pertubation1}), we find the following expression for the comoving curvature perturbations
\begin{eqnarray}\label{curvature-perturbation}
\mathcal{R} = - \frac{{H}}{ \dot \phi} \cos\vartheta\, \Big(\frac{\overline{\delta \sigma}}{a}\Big) \,,
\end{eqnarray} 
and in the same manner and from (\ref{entropic}), we can define the normalized entropy perturbations as follows 
\begin{eqnarray}\label{entropy-perturbation}
\mathcal{S} = - \frac{{H}}{ \dot \phi} \cos\vartheta\, \Big(\frac{\overline{\delta s}}{a}\Big) \,.
\end{eqnarray}
The variables $\mathcal{R}$ and $\mathcal{S}$ in Eqs. (\ref{curvature-perturbation}) and (\ref{entropy-perturbation}) are more closely related to the observable quantities. In practice, however, it is easier to work with the adiabatic and entropy modes 
$\overline{\delta \sigma}$ and $\overline{\delta s}$.  Thus, we perform the computations with respect to $\overline{\delta \sigma}$ and $\overline{\delta s}$ and then translate the results back to the quantities $\mathcal{R}$ and $\mathcal{S}$ at the end.

Substituting the adiabatic and entropy modes defined in Eqs. (\ref{adiabtic}) and (\ref{entropic}) into the quadratic action (\ref{S2-SS}), the quadratic action for the scalar modes in terms of the adiabatic  and entropy modes is given by 
\begin{eqnarray}\label{LSS-sigma-s}
{S}^{\rm SS} &=& \frac{1}{2} \int d\tau \, d^3 x
\Big[ \overline{\delta\sigma}'^2 - (\partial\overline{\delta\sigma})^2 + \frac{2}{\tau^2} (1-4 I ) \overline{\delta\sigma}^2 
+ \overline{\delta{s}}'^2 - (\partial \overline{\delta s})^2 + \frac{2}{\tau^2} ( 1 + 6I ) \overline{\delta s}^ 2\\
\nonumber &&\hspace{.4cm}+ \partial \tilde{U}'^2-(\partial \partial \tilde{U})^2 + \frac{2}{\tau^2} 
\partial \tilde{U}^2 + \frac{8}{\tau^2} \sqrt{I} \overline{\delta\sigma} ( 2 \overline{\delta s} - \tau \overline{\delta s}' )
+ \frac{8}{\tau} \theta ( 2 + I ) \partial \overline{\delta s} \partial {\tilde U}\Big] \, ,
 \end{eqnarray}
 where $\tau = \int dt/a(t)$ is the conformal time and a primes denotes derivative with respect to the conformal time. In the above relation, the superscript ${\rm SS}$ shows that the action is quadratic containing two scalar modes. We use this notation throughout  this paper to label the order and type of modes in expanded actions, Lagrangians, and Hamiltonians. 
 
Before calculating the power spectra  some comments are in order. The adiabatic modes $\overline{\delta\sigma}$ is directly coupled to the entropy modes $\overline{\delta{s}}$ via the exchange vertex interaction of the order of $\sqrt{I}$. This implies that the entropy modes induce corrections proportional to $I$ in the two-point function of adiabatic modes. On the other hand, adiabatic mode $\overline{\delta\sigma}$ is not directly coupled to the isocurvature modes ${\tilde U}$\footnote{Indeed, both of the modes $\overline{\delta{s}}$ and ${\tilde U}$ are isocurvature modes as they are not adiabatic. We, however, call $\overline{\delta{s}}$ as entropy mode while we call ${\tilde U}$ as isocurvature mode. This allows us to discriminate the mode $\overline{\delta{s}}$ which directly interacts with the curvature perturbations from the mode ${\tilde U}$ which does not directly interact with the curvature perturbations.}. This can be understood if we note that ${\tilde U}$, defined in (\ref{perturbations}), corresponds to the perturbations of the magnetic part of the vector fields. The ansatz (\ref{bac-A}) only provides non-vanishing electric field in the background and cannot provide background magnetic field. Then, the scalar mode ${\tilde U}$ does not have background value and this is the reason why it does not directly couple to the adiabatic mode. In this view, ${\tilde U}$ is a pure isocurvature mode. 
However, it couples to the entropy modes through the  parity violating interaction with exchange vertex labeled by $\theta$. Therefore, it would indirectly induce some corrections on the two-point function of adiabatic modes. In other words, ${\tilde U}$ interacts with $ \overline{\delta{s}}$ through exchange vertex $\theta$ while $\overline{\delta{s}}$ interacts with $\overline{\delta{\sigma}}$ through exchange vertex $I$. Then, ${\tilde U}$ induces corrections proportional to $I \theta$ in the two-point function of adiabatic modes $\overline{\delta{\sigma}}$. We will confirm these qualitative arguments in details in the next subsection.

\subsection{Power spectra and cross correlations}
 
As we have already mentioned, the consistency of the background analysis requires 
that $I$ to be small so all interactions that include a factor of $I$ can be treated perturbatively by means of the standard in-in formalism. However, we see that the term $\frac{16 \theta}{\tau} \partial\overline{\delta s} \partial {\tilde U}$ in the quadratic Lagrangian (\ref{LSS-sigma-s}) is independent of $I$. The parity violating interaction does not contribute to the background dynamics so a priori  there is no constraint on the value of $\theta$. We, therefore, cannot treat this term perturbatively at this stage. Since this term induces a mixing between the modes $\overline{\delta s}$ and ${\tilde U}$ we have  to diagonalize the action with respect to the  modes $\overline{\delta s}$ and ${\tilde U}$. 

To perform the diagonalization analysis we go to the Fourier space
\begin{equation}\label{Fourier}
\delta X(\tau,{\bf x}) = \int \frac{d^3k}{(2\pi)^3} \delta X(\tau,\textbf{k}) e^{i {\bf k.x}} \,,
\end{equation}
where $\delta X(\tau,{\bf x})$ is an arbitrary perturbation, which can be either a scalar or a tensor mode,  ${\bf k}$ is the wave vector, $k = |{\bf k}|$ is the magnitude of the wave vector, and $\delta X(\tau,\textbf{k})$ is the corresponding Fourier amplitude. Then  the diagonalization is better expressed in terms of the new entropy fields $\delta{s}_{\pm}$ related linearly to $\overline {\delta s}$ and $\tilde U$ via 
\begin{eqnarray}\label{trans-deltas}
\overline{\delta{s}} \equiv \frac{1}{\sqrt{2}} (\delta{s_+} - \delta{s_-}) \,, \hspace{1cm}
\overline{U} \equiv k\tilde{U} \equiv \frac{1}{\sqrt{2}} (\delta{s_+} + \delta{s_-}) \,,
\end{eqnarray}
where $\overline{U}$ is the canonically normalized field associated to the scalar mode $U$. 
 
Performing the above transformation into the quadratic action (\ref{LSS-sigma-s}) it is straightforward to show that the quadratic action takes the following form
\begin{eqnarray}\label{S2-sigma-s-pm}
&&S^{\rm SS} = \frac{1}{2} \int d\tau d^3 k 
\bigg[ \hspace{.2cm} \overline{\delta\sigma}'^2 - \Big( k^2 - \frac{2}{\tau^2} (1 - 4I ) \Big) {\overline{\delta\sigma}^2}  \nonumber
\\
&&\hspace{3.2cm} + {\delta s'_+}^2 
- \Big( k^2 - \frac{2}{\tau^2} ( 1 + 3I ) 
+ \frac{4}{\tau} k\theta ( 2+ I ) \Big) {\delta s_+^2} 
\nonumber \\
&&\hspace{3.2cm} + {\delta s'_-}^2 
- \Big( k^2 - \frac{2}{\tau^2} ( 1 + 3I ) 
- \frac{4}{\tau}k\theta ( 2 + I ) \Big) {\delta s_-^2} 
\nonumber \\ 
&&\hspace{3.2cm} + \frac{4\sqrt{2}}{\tau^2}\sqrt{I}\,{\overline{\delta \sigma}} 
\big( 2 \delta s_+ - \tau \delta s'_+ - 2 \delta s_- + \tau \delta s'_- \big) 
- \frac{12}{\tau^2} I \delta{s_+} \delta{s_-}
\bigg] \,.
\end{eqnarray}

From the action (\ref{S2-sigma-s-pm}), we can find the free Lagrangian ${ L}_0^{\rm SS}$ given by $S^{\rm SS}|_{I=0} \equiv \int d\tau { L}_0^{\rm SS}$. The quadratic interaction Lagrangian can be defined as $\delta{ L}^{\rm SS} \equiv { L}^{\rm SS}-{ L}_0^{\rm SS}$, where ${ L}^{\rm SS}$ is the total quadratic Lagrangian associated 
with the action  (\ref{S2-sigma-s-pm}), yielding
\begin{eqnarray}\label{L2-int}
\delta{ L}^{\rm SS} & \equiv & \int d^3k \Big[ 
- \frac{4 }{\tau^2} I \overline{\delta \sigma}^2+
\frac{2\sqrt{2}}{\tau^2} \sqrt{I}\, \overline{\delta \sigma} \big(2\delta s_
+ - \tau \delta s'_+ - 2\delta s_-  + \tau  \delta s'_- \big)
\nonumber \\
&-&\frac{2}{\tau} I k\theta (\delta s_+^2 - \delta s_-^2)
+ \frac{3}{\tau^2} I (\delta s_+^2 + \delta s_-^2)
-\frac{6 }{\tau^2} I \delta s_+ \delta s_- \Big] \,,
\end{eqnarray}
which vanishes for $I=0$ as expected. The fact that $I$ is a small parameter allows us to treat the above Lagrangian as small interaction and perform the perturbative expansions in terms of $I$ when computing the correlation functions.

The quantization procedure for the free parts which is represented by  ${ L}_0^{\rm SS}$ goes as usual. We treat all scalar modes in Eq. (\ref{S2-sigma-s-pm}) as quantum operators and then expand the modes in terms of creation and annihilation operators. For the adiabatic mode, we have
\begin{eqnarray}\label{sigma-op}
\overline{\delta\sigma}(\tau,\textbf{k}) = \overline{\delta\sigma}_k(\tau) a_{\textbf{k}} + \overline{\delta\sigma}_k^{*}(\tau) a_{-\textbf{k}}^{\dagger} \,,
\end{eqnarray}
where the mode function $\overline{\delta\sigma}_k(\tau)$ satisfies the Mukhanov-Sasaki equation
\begin{equation}\label{sigma-MSEq}
\overline{\delta\sigma}'' + \Big( k^2 - \frac{2}{\tau^2} \Big) \overline{\delta\sigma} = 0 \,,
\end{equation}
in which  to simplify the notation we have dropped the subscript $k$ for the mode function.  Imposing the Bunch-Davies initial conditions, the positive frequency solution for the above equation is 
\begin{eqnarray}\label{sigma-MF}
\overline{\delta \sigma} (\tau) = \frac{e^{-ik\tau}}{\sqrt{2k}} \Big(1-\frac{i}{k\tau}\Big) \,.
\end{eqnarray}

In a similar way, we can quantize the entropy modes $\delta s_{\pm}$ as
\begin{eqnarray}\label{s-pm-op}
\delta{s}_{\pm}(\tau,\textbf{k}) = \delta{s}_{\pm}(\tau) b^{\pm}_{\textbf{k}} + \delta{s}^{*}_{\pm}(\tau) b_{-\textbf{k}}^{\dagger\pm} \,,
\end{eqnarray}
where the mode functions satisfy the following equations
\begin{eqnarray}\label{s-pm-MF}
\delta{s}''_{\pm} + \Big(k^2 \pm \frac{8}{\tau} k\theta - \frac{2}{\tau^2}\Big)  \delta{s}_{\pm} = 0 \,.
\end{eqnarray}
The  solutions of Eq. (\ref{s-pm-MF}) are the linear combinations of the Whittaker functions $c^{\pm}_1 W(\mp 4i\theta,3/2,2ik\tau) + c^{\pm}_2 M(\mp 4i\theta,3/2,2ik\tau)$. Choosing positive frequency modes deep inside the horizon $k\tau\to-\infty$, we find $c^{\pm}_2=0$ and the other integration constant can be fixed through the Wronskian conditions ${\cal W}[\delta{s}_\pm,\delta{s}^{*}_\pm]=\delta{s}^{*}_\pm \delta{s}'_{\pm} - \delta{s}_\pm \delta{s}'^{*}_{\pm} = i$. Using formula ${\cal W}[W_{\kappa,\varkappa}(z),W_{-\kappa,\varkappa}(e^{\pm{i}\pi}z)] = e^{\pm{i}\kappa\pi}$, we find $c_1^{\pm}= - \frac{e^{\pm{3\pi}}}{\sqrt{2k}}$ and the solution with the  positive frequency Bunch-Davies initial conditions  is given by
\begin{eqnarray}\label{MF-deltas1}
\delta{s}_{\pm} = \frac{e^{\pm{2\pi\theta}}}{\sqrt{2k}} W(\mp 4i\theta,3/2,2ik\tau) \,.
\end{eqnarray}

Finally, note that the quantum operators $a_{\textbf{k}}$, $b^+_{\textbf{k}}$, and $b^-_{\textbf{k}}$ are independent of each other and each satisfies the standard creation and annihilation commutation relations with non-vanishing commutators $[a_{\textbf k},a^{\dagger}_{\textbf{-k}'}]= \delta(\textbf{k}+\textbf{k}')$ and $[b^{\pm}_{\textbf k},b^{\dagger\pm}_{\textbf{-k}'}]=  \delta(\textbf{k}+\textbf{k}')$.

Using Eq. (\ref{curvature-perturbation}), the power spectrum of the curvature perturbations ${\cal P}_{\cal R}$ at the end of inflation $\tau_e$ is given by 
\begin{eqnarray}\label{CF-R2}
\langle {\cal R} (\tau_e,{\bf k})\, {\cal R}(\tau_e,{\bf k}') \rangle =
\frac{1}{a(\tau_e)^2}
\left(\frac{H}{\dot{\phi}}\right)^2 \cos^2\vartheta \,{\big {\langle \overline{\delta\sigma}(\tau_e,{\bf k})
\, \overline{\delta\sigma}(\tau_e,{\bf k}') \big \rangle}}\, 
\equiv \frac{2 \pi^2}{k^{3}} {\cal P}_{\cal R} \, (2 \pi)^3\delta^{(3)}({\bf k}-{\bf k}') \,.
\end{eqnarray}
We therefore need to calculate the two-point function of the adiabatic mode $\overline{\delta \sigma}$. For this purpose, we implement the so-called in-in formalism \cite{Weinberg:2005vy, Chen:2009zp}
 where the expectation value of the quantity $Q$ at the time of end of inflation $\tau_e$ is given by
\begin{equation}\label{IN-IN-def}
\langle Q (\tau_e) \rangle = \big\langle 0 \big| 
\Big[ \overline{T} \exp\Big( i \int_{\tau_0}^{\tau_e} \delta{H}_{I} (\tau) d\tau \Big) \Big] 
Q_I(\tau_e) 
\Big[ T \exp\Big( -i \int_{\tau_0}^{\tau_e} \delta{H}_{I} (\tau) d\tau \Big) \Big]
\big| 0 \big\rangle \,,
\end{equation}
where $| 0 \rangle$ is the vacuum state of the theory which is defined at the far past $\tau_0 \rightarrow -\infty$, $\delta{ H}_{I}$ is the total interaction Hamiltonian in the interaction picture, $Q_I$ is the interaction picture operator associated with $Q$, $T$ and $\overline{T}$ are the time order and anti-time order operators defined as usual.

In the case of two-point function for the adiabatic mode, Eq. (\ref{IN-IN-def}) simplifies to
\begin{eqnarray}\label{IN-IN}
&&\big \langle \overline{\delta\sigma} (\tau_e,{\bf k})  \overline{\delta\sigma} (\tau_e,{\bf k}') \big \rangle 
= \langle 0| \overline{\delta\sigma} (\tau_e,{\bf k})  \overline{\delta\sigma} (\tau_e,{\bf k}') |0\rangle + i
\big\langle 0 \big | \int_{\tau_{0}}^{\tau_{e}} d \tau_{1}\Big[\delta{H}_{I} (\tau_{1}),
\overline{\delta\sigma} (\tau_e,{\bf k})  \overline{\delta\sigma} (\tau_e,{\bf k}')\Big] 
\big | 0 \big \rangle
\nonumber \\
&& \hspace{3cm}
- \big \langle0 \big | \int_{\tau_{0}}^{\tau_{e}} d\tau_{1} \int_{\tau_{0}}^{\tau_{1}} d\tau_{2} 
\Big[ \delta{H}_{{I}} (\tau_{2}), \Big[ \delta{H}_{{I}}(\tau_{1}),
\overline{\delta\sigma} (\tau_e,{\bf k})  \overline{\delta\sigma} (\tau_e,{\bf k}')\Big]\Big]  
\big | 0 \big \rangle + ... \,.
\end{eqnarray}

The first term in the right hand side  of the first line above is the two-point function of the adiabatic mode in the absence of any interaction, which using the free wave function Eq. (\ref{sigma-MF}), turns out to be $\langle 0| \overline{\delta\sigma} (\tau_e,{\bf k})  \overline{\delta\sigma} (\tau_e,{\bf k}') |0\rangle = (2k^3\tau_e^2)^{-1} (2 \pi)^3\delta^{(3)}({\bf k}-{\bf k}')$. Thus from Eq. (\ref{CF-R2}), the power spectrum of curvature perturbations in the absence of interactions  is given by
\begin{eqnarray}\label{CF-R20}
{\cal P}_{\cal R}^{(0)} = \frac{H^2}{8\pi^2 \epsilon } \, ,
\end{eqnarray}
where we have substituted $\big(\frac{H}{\dot{\phi}}\big)^2 \cos^2\vartheta \approx {1}/{2\epsilon}$. 

The corrections to the power spectrum \eqref{CF-R20} coming from the interaction Lagrangian \eqref{L2-int} then can be  computed from the other terms in the right hand side of the formula \eqref{IN-IN}. Indeed, even for the power spectra, as we will see later, we need to compute two-vertex Feynman diagrams like Fig. \ref{fig3} correspond to the interaction Hamiltonians  $\delta{ H}_{{I},i}^{\rm SS}$ with $i=1,..,4$ defined in \eqref{H2-int}. Therefore, we need to expand the formula \eqref{IN-IN-def} up to the second order. In the case of bi-spectra, as we will see, we need to expand it to the cubic order. We relegate the details of the in-in analysis into appendix \ref{app-PS} where it is also shown that the dominant corrections are given by the Feynman diagrams Fig. \ref{fig3}, yielding
\begin{eqnarray} \label{PS-R} 
\mathcal{P}_{\mathcal{R} } 
= {\cal P}^{(0)}_{\cal R} \Big(1 + 16 \cosh({4\pi\theta}) \Theta_1(\theta) I N_{k}^{2}\Big) \,,
\end{eqnarray}
where $N_k = -\ln(-k \tau_e)$ is the number of e-folds when the mode of interest $k$ leaves the horizon till end of inflation. In addition,  we have also defined 
\begin{eqnarray} \label{f}
\Theta_1(\theta) \equiv \frac{1}{1+ 16 \theta^2} \frac{\sinh (4 \pi  \theta )}{4\pi\theta} \,,
\end{eqnarray}
so that $\Theta_1(\theta=0)=1$. 

From Eq. (\ref{PS-R}), we can easily obtain the corrections to the spectral index $\Delta  n_s$, induced by vector field entropy and isocurvature modes, as
\begin{equation}\label{ns}
\Delta n_s  = \Delta \frac{d \ln {\cal P}_{\cal R}}{d \ln k} \Big{|}_* = 
32 \cosh(4\pi\theta) \Theta_1(\theta) I N_k \, ,
\end{equation}
where the subscript $*$ represents the time of horizon crossing for the mode of interest $k$. Demanding a nearly scale invariant power spectrum, $\Delta n_s$ should be of the order of the slow-roll parameters. The function $\Theta_1(\theta)$, defined in Eq. (\ref{f}),  has a minimum at $\theta =0 $ so that $\Theta_1(0)=1$ and therefore  $\Theta_1(\theta)\geq1$. On the other hand, the combination $\cosh(4\pi\theta) \Theta_1(\theta) $ is a growing function and for $\theta \gtrsim 1$ it grows exponentially like $e^{8 \pi \theta}$. Demanding the corrections in spectral index to be at the order of $\epsilon \sim 10^{-2}$ (or smaller), requires roughly that $I \cosh(4\pi\theta) \Theta_1(\theta) \lesssim 10^{-5}$ where we considered $N_k = 60$. This justifies our approximation $I \ll1$ and the corresponding perturbative in-in analysis. In addition, this also implies that $\theta$ can not be large either. Indeed, keeping $\Delta n_s$
at the order of slow parameter imposes  $\theta \lesssim 10^{-1}$ and $I \lesssim 10^{-5}$.

In the setup of anisotropic inflation \cite{Watanabe:2009ct}, the parameter $I$ measures the amplitude of quadrupole statistical anisotropy. The  CMB upper bounds on the amplitude of quadrupole anisotropy implies that in that setup $I \lesssim 10^{-7}$ \cite{Kanno:2010ab,Bartolo:2012sd,Emami:2013bk,Abolhasani:2013zya, Emami:2015qjl}. It is also shown that for the small value of $I \lesssim 10^{-7}$, it is not easy to achieve the attractor solution and larger values $10^{-7} \ll I \lesssim 10^{-2}$ are of more interest \cite{Naruko:2014bxa}. This issue was revisited in \cite{Talebian:2019opf} (see also \cite{Fujita:2017lfu}) by taking into account the stochastic effects of electromagnetic fields perturbations. It is shown in \cite{Talebian:2019opf} that in some corner of parameter space, the classical attractor solution  is replaced by a stationary regime of stochastic dynamics such that the conclusion of  \cite{Watanabe:2009ct} is actually consistent.  In our isotropic setup, we have no constraint from quadrupole anisotropy. Demanding a nearly scale invariant power spectrum only requires $I \lesssim 10^{-5}$. However, as we will show in the next section, constraints  from the NG analysis actually requires smaller values of $I$. 

\begin{figure}
	\begin{center}
		\begin{tikzpicture}[line width=1.5 pt, scale=2]
		\draw[fermionnoarrow] (0:1)--(0,0);
		\begin{scope}[shift={(1,0)}]
		\draw[scalarnoarrow,black] (0:1)--(0,0);
		\end{scope}
		\node at (0.5,0.2){$\overline{\delta \sigma}$};
		\node at (1.5,0.2){$\delta s_{\pm} $};
		\node at (1,0.15){\scriptsize $\sqrt{I }$};
		\node at (2,0.15){\scriptsize $\sqrt{I }$};
		\begin{scope}[shift={(2,0)}]
		\draw[fermionnoarrow] (0:1)--(0,0);
		\end{scope}
		\node at (2.5,0.2){$\overline{\delta \sigma}$}; 
		\end{tikzpicture}
		\caption{Feynman diagrams for the leading corrections to the power spectrum of the adiabatic mode. } \label{fig3}
	\end{center}
\end{figure}

Implementing the in-in formula \eqref{IN-IN} this time for the entropy modes 
$\delta{s}_{\pm}$ we can obtain the power spectra and cross-correlations for the original modes $\overline{\delta{s}}$ and $\overline{U}$ through the Eq. (\ref{trans-deltas}). The details of the calculations are shown in the appendix \ref{app-PS} (see Eqs. \eqref{PS-S1S2} and \eqref{PS-Si}) and here we only present the final results.
The power spectra associated with the normalized entropy modes ${{\cal S}_{\pm}} $
are obtained to be 
\begin{eqnarray}\label{PS-Sib}
{\cal P}_{{\cal S}_{\pm}} 
= e^{\pm{4\pi\theta}} \Theta_1(\theta) {\cal P}_{\cal R}^{(0)}
\Big( 1 - \frac{4}{3} \left(7+312\theta^2\right) e^{\pm{4\pi\theta}} \Theta_1(\theta) I N_k \Big) \, .
\end{eqnarray}
From the above results we see that for positive $\theta$, the power spectrum ${\cal P}_{{\cal S}_{+}}$ is amplified exponentially. This is a well known effect that in the presence of the parity violating term, the vector fields perturbations become chiral and one mode is enhanced exponentially compared to other mode \cite{Barnaby:2010vf, Barnaby:2011vw, Barnaby:2011qe, Sorbo:2011rz}. Here, since we have decomposed the perturbations into the adiabatic and entropy modes, the chirality is translated into the enhancement of the power of the entropy mode ${\cal S}_{+}$ compared to other entropy mode ${\cal S}_{-}$.

The power spectrum of the normalized original entropy mode ${\cal S}$ and normalized isocurvature mode $\mathcal{U} = - \frac{{H}}{ \dot \phi} \cos\vartheta\, \Big(\frac{\overline{ U}}{a}\Big)$ turn out to be
\begin{eqnarray}\label{PS-S}
&&{\cal P}_{{\cal S}} = \cosh({4\pi\theta}) \Theta_1(\theta) {\cal P}_{\cal R}^{(0)}
\Big( 1 - \frac{28}{3} \Theta_1(\theta) \Theta_2(\theta) I N_k \Big) \, ,
\\ \label{PS-U}
&&{\cal P}_{{\cal U}} = \cosh({4\pi\theta}) \Theta_1(\theta) {\cal P}_{\cal R}^{(0)}
\Big( 1 - \frac{28}{3} \Theta_1(\theta) \Theta_3(\theta) I N_k \Big) \, ,
\end{eqnarray}
where we have defined
\begin{eqnarray}\label{Theta2}
&&\Theta_2(\theta) \equiv \frac{1-48 \theta^2}{\cosh({4\pi\theta})} + \frac{312}{7}\theta^2 \,, \\ 
\label{Theta3}
&&\Theta_3(\theta) \equiv \frac{ 324\theta^2}{7\cosh({4\pi\theta})} + \Big(1+\frac{312}{7}\theta^2\Big) 
\frac{ \sinh^2(4\pi\theta)}{\cosh({4\pi\theta})} \,,
\end{eqnarray}
so that $\Theta_2(\theta=0) = 1$ and $\Theta_3(\theta=0) = 0$.

Following Refs. \cite{Komatsu:2008hk,Komatsu:2010fb}, we can define the non-adiabaticity parameter
\begin{equation}\label{alpha-pv}
\frac{{\alpha}}{1-\alpha} \equiv \frac{{\cal P}_{{\cal S}}+{\cal P}_{{\cal U}}}{{\cal P}_{{\cal R}}} \approx
2\cosh(4\pi\theta) \Theta_1(\theta)\,.
\end{equation}

Note that $\theta$ is the parity violating parameter which in our analysis controls 
 the entropy perturbations power spectra and their cross correlation with the curvature perturbation. As such,  we can put constraint on the value of $\theta$ from the CMB observational bounds on the entropy perturbations. However, this
also depends on the mechanism of reheating and how the inflaton and the vector fields transfer their energy to various component of SM particles during reheating 
and afterwards.

We now look at the cross correlation between the curvature perturbations and entropy/isocurvature modes which are obtained to be (see Eq. (\ref{PS-SiR}) in the appendix)
\begin{eqnarray}\label{PS-SR}
&&{\mathcal{C}_{ \mathcal{R}\mathcal{S}}}
= -4 \cosh(4\pi\theta) \Theta_1(\theta)
\sqrt{I} N_k \, {\cal P}_{\cal R}^{(0)} \,, \\
&&{\mathcal{C}_{ \mathcal{R}\mathcal{U}}}
= -4 \sinh(4\pi\theta) \Theta_1(\theta)
\sqrt{I} N_k \, {\cal P}_{\cal R}^{(0)} \,.
\end{eqnarray}
From these cross correlations we can define another observable dimensionless parameter \cite{Komatsu:2008hk,Komatsu:2010fb}
\begin{equation}\label{beta-pv}
{\beta} \equiv - \frac{{\mathcal{C}_{ \mathcal{R}\mathcal{S}}}
+{\mathcal{C}_{ \mathcal{R}\mathcal{U}}}}{\sqrt{ ({\cal P}_{{\cal S}}+{\cal P}_{{\cal U}}) {\cal P}_{{\cal R}} }}
\approx 
2\sqrt{2} \frac{\sqrt{ \Theta_1(\theta)}e^{4\pi\theta}}{\sqrt{\cosh(4\pi\theta)}} \sqrt{I} N_k \,.
\end{equation}
From the above result and Eq. (\ref{ns}), we find
\begin{equation}\label{beta-ns}
{\beta}^2 \approx \frac{1}{4} \frac{e^{8\pi\theta}N_k \Delta n_s}{\cosh^2(4\pi\theta)} \,.
\end{equation}
This is an interesting result which shows that, for $\theta \lesssim 10^{-1}$, curvature perturbations are almost uncorrelated with the entropy perturbations independent of the value of $I$.

We see that the cross-correlation between curvature perturbations and isocurvature mode ${\overline U}$ vanishes for $\theta=0$. As we already mentioned, this can be understood if we look at the action (\ref{LSS-sigma-s}) in terms of the original variables from which we see that ${\tilde U}$ can only indirectly interact with $\overline{\delta\sigma}$ through its interaction with entropy mode $\overline{\delta{s}}$ with vertex $\theta$. It decouples from both the curvature and entropy perturbations in the absence of parity violating term. 

Finally, one can also calculate the cross-correlation between the isocurvature mode ${\overline U}$ and entropy mode ${\mathcal{C}_{ \mathcal{S}\mathcal{U}}}$ which is nonzero in the presence of parity violating term and vanishes for $\theta=0$.


\section{Linear tensor perturbations}
\label{tensor-linear}
 
In this section, we study tensor modes to linear order. Besides the usual tensor modes $\gamma_{ij}$ associated with the metric perturbations, we also have tensor perturbations $t_{ij}$ coming from the matter sector which significantly affect the GWs power spectrum. 
 
Expanding the action (\ref{action}) around background configuration (\ref{bac-A}) and (\ref{FRW}) with tensor modes $\gamma_{ij}$ and $t_{ij}$ given in (\ref{perturbations}), the quadratic action for the tensor modes to leading orders in $I$ and $\epsilon$ is (see appendix \ref{quad-tensor} for the details)
given by
\begin{eqnarray}\label{LTT}
&&{L}^{\rm TT} =\frac{1}{2} \int d^3x
\Big[
\overline{\gamma}'_{ij} \overline{\gamma}'^{ij} - \partial_{i} \overline{\gamma}_{jk} \partial^{i} \overline{\gamma}^{j k} + \frac{2}{\tau^2} (1+I \epsilon ) \overline{\gamma}_{ij} \overline{\gamma}^{ij} 
+ \overline{t}'_{ij}  \overline{t}'^{ij} + \frac{2}{\tau^2} \big( 1-\frac{5}{2}I \epsilon \big) 
\overline{t}_{ij} \overline{t}^{ij} \\ \nonumber 
&& \hspace{2.5cm} - \partial_{k} \overline{t}_{ij} \partial^{k} \overline{t}^{ij} 
+ \partial_{j} \overline{t}_{ik} \partial^{k} \overline{t}^{ij}
+\frac{4}{\tau} \sqrt{I \epsilon}\, \overline{\gamma}_{ij} \overline{t}'^{ij}
-\frac{8}{\tau^2} \sqrt{I \epsilon}\, \overline{\gamma}_{ij} \overline{t}^{ij} 
+ \frac{8}{\tau} \theta \epsilon_{ijk} \overline{t}_m{}^{k} \partial^{j} \overline{t}^{im} \Big] ,
\end{eqnarray} 
where $\overline{\gamma}_{ij}$ and $\overline{t}_{ij}$ are the canonically normalized fields defined as
\begin{equation}\label{tensor-canonical0}
\overline{\gamma}_{ij} \equiv \frac{a}{2} \, \gamma_{ij} \,, \hspace{1cm} 
\overline{t}_{ij} \equiv f \, t_{ij} \,.
\end{equation}

We perform the Fourier transformation \eqref{Fourier} and then decompose the amplitudes in Fourier space in terms of circular polarization tensors $e^{\lambda}_{ij}({\bf k})$ as
\begin{eqnarray}\label{polarization}
{\overline \gamma}_{ij} (\tau,{\bf k}) 
= \sum_{+,\times} \overline{\gamma}^{\lambda}(\tau, {\bf k}) e^{\lambda}_{ij}({\bf k})\,, 
\hspace{1cm} {\overline t}_{ij}(\tau,{\bf k}) = \sum_{+,\times} \overline{t}^{\lambda}(\tau, {\bf k}) e^{\lambda}_{ij}({\bf k}) \,,
\end{eqnarray}
in which the traceless and transverse conditions require 
\begin{equation}\label{eij-TT}
e^\lambda_{ii}({\bf k}) = 0 \,, \hspace{1cm} {\bf k}. e^\lambda_{ij} ({\bf k}) = 0 \, .
\end{equation}

In appendix \ref{quad-tensor} we have presented details of calculations of the quadratic action for the tensor modes in terms of the polarization tensors. The action \eqref{LTT} in terms of the polarization tensors takes the following form
\begin{eqnarray}\label{S2-T-pol}
&&S^{\rm TT} =  \sum_{\lambda} \int d^3 k d\tau \Big[ 
\big( \overline{\gamma}'^\lambda \big)^2 
- \Big( k^2 - \frac{2}{\tau^2} ( 1 +  I \epsilon ) \Big) \big( \overline{\gamma}^\lambda \big)^2
- \frac{4}{\tau^2} \sqrt{I\epsilon} 
\big( 2 \overline{t}^\lambda - \tau\overline{t}'^\lambda \big) \overline{\gamma}^{\lambda} \nonumber \\ 
&& \hspace{3.7cm} + \big( \overline{t}'^\lambda \big)^2 
- \Big( k^2 + \frac{8}{\tau} \lambda k\theta
- \frac{2}{\tau^2} \big( 1 - \frac{5}{2} I \epsilon \big) \Big) 
\big( \overline{t}^\lambda \big)^2
\Big] \,.
\end{eqnarray}
Note that the value of $\lambda$ in the above expression is $+1$  ($-1$) for $+$ ( $\times$ ) polarizations respectively so we deal with four perturbations $\overline{\gamma}^\lambda$ and $\overline{t}^\lambda$ for $\lambda = +,\times$ which are the four real physical degrees of freedom.

From Eq. (\ref{S2-T-pol}) we see that the gravitational tensor perturbations $ {\gamma}_{ij}$ and the vector fields tensor modes ${t}_{ij}$ are directly coupled through a exchange vertex $I$ while there is no  coupling between them via vertex $\theta$. Thus, the power spectra of $ {\gamma}_{ij}$ cannot receive pure $\theta$ corrections without the factor $I$. On the other hand, different polarizations of ${t}_{ij}$ are coupled to each other through the exchange vertex interaction $\theta$ which leads to parity violating correction to the power spectra of ${t}_{ij}$ as we will show below. From Eq. (\ref{S2-T-pol}), we also see that all quadratic non-diagonal terms can be treated perturbatively since $I$ is small. In the case of pure spectator gauge fields (which should be compared to the case of $I=0$ in our model), this is not always the case \cite{Dimastrogiovanni:2016fuu,Fujita:2018vmv,Dimastrogiovanni:2018xnn}. In addition, in previous section we have shown that demanding a nearly scale invariant curvature perturbation  power spectrum requires $\theta$ should be somewhat small. Therefore, the terms including $\theta$ can be treated perturbatively as well. However, comparing the quadratic actions (\ref{S2-sigma-s-pm}) and (\ref{S2-T-pol}), we see that the free wave functions for the different polarizations of the tensor modes ${\overline t}^+$ and ${\overline t}^\times$ have exactly the same functional forms as entropy modes $\delta{s}_+$ and $\delta{s}_-$ respectively. Therefore, we calculate the effects of $\theta$ non-perturbatively and to all orders though we know that it is a small parameter.

The free part of the action  is defined as $S^{\rm TT}|_{I=0} \equiv \int d\tau { L}_0^{\rm TT}$ where ${ L}_0^{\rm TT}$ is the corresponding free Lagrangian. The quadratic interaction Lagrangian will be $\delta{ L}^{\rm TT} \equiv { L}^{\rm TT}-{ L}_0^{\rm TT}$, where ${ L}^{\rm TT}$ is the total quadratic Lagrangian associated with the action (\ref{S2-T-pol}), yielding 
\begin{eqnarray}\label{L2-int-T}
\delta{ L}^{\rm TT}= - \frac{2}{\tau^2} \sqrt{I\epsilon} \sum_{\lambda} \int d^3k
\Big[ 
2 (2 \overline{t}^{\lambda}_k - \tau \overline{t}'^{\lambda}_k) \overline{\gamma}^{\lambda }_k 
+ \sqrt{I\epsilon} \Big( \big(\overline{\gamma}^{\lambda}_k\big)^2
- \frac{5}{2} \big(\overline{t}^{\lambda}_k\big)^2 \Big) \Big] \,.
\end{eqnarray}
It vanishes for $I=0$ by construction. 

\subsection{Power spectra and cross correlations}

In this subsection, we obtain all two-point correlation functions of the  tensor modes and  their cross correlations. 

The dynamics of the free modes of $\overline{\gamma}^{\lambda}(\tau,{\bf k})$ and $\overline{t}^{\lambda}(\tau,{\bf k})$ determine by the free Lagrangian ${ L}_0^{\rm TT}$ and the quantization go as usual. We expand the tensor modes $\overline{\gamma}^{\lambda}(\tau,{\bf k})$ and $\overline{t}^{\lambda}(\tau,{\bf k})$ in terms of the creation and annihilation operators as 
\begin{equation}\label{T-op}
\overline{\gamma}^{\lambda}(\tau,{\bf k}) =  \overline{\gamma}^{\lambda}_k(\tau) a^{\lambda\,}_{\textbf k} + 
\overline{\gamma}^{\lambda*}_k(\tau) a^{\dagger\lambda}_{\textbf{-k}} \,, \hspace{1cm}
\overline{t}^{\lambda}(\tau,\textbf{k}) = \overline{t}^{\lambda}_k(\tau) b^{\lambda}_{\textbf{k}} + \overline{t}^{\lambda *}_k(\tau) b_{-\textbf{k}}^{\dagger\lambda} \,,
\end{equation}
where $a^{\lambda}_{\textbf{k}}$ and $b^{\lambda}_{\textbf{k}} $ are independent operators satisfying the usual commutation relations with non-vanishing commutators $[a^{\lambda}_{\textbf k},a^{\dagger\lambda'}_{-\textbf{k}'}]= \delta^{\lambda\lambda'} \delta(\textbf{k}+\textbf{k}')$ and $[b^{\lambda}_{\textbf k},b^{\dagger\lambda'}_{-\textbf{k}'}]= \delta^{\lambda\lambda'} \delta(\textbf{k}+\textbf{k}')$. Substituting the above relations into the free part of the action $S^{\rm TT}|_{I=0}$ given in Eq. (\ref{S2-T-pol}), we find the equations of motion of the mode functions
\begin{eqnarray}\label{MF-tensors}
\overline{\gamma}^{{\lambda}''}_k + \Big(k^2 - \frac{2}{\tau^2}\Big) \overline{\gamma}^{\lambda}_k = 0 \,, 
\hspace{1cm}
\overline{t}^{{\lambda}''}_k + \Big(k^2 + 8 \lambda \frac{k\theta }{\tau} 
- \frac{2}{\tau^2}\Big) \overline{t}^{\lambda}_k = 0 \,.
\end{eqnarray}

Comparing the equations of different polarizations of  $\overline{t}^{\lambda}$ above with the equations of motion of entropy modes $\delta s_{\pm}$ given in Eq. (\ref{s-pm-MF}) we see that they are exactly the same so that we can identify $\overline{t}^{+}$ and $\overline{t}^{-}$ with $\delta{s}_+$ and $\delta{s}_-$ respectively. We already have found the solution for $\delta{s}_{\pm}$ in Eq. (\ref{MF-deltas1}) so we simply use them here. The wave function for the gravitational tensor modes $\overline{\gamma}^{\lambda}$ have the standard form so we have  the following positive frequency Bunch-Davies wave functions for them, 
\begin{eqnarray}\label{T-MF}
\overline{\gamma}^{\lambda}_k(\tau) = \frac{e^{-i k \tau}}{\sqrt{2 k}} \Big( 1 - \frac{i}{k\tau} \Big) \,,
\hspace{1cm}
\overline{t}^{\lambda}_k(\tau) = \frac{e^{2\pi\lambda\theta}}{\sqrt{2k}} W(-4i\lambda\theta,3/2,2ik\tau) \,.
\end{eqnarray}

In the absence of net polarizations, we define the power spectra for the different polarizations of the gravitational tensor modes $\mathcal{P}^\lambda_{\gamma}$ as
\begin{eqnarray}\label{CF-hh}
\big\langle \gamma^{\lambda}(\tau_e,{\bf k})\, {\gamma}^{\lambda'}(\tau_e,{\bf k}') 
\big\rangle \equiv \frac{2 \pi^2}{k^{3}} 
\mathcal{P}^\lambda_{\gamma} \, \delta_{\lambda\lambda'} \, (2\pi)^3 \delta^{(3)}({\bf k}-{\bf k}') \,.
\end{eqnarray}

The leading corrections from the vector field tensor modes to the power spectra of the different polarizations of the gravitational tensor modes are given by the Fig \ref{fig5}. The details of the in-in analysis are presented in appendix \ref{app-PS}
where it is shown that 
\begin{figure}
	\begin{center}
		\begin{tikzpicture}[line width=0.5 pt, scale=2]
		\draw[vector,black] (0:1)--(0,0);
		\begin{scope}[shift={(1,0)}]
		\draw[gluon,black,line width=0.5 pt] (0:1)--(0,0);
		\end{scope}
		\node at (0.5,0.2){$\overline{\gamma}^{\lambda}$};
		\node at (1.5,0.2){$\overline{t}^{\lambda} $};
		\node at (2.5,0.2){$\overline{\gamma}^{\lambda} $};
		\node at (1,0.15){\scriptsize $\sqrt{I \epsilon }$};
		\node at (2,0.15){\scriptsize $\sqrt{I \epsilon }$};
		\begin{scope}[shift={(2,0)}]
		\draw[vector,black] (0:1)--(0,0);
		\end{scope}
		\end{tikzpicture}
		\caption{Feynman diagrams for the leading corrections to the power spectrum of GWs.} \label{fig5}
	\end{center}
\end{figure}
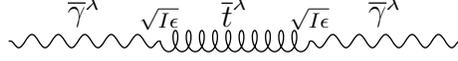
\begin{equation}\label{PS-hh-pol}
\mathcal{P}^\lambda_{\gamma} = 8\mathcal{P}_{\gamma}^{(0)} 
e^{4 \pi \lambda \theta} \Theta_{1}(\theta)I \epsilon N_{k}^{2} \,,
\end{equation}
where $\mathcal{P}_{\gamma}^{(0)} \equiv \frac{2H^2}{\pi^2}$ is the standard tensor power spectrum for GWs in the absence of the interactions with the vector fields tensor modes.

The total power spectrum of GWs to leading order in $I$ then becomes 
\begin{eqnarray}\label{PS-hh-f}
\mathcal{P}_{\gamma} = \sum_{+,\times}\mathcal{P}^\lambda_{\gamma} 
= \mathcal{P}_{\gamma}^{(0)} \Big(1+16 \cosh(4 \pi \theta) \Theta_{1}(\theta)I \epsilon N_{k}^{2} \Big) \,.
\end{eqnarray}
We see that the corrections from vector fields in GWs power spectrum is proportional to $I \epsilon N_k^2 $. For $I \lesssim 10^{-5}, \theta \lesssim 10^{-1}$ and $\epsilon \sim 10^{-2}$ the corrections in tensor power is around few percents.

Because of the  parity violating term the GWs power spectrum is chiral with
\begin{equation}\label{GWs-pol}
\mathcal{P}^+_{\gamma}-\mathcal{P}^\times_{\gamma}=16\mathcal{P}_{\gamma}^{(0)} 
\sinh(4 \pi \theta) \Theta_{1}(\theta)I \epsilon N_{k}^{2} \,.
\end{equation}
In the absence of the parity violating term $\theta=0$, Eq. \eqref{GWs-pol} vanishes and there is no chirality in  GWs while the total power spectrum \eqref{PS-hh-f} still 
receives unpolarized modifications from $t_{ij}$ modes 
proportional to $I$. 

From the power spectrum of the curvature perturbation in Eq. (\ref{PS-R}) and the power spectrum of the GWs in Eq. (\ref{PS-hh-f}), the tensor to scalar ratio turns out to be 
\begin{eqnarray}\label{r-eq}
r = \frac{\mathcal{P}_{\gamma}}{\mathcal{P}_{\cal R}} = 16 \epsilon \bigg( 
\frac{1+16 \cosh(4 \pi \theta) \Theta_{1}(\theta) I \epsilon N_{k}^{2}}{
1+16 \cosh(4 \pi \theta) \Theta_{1}(\theta)I N_{k}^{2}} \bigg) \, .
\end{eqnarray}
Unlike the models of inflation based on scalar field dynamics, the parameter $r$ 
may not  uniquely determine the scale of inflation as there are non-trivial contributions from the parameters $I$ and $\theta$ from the vector fields dynamics. Due to their different natures, the contributions of the scalar and tensor sectors can be disentangled from the CMB observations. The  contribution from the scalar sector, encoded in the total curvature perturbations power spectrum $\mathcal{P}_{\cal R}$,  is independently fixed by the COBE normalization 
while it is possible to separate the tensor power spectrum \eqref{PS-hh-f} into the polarized \eqref{GWs-pol} and unpolarized parts and then look for their  observational features \cite{Thorne:2017jft}.
 
The corrections to the tilt of GWs power spectrum induced by the vector fields  is given by
\begin{equation}\label{nt}
\Delta n_t  = \Delta \frac{d \ln {\cal P}_{\gamma}}{d \ln k}\big |_* = 
32 \cosh(4 \pi \theta) \Theta_{1}(\theta) I \epsilon N_k = \epsilon\, \Delta n_s \,,
\end{equation}
where Eq. (\ref{ns}) has been used in the last step. Since $\Delta n_s$ is of the order of the slow-roll parameters and also its is negative (the power spectrum of curvature perturbations has red tilt), tensor power spectrum in our scenario has very small red tilt such that $\Delta n_t={\cal O}(\epsilon^2)$. 

The dominant contributions to the power spectrum of the vector fields tensor modes are given by the Feynman diagram shown in Fig. \ref{fig6}. In appendix \ref{app-PS} we have calculated these contributions, yielding  the following result for the power spectrum of the vector field tensor modes
\begin{equation}\label{PS-t}
\mathcal{P}_{ t} = \mathcal{P}_{\gamma}^{(0)} \Big(1-8\cosh(8 \pi \theta)   (1+48 \theta^2)\Theta_{1}(\theta)^2 I \epsilon N_{k} \Big) \,.
\end{equation}
In particular,  we see that the terms containing $I\epsilon N_k^2$ cancel one another and therefore there is no $I\epsilon N_k^2$ correction while we have the sub-leading corrections of the order of $I\epsilon N_k$. For $I \sim 10^{-5}$ and $\theta \gtrsim 10^{-1}$ the contributions of the parity violating term is somewhat larger than those from the  parameter $I$.
\begin{figure}
\begin{center}
\begin{tikzpicture}[line width=0.5 pt, scale=2]
\draw[gluon,black,line width=0.5 pt] (0:1)--(0,0);
\begin{scope}[shift={(1,0)}]
\draw[vector,black] (0:1)--(0,0);
\end{scope}
\node at (0.5,0.2){$\overline{t}^{\lambda}$};
\node at (1.5,0.2){$\overline{\gamma}^{\lambda} $};
\node at (2.5,0.2){$\overline{t}^{\lambda} $};
\node at (1,0.15){\scriptsize $\sqrt{I \epsilon }$};
\node at (2,0.15){\scriptsize $\sqrt{I \epsilon }$};
\begin{scope}[shift={(2,0)}]
\draw[gluon,black,line width=0.5 pt] (0:1)--(0,0);
\end{scope}

\end{tikzpicture}
\caption{Feynman diagrams for the corrections to the power spectrum of the vector fields tensor modes. } \label{fig6}
\end{center}
\end{figure}
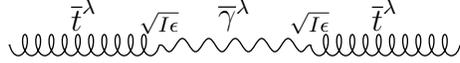

Finally, we obtain the cross-correlation between the GWs and vector fields tensor modes. Since the two different types of tensor modes are uncorrelated, there is no
zeroth order cross-correlation $\langle \overline{\gamma^{\lambda}}\, \overline{t^{\lambda}}\rangle$ while  the leading contribution from the integral like Eq. (\ref{IN-IN}) yields (see Eq. \eqref{PS-ht-pol})
\begin{eqnarray}\label{PS-ht-f} 
\mathcal{C}_{\gamma\,  t}= 4 \mathcal{P}_{\gamma}^{(0)} \cosh(4 \pi \theta) \Theta_{1}(\theta)\sqrt{I \epsilon} N_{k} \,.
\end{eqnarray}

In a sense, the tensor modes associated to the vector fields  are the same as entropy modes in the scalar sector, and in analogy with Eqs. (\ref{alpha-pv}) and (\ref{beta-pv}), we can define the following dimensionless quantities
\begin{eqnarray}\label{alpha-t}
{\alpha}_t &\equiv& \frac{{\cal P}_{t}}{{\cal P}_{t} + {\cal P}_{\gamma}} \approx
\frac{1}{2} \Big( 1- 8 \cosh(4 \pi \theta) \Theta_{1}(\theta) I \epsilon N_k^2 \Big) \,, \\ 
{\beta}_{\gamma{t}} &\equiv& - \frac{{\mathcal{C}_{\gamma{t}}}}{
\sqrt{ {\cal P}_{t} {\cal P}_{\gamma} }} = - 4 \sqrt{\cosh(4 \pi \theta) \Theta_{1}(\theta)I\epsilon} N_k \,.
\end{eqnarray}

From the above result and Eq. (\ref{nt}) we find ${\beta}_{\gamma{t}} ^2 = \frac{1}{2} N_k \Delta n_t$, independent of the value of $I$ and $\theta$. This result is the tensorial counterpart of Eq. (\ref{beta-ns}) which shows that GWs are almost uncorrelated from the vector fields tensor modes. Beside their corrections to GWs power spectrum in (\ref{PS-hh-f}), and depending on the mechanism of reheating, 
the tensor perturbations associated with the vector modes may contribute to the effective number of relativistic degrees of freedom as a dark spin two particles. The effective number of relativistic degrees of freedom is severely constrained \cite{Aghanim:2018eyx}, imposing an upper bound on the energy density of the tensor modes of vector fields after inflation. 


\section{Primordial non-Gaussianities}\label{NG}

Having studied the linear perturbations in  previous sections, in this section we study  NG for the curvature perturbations $\langle {\cal RRR} \rangle$, GWs $\langle\gamma\gamma\gamma\rangle$, and also the three-point cross correlations $\langle {\cal RR} \gamma \rangle$ and $\langle {\cal R} \gamma \gamma \rangle$ between curvature perturbations and GWs. 

It is well known that NG of the curvature perturbations are enhanced in multiple fields scenarios \cite{Wands:2002bn,Seery:2005gb,Bassett:2005xm,Sasaki:2006kq,Langlois:2008qf,Chen:2010xka}. Since our model is in essence a multiple fields setup, we expect non-trivial NG in our model, both in scalar and tensor sectors.
The scalar modes associated with the vector fields  behave as mediator particles and enhance scalar NG accordingly \cite{Chen:2009zp}. Similarly, the tensor modes associated with the  vector field ($t_{ij}$) would enhance NG of GWs.  

As we will explicitly show in this section, all bispectra in our setup peak in the squeezed limit. Therefore, for the three-point function of mode $X$ 
we employ the following parameterization based on the local shape  NG,
\begin{equation}\label{fNL}
\langle X({\bf k}_1) X({\bf k}_2) X({\bf k}_3) \rangle \equiv 
(2\pi)^7 \delta^3({\bf k}_1+{\bf k}_2+{\bf k}_3)
\frac{\sum_i k_i^3}{\Pi_i k_i^3} \Big(\frac{3}{10} f^{X}_{\rm NL} \Big) {\mathcal P}_{X}^2 \,,
\end{equation}
where ${\mathcal P}_{X}$ is the dimensionless power spectrum of the mode $X$ and the dimensionless quantity $f_{\rm NL}^{X}$ characterizes the amplitude of the three-point functions which may be constrained from the cosmological observations. All the external legs are computed at the time of end of inflation $\tau_e$ and, 
from now on, for the sake of simplicity of notation, we do not explicitly show their time dependence. One may also define the dimensionless quantity ${\mathcal S}_{X}(k_1,k_2,k_3) \equiv 
\frac{\sum_i k_i^3}{\Pi_i k_i} \Big(\frac{3}{10} f^{X}_{\rm NL} \Big)$ which determines the shape and running of the three-point function of the mode $X$. As we mentioned above, in our setup all NGs peak in the squeezed limit so we can use $f_{\rm NL}^{X}$ to constrain the free parameters of the model such as $\theta$ and $I$.

In order to find the bispectra we need the cubic actions associated with Eq. (\ref{action}) constructed from the leading interaction terms. In appendix \ref{app-A}, we have computed the cubic interaction Lagrangians of the form scalar-scalar-scalar $\delta{ L}^{\rm SSS}$ in Eq. (\ref{cubic-LSSS}), scalar-scalar-tensor $\delta{ L}^{\rm SST}$ in Eq. (\ref{cubic-LSST}), scalar-tensor-tensor $\delta{ L}^{\rm STT}$ in Eq. (\ref{cubic-LSTT}), and tensor-tensor-tensor $\delta{ L}^{\rm TTT}$ in Eq. (\ref{cubic-LTTT})  to  leading orders  in slow-roll parameter $\epsilon$ and the parameter $I$. The corresponding cubic interaction Hamiltonians in interaction picture $\delta{ H}^{\rm SSS}_{I}$, $\delta{ H}^{\rm SST}_{I}$, $\delta{ H}^{\rm STT}_{I}$, and $\delta{ H}^{\rm TTT}_{I}$ are then obtained in the appendix \ref{app-B} in Eqs. (\ref{H-int-SSS}), (\ref{H-int-SST}), (\ref{H-int-STT}), and (\ref{H-int-TTT}) respectively. Having all interaction Hamiltonians in hand, we can calculate various three-point correlations. 


\subsection{Curvature perturbations $ \langle {\cal RRR} \rangle$}

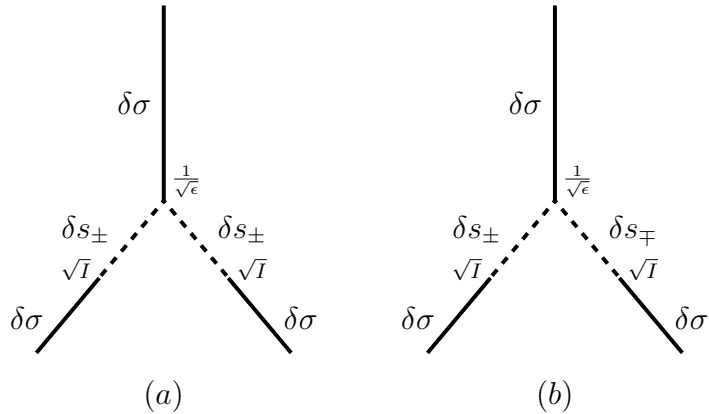
\begin{figure}
	\begin{center}
		\begin{tikzpicture}[line width=1.5 pt, scale=1.3]
		\begin{scope}[rotate=90]
		\draw[scalarnoarrow,black] (-140:1)--(0,0);
		\draw[scalarnoarrow,black] (140:1)--(0,0);
		\draw[fermionnoarrow] (0:2)--(0,0);
		\node at (-1.2,-1.4) {$\delta \sigma$};
		\node at (-1.2,1.4) {$\delta \sigma$};
		\node at (-0.3,0.8) {$\delta s_{\pm}$};
		\node at (-0.3,-0.8) {$\delta s_{\pm}$};
		\node at (1,.3)  {$\delta \sigma$};	
		\node at (-2,0)  {$(a)$};	
		\node at (-0.7,0.9) {\scriptsize$\sqrt{I}$};
		\node at (-0.7,-0.9) {\scriptsize$\sqrt{I}$};
		\node at (0.2,-0.23) {\scriptsize$\frac{1}{\sqrt{\epsilon}}$};
		\begin{scope}[shift={(-0.79,0.66)}]
		\draw[fermionnoarrow] (140:1)--(0,0);
		\end{scope}
		\begin{scope}[shift={(-0.79,-0.66)}]
		\draw[fermionnoarrow] (-140:1)--(0,0);
		\end{scope}
		\end{scope}
		\begin{scope}[shift={(4,0)}]
		\begin{scope}[rotate=90]
		\draw[scalarnoarrow,black] (-140:1)--(0,0);
		\draw[scalarnoarrow,black] (140:1)--(0,0);
		\draw[fermionnoarrow] (0:2)--(0,0);
		\node at (-1.2,-1.4) {$\delta \sigma$};
		\node at (-1.2,1.4) {$\delta \sigma$};
		\node at (-0.3,0.8) {$\delta s_{\pm}$};
		\node at (-0.3,-0.8) {$\delta s_{\mp}$};
		\node at (1,.3)  {$\delta \sigma$};	
		\node at (-2,0)  {$(b)$};	
		\node at (-0.7,0.9) {\scriptsize$\sqrt{I}$};
		\node at (-0.7,-0.9) {\scriptsize$\sqrt{I}$};
		\node at (0.2,-0.23) {\scriptsize$\frac{1}{\sqrt{\epsilon}}$};
		\begin{scope}[shift={(-0.79,0.66)}]
		\draw[fermionnoarrow] (140:1)--(0,0);
		\end{scope}
		\begin{scope}[shift={(-0.79,-0.66)}]
		\draw[fermionnoarrow] (-140:1)--(0,0);
		\end{scope}
		\end{scope}
		\end{scope}
		\end{tikzpicture}
		\caption{The  three-vertex diagrams for the NG of the curvature perturbations. These diagrams give the dominant contribution to the scalar NG. 
		} \label{fig88}
	\end{center}
\end{figure}

There are different contributions to the NG of the curvature perturbations. In appendix \ref{app-RRR}, we have shown that the dominant contributions are given by the three-vertex Feynman diagrams Fig. \ref{fig88}, leading to the following result
\begin{equation}\label{RRR-3V}
\langle \mathcal{R}(\textbf{k}_{1})\mathcal{R}(\textbf{k}_{2})\mathcal{R}(\textbf{k}_{3})\rangle|_{\rm Fig. \ref{fig88}}
= \frac{ 3 H^4 }{2 \epsilon^2} (1+9 \cosh(8 \pi \theta)) \Theta_1(\theta)^{2} \, I N_K^3
\frac{\sum_i k_{i}^3}{\Pi_ik_{i}^3} \, 
(2\pi)^3 \delta^{3}(\textbf{k}_{1}+\textbf{k}_{2}+\textbf{k}_{3}) \,,
\end{equation}
where $K \equiv \frac{1}{3}(k_1+k_2+k_3)$ is a reference momentum and $N_K=-\ln(-K\tau_e)$ is the number of e-folds associated to $K$ till the end of inflation.
Comparing Eq. (\ref{RRR-3V}) with the definition (\ref{fNL}), we see that curvature perturbations NG has the local shape and $f_{\rm NL}^{\cal R}$ is obtained to be 
\begin{equation}\label{fNL-RRR}
f_{\rm NL} = 20 (1+9 \cosh(8 \pi \theta))\Theta_1(\theta)^{2} I N_{K}^3 \,,
\end{equation}
where have dropped the superscript ${\cal R} $ in this case to keep the notation simple. From the linear perturbations analysis  we know that both parameters 
 $I$ and $\theta$ are small. Expanding the above result for small $\theta$ we then find
\begin{equation}\label{fNL-RRR-expand}
f_{\rm NL} \simeq 200 I N_{K}^3 \Big[ 1 + \Big( \frac{512}{15} \pi^2-32 \Big) \theta^2 \Big] \,.
\end{equation}

The result (\ref{fNL-RRR}) is interesting since it puts stronger 
constraints on the model parameter $I$ than the power spectrum does. Taking $N_K\sim 60$ and $f_{\rm NL}\sim 1-10$ from the observational bound
on local type non-Gaussianity, we find $I={\cal O}(10^{-7})$. This is about two orders of magnitude stronger than the bound $I={\cal O}(10^{-5})$  obtained from the spectral tilt of curvature perturbations power spectrum Eq. (\ref{ns}). 

We note that in power spectrum of curvature perturbations \eqref{PS-R}, we should have $I N_k^2\ll1$ to keep the setup perturbative which for $I={\cal O}(10^{-7})$ gives the upper bound $N_k\ll {\cal O}(10^{3})$ on the number of e-folds. Of course, we can consider very small values for $I$ to have larger values for $N_k$. Similarly, if we demand $f_{\rm NL}<100$, we find stronger upper bound $N_K\ll {\cal O}(10^{2})$ for $I={\cal O}(10^{-7})$. Thus, to keep our setup perturbative and assuming $I$ to be not very small $I={\cal O}(10^{-7})$, we find an upper bound on the number of e-folds in our scenario.

Moreover, based on the calculations in this subsection and also appendix \ref{app-RRR}, we can estimate the order of magnitude of the trispectrum for the curvature perturbations. Looking at the corresponding Feynman diagrams, we find that the dominant contribution is given by $g_{\rm NL} \propto I N_{k_*}^4$, where $g_{\rm NL}$ is the amplitude of the trispectrum and $k_*$ is a typical momentum which will be determined by a combination of the momenta of the external legs (see also Ref. \cite{Abolhasani:2013zya}). From this estimation we find that analysis of bispectra may put even stronger bound on the parameters $I$ and $\theta$ which is beyond the scope of this paper.

\subsection{Gravitational waves $\langle \gamma\gamma\gamma\rangle$}

To calculate NG for GWs, we need to find $\langle\gamma^{\lambda_1}\gamma^{\lambda_2}\gamma^{\lambda_3}\rangle$ where $\lambda_i$ can be either $+$ or $\times$ polarizations. We thus need to compute one by one all non-vanishing three-point functions for example $\langle \gamma^{+}\gamma^{+}\gamma^{+}\rangle$, $\langle\gamma^{+}\gamma^{+}\gamma^{-}\rangle$, and so on. However, here we only present the details of the calculations for one case which is enough for our purpose to estimate the order of magnitude of the GWs NG. The calculations for  other cases are similar and straightforward. 

The dominant contributions for the three-point functions $\langle \gamma^{\lambda}\gamma^{\lambda}\gamma^{\lambda}\rangle$ come from the three-vertex Feynman diagrams that are shown in Fig \ref{fig16}. It is straightforward to show that the contribution coming from this diagram yields 
\begin{eqnarray}\label{GGG}
&& \langle \gamma^{\lambda}(\textbf{k}_{1})\gamma^{\lambda} ({\textbf{k}_{2}}) \gamma^{\lambda}({\textbf{k}_{3}})\rangle|_{\rm Fig. (\ref{fig16})}
=- 192  I \epsilon H^4 N_K^3 e^{8 \pi \lambda \theta} \Theta_{1}(\theta)^2 
\nonumber \\ \nonumber &&\hspace{5cm} 
\times e_{ij}^{\lambda}(\textbf{k}_{1}) e_{mj}^{\lambda}(\textbf{k}_{2})
e_{mi}^{\lambda}(\textbf{k}_{3})
\frac{\sum_i k_i^3}{\Pi_{i}k_i^3}
(2 \pi)^3 \delta^{3}(\textbf{k}_{1}+\textbf{k}_{2}+\textbf{k}_{3}) \,,
\end{eqnarray}
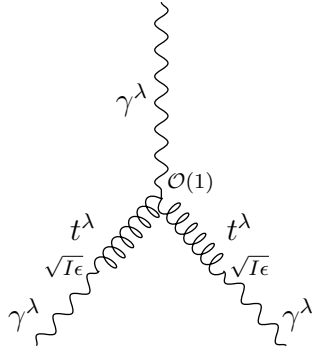
\begin{figure}
	\begin{center}
		\begin{tikzpicture}[line width=1.5 pt, scale=1.3]
		\begin{scope}[rotate=90]
		\draw[gluon,black,line width=0.5 pt] (-140:1)--(0,0);
		\draw[gluon,black,line width=0.5 pt] (140:1)--(0,0);
		\draw[vector,black,line width=0.5 pt] (0:2)--(0,0);
		\node at (1,.3)  {$\gamma^{\lambda}$};
		\node at (-0.7,-0.9) {\scriptsize $\sqrt{I \epsilon}$};
		\node at (-0.7,1) {\scriptsize $\sqrt{I \epsilon}$};
		\node at (-1.2,1.4) {$\gamma^{\lambda}$};
		\node at (-0.3,0.8) {$t^{\lambda}$};
		\node at (-1.2,-1.4) {$\gamma^{\lambda}$};
		\node at (-0.3,-0.8) {$t^{\lambda}$};
		\node at (0.15,-0.3)  {\scriptsize$\mathcal{O}(1)$};
		\begin{scope}[shift={(-0.75,0.635)}]
		\draw[vector,black,line width=0.5 pt] (140:1)--(0,0);
		\end{scope}
		\begin{scope}[shift={(-0.75,-0.635)}]
		\draw[vector,black,line width=0.5 pt] (-140:1)--(0,0);
		\end{scope}
		\end{scope}
		\end{tikzpicture}
		\caption{ Dominant diagrams for the NG $\langle \gamma^{\lambda}(\textbf{k}_{1})
			\gamma^{\lambda} ({\textbf{k}_{2}}) \gamma^{\lambda}({\textbf{k}_{3}})\rangle$. }
		\label{fig16}
	\end{center}
\end{figure}
where taking the same $\lambda$ in both sides means that it is only applicable for two cases of $\langle \gamma^{+}\gamma^{+}\gamma^{+}\rangle$ and $\langle \gamma^{-}\gamma^{-}\gamma^{-}\rangle$. We keep this notation throughout this paper. First of all we see that $\langle \gamma^{+}\gamma^{+}\gamma^{+}\rangle\neq\langle \gamma^{-}\gamma^{-}\gamma^{-}\rangle$ which is the direct feature of the parity violating interaction. In the absence of the  parity violating  interaction ($\theta=0$) these two three-point functions coincide as a result of the parity symmetry. We also see that $\langle \gamma^{+}\gamma^{+}\gamma^{+}\rangle$ is exponentially enhanced compared to $\langle \gamma^{-}\gamma^{-}\gamma^{-}\rangle$. From now on, we only focus on the case of $\langle \gamma^{+}\gamma^{+}\gamma^{+}\rangle$ to estimate the order of magnitude for the GWs NG. In this case, after substituting from Eq. (\ref{eij-cubic-3}) for the contractions of  the products of three polarization tensors, and comparing the result with definition (\ref{fNL}), we find 
\begin{equation}\label{fNL-GGG}
f^{+++}_{\rm NL} 
= \frac{5}{32} e^{8 \pi \theta} \Theta_{1}(\theta)^2 I \epsilon N_{K}^3 
\frac{\big( 1-(x_2+x_3)^2 \big) \big( 1-(x_2-x_3)^2 \big) (1+x_2+x_3)^2}{x_2^2 x_3^2} \,,
\end{equation}
where we have normalized the momenta $k_2$ and $k_3$ with respect to the momentum $k_1$ as follows
\begin{equation}\label{x-def}
x_2 \equiv \frac{k_2}{k_1} \,, \hspace{1cm} x_3 \equiv \frac{k_3}{k_1} \,.
\end{equation}

The dimensionless quantity $f^{\gamma}_{\rm NL} $ determines the amplitude of the NG for GWs. It peaks in the squeezed limit of $k_3 \ll k_2 \approx k_1$ or equivalently $x_2\to 1$ and $x_3\to 0$, yielding 
\begin{equation}\label{fNL-gamma}
f^{+++}_{{\rm NL},{\rm sq}} \approx - \frac{5}{2} I \epsilon N_{K}^3 
( 1 + 8 \pi \theta ) \,.
\end{equation}

We have only considered the case of all $+$ polarizations while we know that 
the contributions from different polarizations  to the total three-point functions (which is the sum over all polarizations) are of the order of $I \epsilon N_{K}^3$. Therefore, the order of magnitude of the total three-point function is  $f^{\gamma}_{{\rm NL},{\rm sq}}\propto {I} \epsilon N_{K}^3 $. Comparing the above result with Eq. (\ref{fNL-RRR-expand}) we see that $f^{\gamma}_{\rm NL} $ is smaller than $ f_{\rm NL}$  by a factor of slow-roll parameter $\epsilon$. However, $f^{\gamma}_{{\rm NL},{\rm loc}}$ is larger than its counterpart coming from the gravitational vacuum fluctuations \cite{Maldacena:2002vr,Maldacena:2011nz,Soda:2011am,Bordin:2020eui}. A similar result is obtained if one considers some spectator gauge fields (correspond to $I=0$ in our case) \cite{Agrawal:2018mrg}. However, the mechanism of  enhancement of $f^{\gamma}_{\rm NL} $ is different here i.e. the NG for the pure spectator gauge fields peaks in equilateral limit while in our model with $I\neq0$ it peaks in the squeezed limit.


\subsection{Mixed bispectra $\langle{\cal RR}\gamma\rangle$ and $\langle{\cal R}\gamma\gamma\rangle$}

In this subsection, we calculate the  mixed NG between curvature perturbations and GWs. From the results in previous subsections we expect that these types of three-point functions to be enhanced as well. 

We first look at the  three-point cross correlation between two curvature perturbations and one gravitational tensor mode. The dominant Feynman diagram is shown in Fig. \ref{figSST}. The Hamiltonian interaction responsible for the cubic vertex in this diagram is given by Eq.  (\ref{H-int-SST}) which is 
$\delta{H}_{I}^{\rm SST} = -\frac{4\theta}{\sqrt{\epsilon}} \tau^3 H \int d^3x(\overline{\delta\sigma}
+\sqrt{I}\,\overline{\delta s}) (\tau^{-2}{\overline t}_{ij})' \partial^i\partial^j \tilde{U}$. Rewriting these interactions in terms of $\delta{s}_{\pm}$ from Eq. (\ref{trans-deltas}) and then going to the Fourier space, it is straightforward to find
\begin{eqnarray}\label{NG-RRT}
&& \langle \mathcal{R}(\textbf{k}_{1})\mathcal{R}(\textbf{k}_{2})
\gamma^{\lambda}(\textbf{k}_{3})\rangle|_{\rm Fig. \ref{figSST}}
=12 I\theta^2 \frac{H^4}{\epsilon} N_K^2 
e^{4 \pi \lambda \theta} \Theta_{1}(\theta)^2 
\Big( x_2 \cosh(4 \pi \theta) + \frac{5}{9} \lambda x_{3}^2 
\sinh(4 \pi \theta)\Big)
 \\ \nonumber
&& \hspace{5cm} \times 
\frac{1+x_2}{x_2 x_3^2} \Big(\big(1-x_2^2+x_3^2 \big)^2 - 4x_3^2\Big) 
\frac{k_1^3}{\Pi_ik_{i}^3}
(2\pi)^3 \delta^{3}(\textbf{k}_{1}+\textbf{k}_{2}+\textbf{k}_{3}) .
\end{eqnarray}
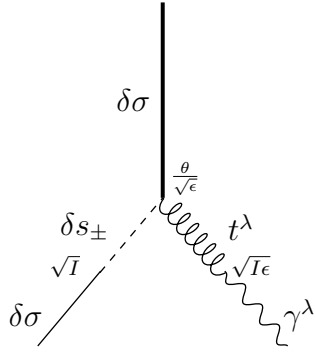
\begin{figure}
	\begin{center}
		\begin{tikzpicture}[line width=1.5 pt, scale=1.3]
		\begin{scope}[rotate=90]
		\draw[gluon,black,line width=0.5 pt] (-140:1)--(0,0);
		\draw[scalarnoarrow,black,line width=0.5 pt] (140:1)--(0,0);
		\draw[fermionnoarrow] (0:2)--(0,0);
		\node at (1,.3)  {$\delta \sigma$};
		\node at (0.2,-0.23)  {\scriptsize$\frac{
				\theta}{\sqrt{\epsilon}}$};
		\node  at (-0.7,-0.9) {\scriptsize $\sqrt{I \epsilon}$};
		\node  at (-0.7,1) {\scriptsize $\sqrt{I }$};
		\node at (-1.2,1.4) {$\delta \sigma$};
		\node at (-0.3,0.8) {$\delta s_{\pm}$};
		\node at (-1.2,-1.4) {$\gamma^{\lambda}$};
		\node at (-0.3,-0.8) {$t^{\lambda}$};
		\begin{scope}[shift={(-0.75,0.635)}]
		\draw[fermionnoarrow,black,line width=0.5 pt] (140:1)--(0,0);
		\end{scope}
		\begin{scope}[shift={(-0.75,-0.635)}]
		\draw[vector,black,line width=0.5 pt] (-140:1)--(0,0);
		\end{scope}
		\end{scope}
		\end{tikzpicture}
		\caption{ Dominant diagrams for mixed NG $\langle \delta \sigma(\textbf{k}_{1})\delta \sigma ({\textbf{k}_{2}}) \gamma^{\lambda}({\textbf{k}_{3}})\rangle$. 
		} \label{figSST}
	\end{center}
\end{figure}
In obtaining the above result we have used Eq. (\ref{eij-cubic-2kk}) to simplify expressions containing the contractions between polarization tensors and the wave vectors. The three-point function (\ref{NG-RRT}) is symmetric with respect to the exchange of momenta ${\bf k}_1\leftrightarrow{\bf k}_2$ and, therefore, we normalized the result with respect to ${\bf k}_1$. 

Expanding (\ref{NG-RRT}) for small $\theta$ we find 
\begin{eqnarray}\label{NG-RRT-approx}
\langle \mathcal{R}(\textbf{k}_{1})\mathcal{R}(\textbf{k}_{2})
\gamma^{\lambda}(\textbf{k}_{3})\rangle
= 12 I\theta^2 \frac{H^4}{\epsilon} N_K^2
\frac{1+x_2}{x_3^2} \Big(\big(1-x_2^2+x_3^2 \big)^2 - 4x_3^2\Big)
\frac{k_1^3}{\Pi_ik_{i}^3} 
(2\pi)^3 \delta^{3}(\textbf{k}_{1}+\textbf{k}_{2}+\textbf{k}_{3}).\nonumber \\  
\end{eqnarray}
These types of mixed NG can generate clustering fossils 
from one long mode of tensor ($x_3 \rightarrow 0$) on the power spectrum of  two 
scalar modes \cite{Dai:2013kra, Emami:2015uva,Dimastrogiovanni:2014ina, Ricciardone:2017kre,  Dimastrogiovanni:2019bfl, Akhshik:2014bla}. However, we see that it is proportional to $I \theta^2$ so unfortunately  it is very small in comparison with other three-point functions. Moreover, we note that it vanishes for $\theta=0$ and, therefore, in the absence of the  parity violating term, our model cannot provide any significant cross-correlation between two curvature perturbations and one GWs tensor mode.

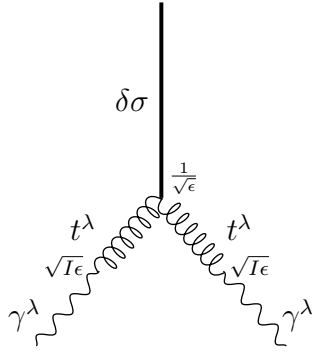
\begin{figure}
	\begin{center}
		\begin{tikzpicture}[line width=1.5 pt, scale=1.3]
		\begin{scope}[rotate=90]
		\draw[gluon,black,line width=0.5 pt] (-140:1)--(0,0);
		\draw[gluon,black,line width=0.5 pt] (140:1)--(0,0);
		\draw[fermionnoarrow] (0:2)--(0,0);
		\node at (1,.3)  {$\delta \sigma$};
		\node at (0.2,-0.23)  {\scriptsize$\frac{
				1}{\sqrt{\epsilon}}$};
		\node  at (-0.7,-0.9) {\scriptsize $\sqrt{I \epsilon}$};
		\node  at (-0.7,1) {\scriptsize $\sqrt{I \epsilon}$};
		\node at (-1.2,1.4) {$\gamma^{\lambda}$};
		\node at (-0.3,0.8) {$t^{\lambda}$};
		\node at (-1.2,-1.4) {$\gamma^{\lambda}$};
		\node at (-0.3,-0.8) {$t^{\lambda}$};
		\begin{scope}[shift={(-0.75,0.635)}]
		\draw[vector,black,line width=0.5 pt] (140:1)--(0,0);
		\end{scope}
		\begin{scope}[shift={(-0.75,-0.635)}]
		\draw[vector,black,line width=0.5 pt] (-140:1)--(0,0);
		\end{scope}
		\end{scope}
		\end{tikzpicture}
		\caption{Dominant diagrams for mixed NG $\langle \delta \sigma(\textbf{k}_{1})\gamma^{\lambda} ({\textbf{k}_{2}}) \gamma^{\lambda}({\textbf{k}_{3}})\rangle$. 
		} \label{fig13}
	\end{center}
\end{figure}

Now, we look at the mixing between one curvature mode and two GWs tensor modes. The dominant contribution comes from the Feynman diagram  shown in 
Fig. \ref{fig13}, which after performing calculations, results in
\begin{eqnarray}\label{RGG}
&& \nonumber \langle \mathcal{R}(\textbf{k}_{1})  
\gamma^{\lambda}({\textbf{k}_{2}})\gamma^{\lambda} ({\textbf{k}_{3}})
 \rangle|_{\rm Fig. (\ref{fig13})}
= 96 I H^4
e^{8 \pi \lambda \theta} \Theta_{1}(\theta)^2 
\Big( N_{K}^3 + \frac{8}{3} \lambda N_{K}^{2}\theta^2 
(x_{3}+x_{2}) (x_{3}^2+x_{2}^2) \Big) \\
&& \hspace{5cm} \times
e_{ij}^{\lambda}(\textbf{k}_{2}) e_{ij}^{\lambda}(\textbf{k}_{3}) 
 \frac{k_{1}^{3}}{\Pi_{i}k_i^3} 
(2 \pi)^3 \delta^{3}(\textbf{k}_{1}+\textbf{k}_{2}+\textbf{k}_{3}) \,,
\end{eqnarray}
where we have normalized the results with respect to the momentum $k_1$ in the scalar sector and also again we have restricted our calculations to the case of the same polarizations $\langle\mathcal{R} \gamma^{+}\gamma^{+}\rangle$ and $\langle\mathcal{R} \gamma^{-}\gamma^{-}\rangle$. Using Eq. (\ref{eij-cubic-2}) for the contraction of two polarization tensors and comparing the result with the definition Eq. (\ref{fNL}), the amplitude of NG in the squeezed limit $x_3\to0$ for the case of $\langle\mathcal{R} \gamma^{+}\gamma^{+}\rangle$ is given by
\begin{equation}\label{fNL-RTT}
f^{{\mathcal R}++}_{NL, sq} \simeq 
160 e^{ 8 \pi \theta} \Theta_{1}(\theta)^2 \frac{(1-x_2+x_3)^2}{x_3^2} I \epsilon^2 N_{K}^3  \,.
\end{equation}
In obtaining  Eq. (\ref{fNL-RTT}) the normalization is performed with respect to the power spectrum of the curvature perturbations which induces the factor $\epsilon^2$. We see that while Eq. (\ref{NG-RRT-approx}) vanishes for $\theta=0$ but Eq. (\ref{fNL-RTT}) does not vanish.

Mixed NGs between curvature perturbations and GWs are recently studied in Refs. \cite{Dimastrogiovanni:2018xnn,Bordin:2018pca,Fujita:2018vmv}. In the squeezed limit, they can be thought as the modulation of the power spectra which makes it possible to look for their observational effects. It is also worth mentioning that in the case of spectator gauge fields which do not contribute to the background dynamics with axionic-like interaction, the three-point functions for mixing between curvature perturbations and GWs cannot be computed by means of the perturbative in-in formalism since some non-perturbative effects show up at the quadratic level. In our model, however, the interactions have different nature as the vector fields are not spectator fields ($I\neq0$). Consequently, the small parameter $I$, measuring the fractional energy density of the vector fields, appears in all of our quadratic and cubic interactions. We, therefore, can treat all quadratic and cubic interactions perturbatively and compute the corresponding three-point functions by means of the in-in formalism.

\section{Bispectra of vector fields perturbations}\label{NG-GF}

In the previous section, we studied the NGs in curvature perturbations and GWs where the vector fields scalar and tensor modes played the roles of mediator particles to enhance the bispectra at tree-level. For the sake of completeness, here we calculate the three-point functions between the vector fields modes and curvature perturbations and/or GWs. 

We have shown in section \ref{sec-scalar} that the scalar modes associated to the three vector fields  can be decomposed into entropy mode $\delta{s}$ and pure isocurvature mode $U$. Thus, depending on the reheating scenario and the expansion history of Universe, one may use the observational bounds on entropy and isocurvature modes to put constraints on the observables associated to these quantities. For mixing NG between curvature perturbations ${\cal R}$ and entropy modes $\delta{s}_{\pm}$, if we pick the relevant interaction Hamiltonians from Eq. (\ref{H-int-SSS}) and then perform the direct calculations, we find the following results to the leading orders
\begin{equation}
\langle {\cal R}({\bf k}_1) {\cal R}({\bf k}_2) {\cal S}_{\pm}({\bf k}_3) \rangle
\sim \langle {\cal R}({\bf k}_1) {\cal S}_{\pm}({\bf k}_2) {\cal S}_{\pm}({\bf k}_3)\rangle
\sim \langle {\cal R}({\bf k}_1) {\cal S}_{\pm}({\bf k}_2) {\cal S}_{\mp}({\bf k}_3)\rangle
\sim  \mathcal{O}(I^{\frac{3}{2}} N_K^3) .
\end{equation} 
The above results are suppressed in comparison with the $I N_K^3$ correction appearing in the three-point function of the curvature perturbations. This shows that although NG of curvature perturbations receives correction from the entropy and isocurvature modes (\ref{RRR-3V}), the superhorizon curvature perturbations are almost uncorrelated with the entropy perturbations at the nonlinear level. This is also consistent with the result (\ref{beta-ns}) which was found previously at the level of linear perturbations.

The  tensor modes associated to the vector fields can also be thought as entropy modes for the GWs. They can be converted to the GWs during the reheating  or even survive after the reheating similar to the primordial GWs. We therefore find the bispectra for mixing between these  tensor modes and curvature perturbations and/or GWs with the hope that it may become possible to  indirectly constrain them with some observable quantities in future.

\begin{figure}
	\begin{center}
		\begin{tikzpicture}[line width=1.5 pt, scale=1.3]
		\begin{scope}[rotate=90]
		\draw[gluon,black,line width=0.5 pt] (-140:2)--(0,0);
		\draw[gluon,black,line width=0.5 pt] (140:2)--(0,0);
		\draw[fermionnoarrow] (0:2)--(0,0);
		\node at (-0.5,-1) {$t^{\lambda}$};
		\node at (-0.5,1) {$t^{\lambda}$};
		\node at (1,.3)  {$\delta \sigma$};
		\node at (0,-1.5)  {$\equiv$};
		\node  at (0,-4.5) {$+ 2$};
		\node  at (0,-7.5) {$+$};	
		\draw[fill=black] (0,0) circle (.3cm);
		\draw[fill=white] (0,0) circle (.29cm);
		\begin{scope}
		\clip (0,0) circle (.3cm);
		\foreach \x in {-.9,-.8,...,.3}
		\draw[line width=1 pt] (\x,-.3) -- (\x+.6,.3);
		\end{scope}
		\end{scope}
		\begin{scope}[shift={(3,0)}]
		\begin{scope}[rotate=90]
		\draw[gluon,black,line width=0.5 pt] (-140:2)--(0,0);
		\draw[gluon,black,line width=0.5 pt] (140:2)--(0,0);
		\draw[fermionnoarrow] (0:2)--(0,0);
		\node at (-0.5,-1) {$t^{\lambda}$};
		\node at (-0.5,1) {$t^{\lambda}$};
		\node at (1,.3)  {$\delta \sigma$};
		\node at (0.2,-0.23)  {\scriptsize $\frac{1}{\sqrt{\epsilon}}$};
		\end{scope}
		\end{scope}
		\begin{scope}[shift={(9,0)}]
		\begin{scope}[rotate=90]
		\draw[gluon,black,line width=0.5 pt] (-140:2)--(0,0);
		\draw[gluon,black,line width=0.5 pt] (140:2)--(0,0);
		\draw[scalarnoarrow,black] (0:1)--(0,0);
		\node at (-0.8,-1.2) {$t^{\lambda}$};
		\node at (-0.8,1.2) {$t^{\lambda}$};
		\node at (0.5,.3)  {$\delta s_{\pm}$};
		\node at (1.5,.3)  {$\delta \sigma$};
		\node at (1,.25)  {\scriptsize $\sqrt{I}$};
		\node at (0.2,-0.23)  {\scriptsize $\sqrt{\frac{I}{\epsilon}}$};
		\begin{scope}[shift={(1,0)}]
		\draw[fermionnoarrow] (0:1)--(0,0);
		\end{scope}
		\end{scope}
		\end{scope}
		\begin{scope}[shift={(6,0)}]
		\begin{scope}[rotate=90]
		\draw[gluon,black,line width=0.5 pt] (-140:2)--(0,0);
		\draw[vector,black,line width=0.5 pt] (140:1)--(0,0);
		\draw[fermionnoarrow] (0:2)--(0,0);
		\node at (-0.8,-1.2) {$t^{\lambda}$};
		\node at (-1.2,1.35) {$t^{\lambda}$};
		\node at (-0.2,0.6) {$\gamma^{\lambda}$};
		\node at (1,.3)  {$\delta \sigma$};
		\node  at (-0.7,1) {\scriptsize$\sqrt{I \epsilon}$};
		\node at (0.2,-0.23)  {\scriptsize$\frac{1}{\sqrt{\epsilon}}$};
		\begin{scope}[shift={(-0.75,0.635)}]
		\draw[gluon,black,line width=0.5 pt] (140:1)--(0,0);
		\end{scope}
		\end{scope}
		\end{scope}
		\end{tikzpicture}
		\caption{ Feynman diagrams that contribute to the three-point function $\langle \delta \sigma(\textbf{k}_{1})t^{\lambda} ({\textbf{k}_{2}}) t^{\lambda}({\textbf{k}_{3}})\rangle$ 
		} \label{fig10}
	\end{center}
\end{figure}
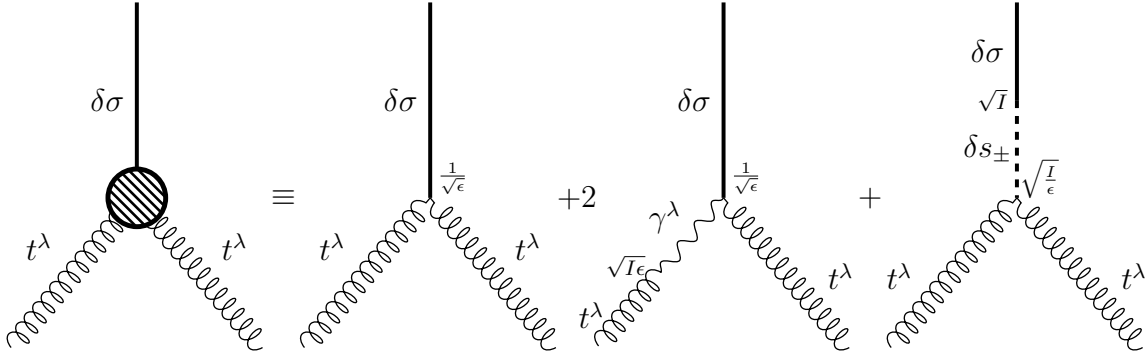

For the mixing between the curvature perturbations and two tensor modes of vector fields, the dominant Feynman diagrams are illustrated in Fig. \ref{fig10} which result in
\begin{eqnarray}\label{RTT}
&&\langle \mathcal{R}(\textbf{k}_{1}) 
t^{\lambda} ({\textbf{k}_{2}}) t^{\lambda}({\textbf{k}_{3}})
\rangle|_{\rm Fig. (\ref{fig10})} 
= \frac{3H^4}{2 \epsilon} N_{K}
e^{8 \pi {\lambda} \theta}\Theta_{1}(\theta)^2
\Big( 1 - \frac{16}{3} \lambda \theta^2 (x_{2}+x_{3}) ( x_{2} - x_{3} )^2 \Big)
\\ \nonumber && \hspace{7cm} \times
e_{ij}^{ \lambda}(\textbf{k}_{1}) e_{ij}^{\lambda}(\textbf{k}_{2}) \frac{k_{1}^3}{\Pi_{i}k_{i}^{3}} 
(2 \pi)^3 \delta^{3}(\textbf{k}_{1}+\textbf{k}_{2}+\textbf{k}_{3}) \,,
\end{eqnarray}
which is computed only for the case that all $\lambda$'s on  both sides are the same. From the Feynman diagrams in Fig. \ref{fig10}, we see that there would also be  some contributions proportional to $N_K^2$ in the above result.  These terms, however, neatly cancel each other and we are left only with the linear contribution of $N_K$. We can estimate the order of bispectra (\ref{RTT}) by normalizing it with the power spectrum of the curvature perturbations through the definition (\ref{fNL}) which yields $f_{\rm NL}^{tt{\cal R}}\sim {\cal O}(1)\epsilon N_K$.

\begin{figure}
	\begin{center}
		\begin{tikzpicture}[line width=1.5 pt, scale=1.3]
		\begin{scope}[rotate=90]
		\draw[gluon,black,line width=0.5 pt] (-140:2)--(0,0);
		\draw[vector,black,line width=0.5 pt] (140:2)--(0,0);
		\draw[fermionnoarrow] (0:2)--(0,0);
		\node at (-0.5,-1) {$t^{\lambda}$};
		\node at (-0.5,1) {$\gamma^{\lambda}$};
		\node at (1,.3)  {$\delta \sigma$};
		\node at (0,-1.5)  {$\equiv$};
		\node  at (0,-4.5) {$+$};
		\draw[fill=black] (0,0) circle (.3cm);
		\draw[fill=white] (0,0) circle (.29cm);
		\begin{scope}
		\clip (0,0) circle (.3cm);
		\foreach \x in {-.9,-.8,...,.3}
		\draw[line width=1 pt] (\x,-.3) -- (\x+.6,.3);
		\end{scope}
		\end{scope}
		\begin{scope}[shift={(6,0)}]
		\begin{scope}[rotate=90]
		\draw[gluon,black,line width=0.5 pt] (-140:2)--(0,0);
		\draw[gluon,black,line width=0.5 pt] (140:1)--(0,0);
		\draw[scalarnoarrow,black] (0:1)--(0,0);
		\node at (-0.8,-1.2) {$t^{\lambda}$};
		\node at (-1.2,1.3) {$\gamma^{\lambda}$};
		\node at (-0.2,0.6) {$t^{\lambda}$};
		\node at (0.5,.3)  {$\delta s_{\pm}$};
		\node at (1.5,.3)  {$\delta \sigma$};
		\node at (1,.25)  {\scriptsize$\sqrt{I}$};
		\node at (0.2,-0.23)  {\scriptsize $\sqrt{\frac{I}{\epsilon}}$};
		\node  at (-0.7,1) {\scriptsize $\sqrt{I \epsilon}$};
		\begin{scope}[shift={(-0.75,0.635)}]
		\draw[vector,black,line width=0.5 pt] (140:1)--(0,0);
		\end{scope}
		\begin{scope}[shift={(1,0)}]
		\draw[fermionnoarrow] (0:1)--(0,0);
		\end{scope}
		\end{scope}
		\end{scope}
		\begin{scope}[shift={(3,0)}]
		\begin{scope}[rotate=90]
		\draw[gluon,black,line width=0.5 pt] (-140:2)--(0,0);
		\draw[gluon,black,line width=0.5 pt] (140:1)--(0,0);
		\draw[fermionnoarrow] (0:2)--(0,0);
		\node at (-0.8,-1.2) {$t^{\lambda}$};
		\node at (-1.2,1.3) {$\gamma^{\lambda}$};
		\node at (-0.2,0.6) {$t^{\lambda}$};
		\node at (1,.3)  {$\delta \sigma$};
		\node  at (-0.7,1) {\scriptsize $\sqrt{I \epsilon}$};
		\node at (0.2,-0.23)  {\scriptsize$\frac{1}{\sqrt{\epsilon}}$};
		\begin{scope}[shift={(-0.75,0.635)}]
		\draw[vector,black,line width=0.5 pt] (140:1)--(0,0);
		\end{scope}
		\end{scope}
		\end{scope}
		\end{tikzpicture}
		\caption{ Feynman diagrams that contribute to $\langle \delta \sigma(\textbf{k}_{1})\gamma^{\lambda} ({\textbf{k}_{2}}) t^{\lambda}({\textbf{k}_{3}})\rangle$.} \label{fig12}
	\end{center}
\end{figure}
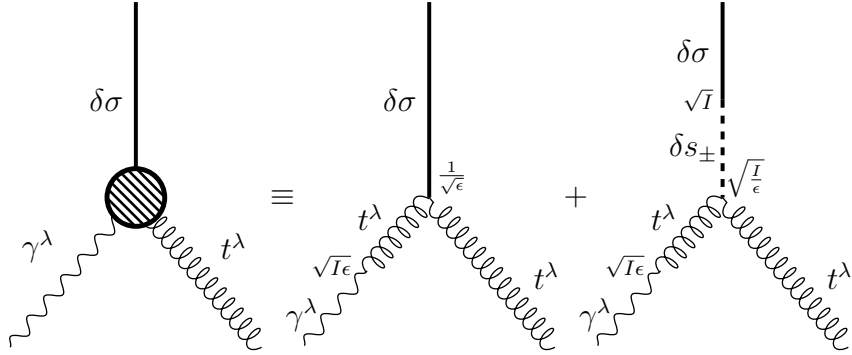

The next case is the mixing between curvature perturbations, gravitational tensor modes, and tensor modes of vector fields. The relevant Feynman diagrams are shown in Fig \ref{fig12} which after direct calculations, give the following three-point function
\begin{eqnarray}\label{RGT}
&& \langle \mathcal{R}(\textbf{k}_{1})\gamma^{\lambda} ({\textbf{k}_{2}}) 
t^{\lambda}({\textbf{k}_{3}})\rangle|_{\rm Fig. (\ref{fig12})}
= -12 H^4 \sqrt{\frac{I}{\epsilon}} N_K^2
e^{8 \pi \lambda \theta} \Theta_{1}(\theta)^2 
\Big(1 + \frac{16}{3} \theta^2 x_{3}^2 \big( 2 x_{2}-\lambda(x_{2}+x_{3}) \big) \Big)  \nonumber \\
&& \hspace{7.5cm} \times 
e_{ij}^{\lambda}(\textbf{k}_{2}) e_{ij}^{ \lambda}(\textbf{k}_{3}) \frac{k_{1}^3}{\Pi_{i}k_i^3} 
(2 \pi)^3 \delta^{3}(\textbf{k}_{1}+\textbf{k}_{2}+\textbf{k}_{3}) \,,
\end{eqnarray}
where Eq.  (\ref{eij-cubic-2})  has been used  for $e_{ij}^{\lambda}(\textbf{k}_{2}) e_{ij}^{\lambda}(\textbf{k}_{3})$. Note that, as before,  we have concentrated to the case where   all $\lambda$'s on both sides are the same.  This bispectrum does not have any symmetry for the external legs. Similar to the previous case, the order of magnitude of the bispectrum (\ref{RGT}) is $f_{\rm NL}^{{\cal R}\gamma{t}}\sim {\cal O}(1) \sqrt{I} \epsilon^{3/2} N_K^2$.

For the last cases in this subsection, we consider the three-point cross correlation 
of the form $\langle  \gamma t t \rangle$  and $\langle  \gamma \gamma t \rangle$.  The corresponding Feynman diagrams  are shown in Figs. \ref{fig14} and \ref{fig15} which yield
\begin{eqnarray}\label{GTT}
&&\nonumber \langle \gamma^{\lambda}(\textbf{k}_{1})t^{\lambda} ({\textbf{k}_{2}}) t^{\lambda}({\textbf{k}_{3}})\rangle|_{\rm Fig. (\ref{fig14})} 
= - H^4 N_K e^{8\lambda \pi \theta} \Theta_{1}(\theta)^2 e_{ij}^{\lambda}(\textbf{k}_{1})e_{mj}^{\lambda}(\textbf{k}_{2})
e_{mi}^{\lambda}(\textbf{k}_{3}) \\
&& \hspace{1cm} \times
\Big[ 3 + 8 \theta^2 (x_{2}+x_{3}) ( 4 x_{2}x_{3} 
- \lambda (3 x_{2}^{2} - 2 x_{2} x_{3}+3 x_{3}^2)) \Big]
\frac{k_1^3}{ \Pi_{i}k_i^3}
(2 \pi)^3 \delta^{3}(\textbf{k}_{1}+\textbf{k}_{2}+\textbf{k}_{3}) ,
\end{eqnarray}
and 
\begin{eqnarray}\label{GGT}
&& \langle \gamma^{\lambda}(\textbf{k}_{1})\gamma^{\lambda} ({\textbf{k}_{2}}) t^{\lambda}({\textbf{k}_{3}})\rangle|_{\rm Fig. (\ref{fig15})} 
= 24 \sqrt{I \epsilon} H^4  N_K^2
e^{8 \pi \lambda \theta} \Theta_{1}(\theta)^2
e_{ij}^{\lambda}(\textbf{k}_{1}) e_{mj}^{\lambda}(\textbf{k}_{2})
e_{mi}^{\lambda}(\textbf{k}_{3}) \\ \nonumber
&&\hspace{1.5cm} \times
\Big[ ( 1 + x_{2}^3) + \frac{8}{3} (4+ \lambda) \theta^2 x_{3}^2(x_{1}+x_{2})
+ 16 \lambda\theta^2 x_{3}^3 \Big]
 \frac{k_{1}^3}{ \Pi_i k_i^3}
(2 \pi)^3 \delta^{3}(\textbf{k}_{1}+\textbf{k}_{2}+\textbf{k}_{3})\,.
\end{eqnarray}
\begin{figure}
	\begin{center}
		\begin{tikzpicture}[line width=1.5 pt, scale=1.3]
		\begin{scope}[rotate=90]
		\draw[gluon,black,line width=0.5 pt] (-140:2)--(0,0);
		\draw[gluon,black,line width=0.5 pt] (140:2)--(0,0);
		\draw[vector,black,line width=0.5 pt] (0:2)--(0,0);
		\node at (-0.5,-1) {$t^{\lambda}$};
		\node at (-0.5,1) {$t^{\lambda}$};
		\node at (1,.3)  {$\gamma^{\lambda}$};
		\node at (0.15,-0.4)  {\scriptsize$\mathcal{O}(1)$};
		\end{scope}
		\end{tikzpicture}
		\caption{ Dominant diagram for the three-point cross correlation $\langle \gamma^{\lambda}(\textbf{k}_{1})t^{\lambda} ({\textbf{k}_{2}}) t^{\lambda}({\textbf{k}_{3}})\rangle$.
		} \label{fig14}
	\end{center}
\end{figure}
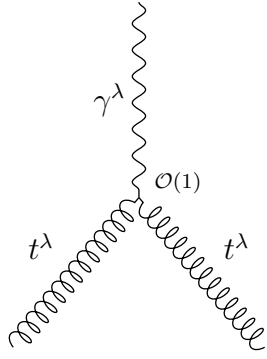
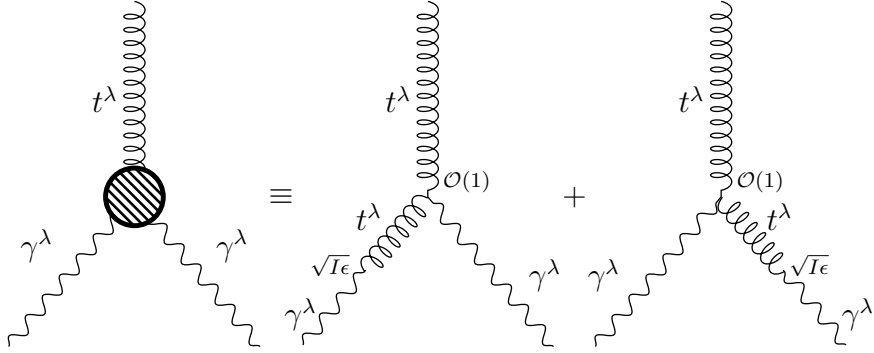
\begin{figure}
	\begin{center}
		\begin{tikzpicture}[line width=1.5 pt, scale=1.3]
		\begin{scope}[rotate=90]
		\draw[vector,black,line width=0.5 pt] (-140:2)--(0,0);
		\draw[vector,black,line width=0.5 pt] (140:2)--(0,0);
		\draw[gluon,black,line width=0.5 pt] (0:2)--(0,0);
		\node at (-0.5,-1) {$\gamma^{\lambda}$};
		\node at (-0.5,1) {$\gamma^{\lambda}$};
		\node at (1,.3)  {$t^{\lambda}$};
		\node at (0,-1.5)  {$\equiv$};
		\node  at (0,-4.5) {$+$};	
		\draw[fill=black] (0,0) circle (.3cm);
		\draw[fill=white] (0,0) circle (.29cm);
		\begin{scope}
		\clip (0,0) circle (.3cm);
		\foreach \x in {-.9,-.8,...,.3}
		\draw[line width=1 pt] (\x,-.3) -- (\x+.6,.3);
		\end{scope}
		\end{scope}
		\begin{scope}[shift={(3,0)}]
		\begin{scope}[rotate=90]
		\draw[vector,black,line width=0.5 pt] (-140:2)--(0,0);
		\draw[gluon,black,line width=0.5 pt] (140:1)--(0,0);
		\draw[gluon,black,line width=0.5 pt] (0:2)--(0,0);
		\node at (-0.8,-1.2) {$\gamma^{\lambda}$};
		\node at (-1.2,1.3) {$\gamma^{\lambda}$};
		\node at (-0.2,0.6) {$t^{\lambda}$};
		\node at (0.15,-0.4)  {\scriptsize$\mathcal{O}(1)$};
		\node at (1,.3)  {$t^{\lambda}$};
		\node  at (-0.7,1) {\scriptsize$\sqrt{I \epsilon}$};
		\begin{scope}[shift={(-0.75,0.635)}]
		\draw[vector,black,line width=0.5 pt] (140:1)--(0,0);
		\end{scope}
		\end{scope}
		\end{scope}
		\begin{scope}[shift={(6,0)}]
		\begin{scope}[rotate=90]
		\draw[gluon,black,line width=0.5 pt] (-140:1)--(0,0);
		\draw[vector,black,line width=0.5 pt] (140:2)--(0,0);
		\draw[gluon,black,line width=0.5 pt] (0:2)--(0,0);
		\node at (-0.8,1.2) {$\gamma^{\lambda}$};
		\node at (-1.2,-1.4) {$\gamma^{\lambda}$};
		\node at (-0.2,-0.6) {$t^{\lambda}$};
		\node at (0.15,-0.4)  {\scriptsize$\mathcal{O}(1)$};
		\node at (1,.3)  {$t^{\lambda}$};
		\node  at (-0.7,-0.9) {\scriptsize $\sqrt{I \epsilon}$};
		\begin{scope}[shift={(-0.75,-0.635)}]
		\draw[vector,black,line width=0.5 pt] (-140:1)--(0,0);
		\end{scope}
		\end{scope}
		\end{scope}
		\end{tikzpicture}
		\caption{ Feynman diagrams that contribute to the three-point function $\langle \gamma^{\lambda}(\textbf{k}_{1})\gamma^{\lambda} ({\textbf{k}_{2}}) t^{\lambda}({\textbf{k}_{3}})\rangle$. 
		} \label{fig15}
	\end{center}
\end{figure}

The explicit relation for the contraction between three polarization tensors is obtained in Eq. (\ref{eij-cubic-3}). In Eqs. (\ref{GTT}) and (\ref{GGT}) we only deal with tensor modes and therefore it makes sense to normalize these bispectra with the power spectrum of the GWs (\ref{PS-hh-f}). Doing so, from the definition (\ref{fNL}) we  estimate the order of magnitude as $f_{\rm NL}^{\gamma{tt}} \sim {\cal O}(1) N_K$ and $f_{\rm NL}^{t\gamma\gamma} \sim {\cal O}(1) \sqrt{I\epsilon} N_K^2$.

\section{Comparison to other models}
\label{compare}

Our setup with three vector fields with internal global $O(3)$ symmetry has some similarities/differences with the so-called anisotropic inflation \cite{Watanabe:2009ct} in one side and inflationary models which deal with non-Abelian $SU(2)$ gauge fields in the other side. In some sense, our model lies between these two types of models. Therefore, in this section  we compare our setup with these types of models.

In anisotropic inflation, a vector field with the Abelian $U(1)$ symmetry is non-minimally coupled to the inflaton field. Indeed, considering one copy of $U(1)$ symmetry in the action of our model (\ref{action}) and setting $\theta=0$, we recover the action of anisotropic inflation. At the level of background, similar to the anisotropic inflation, the vector fields have vev in our model while in contrast to the anisotropic inflation, our model provides isotropic background thanks to the internal $O(3)$ symmetry of vector fields. At the level of perturbations, in the anisotropic inflation setup, the two vector modes of the vector field couple to the scalar mode at the linear level and also to the tensor modes at the nonlinear level  \cite{  Kanno:2010ab,  Emami:2013bk}.  In our model with three vector fields and isotropic background, these perturbations do not mix at the linear level and, more importantly, the $O(3)$ symmetry of the vector fields provides two scalar modes and two tensor modes in addition of two vector modes. The most important difference between our setup and the setup of anisotropic inflation is that the gravitational tensor modes, as we have shown, are affected by the tensor modes of vector fields at the linear level.

In the case of inflationary models which deal with $SU(2)$ gauge fields, the most relevant model to our setup is the so-called chromo-natural inflation where inflation is driven by an axionic field $\chi$ which is coupled to three $SU(2)$ gauge fields through the well-known parity violating interaction $\chi{F}{\tilde F}$\cite{Adshead:2012kp}. Similar to our setup the non-Abelian gauge fields have vev and contribute to the isotropic background. Due to the non-vanishing vev of the gauge fields, not only the tensor sector but also the scalar sector receive some corrections at the level of perturbations \cite{Adshead:2013nka}. The situation is the same as in our model where, as we have seen, the power spectra and bispectra of curvature perturbations are modified. However, the way that the scalar sector is modified is different in our scenario. The reason is that the coupling of inflaton to the gauge field sector is different. More precisely, the coupling $\chi{F}{\tilde F}$ in chromo-natural inflation cannot prevent the gauge field vev to decay while in our model the coupling is chosen as in Eq. (\ref{a1}) which prevents vector field to decay. This choice, which is suggested in anisotropic inflation scenario, has significant impact on the perturbations so that, in contrast to the chromo-natural model, the dominant corrections to the correlation functions depend on the number of e-folds as can be seen in Eqs. (\ref{PS-R}) and (\ref{PS-hh-f}) for the power spectra of curvature perturbations and GWs and in Eqs. (\ref{RRR-3V}) and (\ref{GGG}) for NG of curvature perturbations and GWs respectively. The dominant contribution to the three-point function of GW in these types of models are proportional to the gauge coupling constant $g$ (see for instance Refs. \cite{Agrawal:2017awz,Agrawal:2018mrg}). But, since our model can be realized from the global limit $g\to0$ of $SU(2)$ gauge symmetry, therefore, these types of vertices are absent in our scenario and we have some other types of vertices instead. 

Finally we comment on Ref. \cite{Yamamoto:2012sq} which is very relevant to work here.  Indeed, our model reduces to the model of the Ref. \cite{Yamamoto:2012sq} if we set $\theta=0$ in the action (\ref{action}).  Compared to \cite{Yamamoto:2012sq} we have decomposed the scalar modes into the adiabatic and entropy modes and  obtained their power spectra while in Ref. \cite{Yamamoto:2012sq} the entropy modes are not studied. At the quadratic level, we have found that the isocurvature mode $U$ sources the entropy mode $\delta{s}$ through the parity violating term. We then had to diagonalize the quadratic action since we did not take $\theta$ to be small from the beginning. On the other hand, in the case of NG, only the three-point function of curvature perturbations with the $I N_k^3$ correction is obtained in Ref. \cite{Funakoshi:2012ym}. This result can be recovered from our general result (\ref{RRR-3V}) by setting $\theta=0$. 
However, the NGs for the tensor modes, whether from $\gamma_{ij} $ or $t_{ij}$, 
are  not studied in \cite{Yamamoto:2012sq}.  Here, we have found the bispectra of GWs and also the mixing three-point functions between GWs and curvature perturbations. Moreover, we found the mixing of $t_{ij}$ with GWs and curvature perturbations.

\section{Summary and conclusions}
\label{summary}

We have studied the isotropic extension of the so-called anisotropic inflation in the presence of  parity violating interaction defined by the action (\ref{action}). The vector fields enjoy $O(3)$ internal symmetry and the setup admits isotropic background with non-vanishing time-dependent vev for the vector fields. The ratio of energy density of vector fields to the total background energy density is given by parameter $I$ defined in (\ref{I}). This parameter is small in order to allow for the attractor near de Sitter background. We studied cosmological perturbations in this scenario. Vector fields provide two dynamical scalar modes, of which one of them contributes to the  curvature perturbations with the  contribution proportional to $I$ as expected. Therefore, we decomposed the linear scalar perturbations into the adiabatic and entropy modes. The other scalar mode is an isocurvature mode in the absence of parity violating term while it sources the entropy mode in the presence of parity violating term. The strength of this coupling is determined by the parity violating parameter $\theta$ which we did not treat as a small parameter from the beginning. We, therefore diagonalized the quadratic action for the scalar perturbations and found the power spectra for the all scalar modes including curvature perturbations and entropy modes. From the observational bound on the power spectrum of curvature perturbations, we then found the bounds $\theta\lesssim 10^{-1}$ and ${I}\lesssim {10}^{-5}$. The isotropic configuration of the vector fields also provides tensor modes which source the gravitational tensor modes at the level of linear perturbations. The power spectrum of the GWs then receive both polarized and unpolarized contributions from the tensor modes of the vector fields. The chiral part of the GWs originating from the parity violating interaction provides distinct observational feature of the model. 

In the next step, we studied nonlinear perturbations for all scalar and tensor modes. We have found that vector field particles enhance the three-point functions as mediator particles. All NGs peak in the squeezed limit and from the observational bounds on the NG of curvature perturbations we have found the stronger bound  ${I}\lesssim {10}^{-7}$. We also computed the mixed NGs between curvature perturbations and GWs. Finally, a brief comparison of our setup and results compared to the previous works in literature is presented.

\vspace{1cm}


{\bf Acknowledgments:} M.A.G thanks Ghadir Jafari for his kind assistances with the xTras package \cite{Nutma:2013zea} which was used for tensorial calculations. The work of M.A.G. was supported by Japan Society for the Promotion of Science Grants-in-Aid for international research fellow No. 19F19313.

\vspace{0.7cm}

\appendix

\section{Circular polarization tensors}\label{app}
\setcounter{equation}{0}
\renewcommand{\theequation}{A\arabic{equation}}

In this appendix we present some identities and formula for the circular polarization tensors $e^\lambda_{ij}({\bf k})$ which we use in the main text.

When computing the quadratic action for the tensor perturbations, we deal with two contracted polarization tensors in which the following identity is being used, 
\begin{equation}\label{eij-quadratic}
e^\lambda_{ij} ({\bf k}) e^{\lambda'*}_{ij}({\bf k}') 
= \frac{1}{4} \big( 1 + \lambda\lambda' \cos\chi \big)^2 \,,
\end{equation}
where $\chi = \cos^{-1}(k_ik'_i/kk')$ is the angle between two wave vectors ${\bf k}$ and ${\bf k}'$. Note that the repeated indices of the tensor components $i, j,...$ are summed over.  In particular,  the conservation of momentum fixes the momenta of two circular polarization tensors in quadratic action as ${\bf k}'=-{\bf k}$ which yields  the well-known formula $e^\lambda_{ij} ({\bf k}) e^{\lambda'}_{ij}(-{\bf k}) = \delta^{\lambda\lambda'}$ where we have used $e^{\lambda*}_{ij}({\bf k}) = e^{\lambda}_{ij}(-{\bf k})$. Moreover, in simplifying the parity violating terms, the following identity has been used
\begin{equation}\label{eij-pv-identity}
i \epsilon_{ijk} k_j e^\lambda_{kl} ( {\bf k}) = \lambda  {k} e^\lambda_{il}( {\bf k}) \,,
\end{equation}
where  the value of $\lambda$ is $+1$ ($-1$) for $+$($\times$) polarization.

In the case of three-point correlation functions, we deal with polarizations tensors which are contracted either with each other or with wave vectors. In order to make the calculations simple, we use the conservation of momentum: there are three external legs with different momenta ${\bf k}_1$, ${\bf k}_2$, and ${\bf k}_3$ which satisfy ${\bf k}_1+{\bf k}_2+{\bf k}_3=0$. Therefore, they should be in a plane and the circular polarization tensor simplifies to \cite{Agrawal:2018mrg,Soda:2011am}
\begin{eqnarray}\label{eij}
e^{\lambda}_{ij}({\bf k}_J)\Big{|}_{\mbox{plane}} := \frac{1}{2}
\begin{pmatrix}
- \sin^2\varphi_J & \sin\varphi_J \cos\varphi_J & i\lambda \sin\varphi_J \\
\sin\varphi_J \cos\varphi_J & -\cos^2\varphi_J & -i\lambda \cos\varphi_J \\
i\lambda \sin\varphi_J & -i\lambda \cos\varphi_J& 1
\end{pmatrix} \,,
\end{eqnarray}
where $\varphi_J=\{\varphi_1,\varphi_2,\varphi_3\}$ are the azimuthal angles of ${\bf k}_J=\{{\bf k}_1,{\bf k}_2,{\bf k}_3\}$. In this plane, every momentum has its own magnitude $k_i = |{\bf k}_i|$ and their direction can be completely fixed through two relative angles $\varphi_2-\varphi_1$ and $\varphi_3-\varphi_1$. Therefore, without loss of generality we choose the origin so that $\varphi_1=0$ and from the conservation of momentum we find $\varphi_2 = \cos^{-1}((1+x_2^2-x_3^2)/2x_2)$ and $\varphi_3 = \cos^{-1}((1-x_2^2+x_3^2)/2x_3)$ where we have defined the wave number ratios $x_{2}\equiv k_2/k_1$ and $x_{3}\equiv k_3/k_1$. 

Having Eq. (\ref{eij}) in hand, we can compute any contractions of the polarizations tensors with themselves or with wave vectors in terms of the wave vector ratios $x_i$ and polarizations $\lambda_i$. In particular we calculate the following contractions which are used in the main text, 
\begin{eqnarray}\label{eij-cubic-2}
e_{ij}^{\lambda}( {\bf k}_{2}) e_{ij}^{ \lambda}({\bf k}_{3})
&=& \frac{(1-x_2+x_3)^2(1+x_2-x_3)^2}{16 x_2^2 x_3^2}\,, \\
\label{eij-cubic-3}
e_{ij}^{\lambda}(\textbf{k}_{1}) e_{mj}^{\lambda}(\textbf{k}_{2})e_{mi}^{\lambda}(\textbf{k}_{3}) 
&=&- \frac{\big( 1-(x_2+x_3)^2 \big) \big( 1-(x_2-x_3)^2 \big) (1+x_2+x_3)^2}{64 x_2^2 x_3^2} \,,
\\ \label{eij-cubic-2kk}
e_{ij}^{\lambda}(\textbf{k}_{3}) {\bf k}_1^i {\bf k}_1^j 
&=& e_{ij}^{\lambda}(\textbf{k}_{3}) {\bf k}_2^i {\bf k}_2^j
= \frac{k_1^2}{8x_3^2} \Big( \big(1-x_2^2+x_3^2 \big)^2 - 4x_3^2 \Big) \,,
\end{eqnarray}
where all $\lambda$'s in the left hand side are the same and could be either $+$ or $\times$. In other words, the first two expressions in the above relations cannot be used for the case of mixed combinations of $+$ and $-$.


\section{Quadratic actions  }\label{quad-action}
In this appendix, we present details of calculations of the quadratic actions for the scalar and tensor perturbations.

\subsection{Quadratic actions for scalar perturbations }\label{quad-scalar}
\setcounter{equation}{0}
\renewcommand{\theequation}{B\arabic{equation}}

As explained in the main text we have six scalar modes $(Y,\delta{Q},U,\alpha,\beta,\delta\phi)$ defined in Eq. (\ref{perturbations}) among which $(Y, \alpha, \beta)$ are non-dynamical. The non-dynamical modes of the metric perturbations $\alpha$ and $\beta$ turned out to be proportional to the slow-roll parameter and, neglecting gravitational back-reactions to leading order in slow-roll parameter, we can ignore them as well \cite{Emami:2013bk}. However, the non-dynamical mode $Y$ from  the vector fields is not slow-roll suppressed and we cannot ignore it.

Expanding the action (\ref{action}) around the background configuration Eqs. (\ref{bac-A}) and (\ref{FRW}) up to the second order of scalar perturbations defined in Eq. \eqref{perturbations}, and performing some integration by parts, we find the following quadratic action
\begin{eqnarray}\label{S2-S0}
S^{\rm SS} &=&\frac{1}{2}\int d\tau d^3 x \Big[a^2 \delta \phi'^2-a^2 (\partial \delta \phi)^2+ \big(3f f_{, \phi \phi} A'^2+ 3f_{,\phi}^2 A'^2 -a^4 V_{, \phi \phi} \big) \delta\phi^2  \\ \nonumber
&+& 3 f^2 \delta Q'^2 - 2 f^2 (\partial \delta Q)^2 + 12 f f_{,\phi} A' \delta \phi \delta Q' + f^2 \partial^2 Y \big(\partial^2 Y - 4 f^{-1}f_{,\phi} A' \delta\phi - 2 \delta Q' \big) \\
\nonumber &+& 2 f^2 (\partial U')^2 - 2 f^2 (\partial^2 U)^2 - 4 \theta f^2 \big( 2 \partial \delta Q' \partial U + 2 \partial U' \partial \delta Q + 4 f^{-1}f_{,\phi} A' \partial \delta \phi \partial U\big) \Big] \,,
\end{eqnarray}
where the subscript ``${,\phi}$" shows derivative with respect to $\phi$, $\tau = \int dt/a(t)$ is the conformal time, and a primes denotes derivative with respect to the conformal time. 

As we already mentioned, from the above action, we see that the mode $Y$ appears with no time derivative which shows that it is non-dynamical. Varying the above action with respect to $Y$ yields 
\begin{equation}\label{Y0}
\partial^i\partial^j \Big[\partial_i \partial_j Y - (\delta Q' + 2 A' f^{-1}f_{,\phi} \delta \phi) \delta_{ij} \Big] = 0 \,,
\end{equation}
which has the following algebraic  solution
\begin{equation}\label{Y}
\partial^2 Y= \delta Q'+2 A' f^{-1}f_{,\phi} \delta \phi \,.
\end{equation}
Plugging the above solution into (\ref{S2-S0}) and then expanding it in terms of  small parameters $\epsilon$ and $I$, we obtain the quadratic action for the remaining dynamical scalar modes given by Eq. \eqref{S2-SS}. 

\subsection{Quadratic actions for tensor perturbations }
\label{quad-tensor}

Here we present the quadratic action for the tensor perturbations $\gamma_{ij}$
and $t_{ij}$. Expanding the action (\ref{action}) around background configuration (\ref{bac-A}) and (\ref{FRW}) with tensor modes $\gamma_{ij}$ and $t_{ij}$ yields 
\begin{eqnarray}
&&S^{\rm TT} = \frac{1}{8} \int d\tau d^3 x \Big[
a^2 \gamma'_{ij} \gamma'^{ij} - a^2 \partial_{i} \gamma_{jk} \partial^{i} \gamma^{j k} 
+ 2 A'^2 f^2 \gamma_{ij} \gamma^{ij} - 8f^2 A' \gamma_{ij} t'^{ij} \\ \nonumber 
&& \hspace{3.3cm} + 4 f^2  t'_{ij}  t'^{ij} - 4 f^2 \partial_{k} t_{ij} \partial^{k} t^{ij} 
+ 4 f^2 \partial_{j} t_{ik} \partial^{k} t^{ij} - 16 \theta f^2 \epsilon_{ijk} {t'}_m{}^{k} \partial^{j} t^{im} \Big] \,,
\end{eqnarray}
where we have used the traceless and transverse conditions Eq. (\ref{traceless-transverse}) along with some integration by parts. By expanding the above quadratic action to linear order in $I$ and $\epsilon$, we find the following quadratic Lagrangian for the tensor modes,
\begin{eqnarray}\label{LTT-app}
&&{L}^{\rm TT} =\frac{1}{2} \int d^3x
\Big[
 \overline{\gamma}'_{ij} \overline{\gamma}'^{ij} - \partial_{i} \overline{\gamma}_{jk} \partial^{i} \overline{\gamma}^{j k} + \frac{2}{\tau^2} (1+I \epsilon ) \overline{\gamma}_{ij} \overline{\gamma}^{ij} 
+ \overline{t}'_{ij}  \overline{t}'^{ij} + \frac{2}{\tau^2} \big( 1-\frac{5}{2}I \epsilon \big) 
\overline{t}_{ij} \overline{t}^{ij} \\ \nonumber 
&& \hspace{2.5cm} - \partial_{k} \overline{t}_{ij} \partial^{k} \overline{t}^{ij} 
+ \partial_{j} \overline{t}_{ik} \partial^{k} \overline{t}^{ij}
+\frac{4}{\tau} \sqrt{I \epsilon}\, \overline{\gamma}_{ij} \overline{t}'^{ij}
 -\frac{8}{\tau^2} \sqrt{I \epsilon}\, \overline{\gamma}_{ij} \overline{t}^{ij} 
+ \frac{8}{\tau} \theta \epsilon_{ijk} \overline{t}_m{}^{k} \partial^{j} \overline{t}^{im} \Big] ,
\end{eqnarray} 
where we have defined the following canonically normalized fields
\begin{equation}\label{tensor-canonical}
\overline{\gamma}_{ij} \equiv \frac{a}{2} \, \gamma_{ij} \,, \hspace{1cm} 
\overline{t}_{ij} \equiv f \, t_{ij} \,.
\end{equation}

Now going to Fourier space, the quadratic action to leading order in terms of the small parameters $I$ and $\epsilon$ is given by 
\begin{eqnarray}\label{S2-T0}
&&S^{\rm TT} = \frac{1}{2}\int d^3 k d\tau \Big[
{\overline{\gamma}'_{ij}}^2 - \Big( k^2 - \frac{2}{\tau^2} ( 1 +  I \epsilon ) \Big) \overline{\gamma}_{ij}^2
+ {\overline{t}'_{ij}}^2 - \Big( k^2 - \frac{2}{\tau^2} 
\big( 1 - \frac{5}{2} I \epsilon \big) \Big) \overline{t}_{ij}^2  \\ \nonumber 
&& \hspace{3.3cm} - \frac{4}{\tau^2} \sqrt{I\epsilon} \big( 2 \overline{t}_{ij} 
- \tau\overline{t}'_{ij} \big) \overline{\gamma}^{ij}
- \frac{8}{\tau} \theta i \epsilon_{ijk} k^{j} \overline{t}^{kl} {\overline t}^{{i}}_{l}
 \Big] \,,
\end{eqnarray}
where $\overline{\gamma}_{ij}(\tau,{\bf k})$ and $\overline{t}_{ij}(\tau,{\bf k})$ are the amplitudes in Fourier space  satisfying the traceless and transverse conditions (\ref{traceless-transverse}) as ${\overline \gamma}_{ii} = k^i \overline{\gamma}_{ij} = {\overline t}_{ii} = k^i {\overline t}_{ij} = 0$. It is convenient to express tensor modes in terms of circular polarization tensors $e^{\lambda}_{ij}({\bf k})$ as ${\overline \gamma}_{ij} (\tau,{\bf k}) = \sum_{+,\times} \overline{\gamma}^{\lambda}(\tau) e^{\lambda}_{ij}({\bf k})$ and ${\overline t}_{ij}(\tau,{\bf k}) = \sum_{+,\times} \overline{t}^{\lambda}(\tau) e^{\lambda}_{ij}({\bf k})$, then the traceless and transverse conditions require $e^\lambda_{ii}({\bf k}) = 0$, and ${\bf k}. e^\lambda_{ij} ({\bf k}) = 0$. The properties of the circular polarization tensor are presented in appendix \ref{app}. Expanding (\ref{S2-T0}) in terms of the circular polarization tensors and then using the identities (\ref{eij-quadratic}) and (\ref{eij-pv-identity}), we find the quadratic action Eq. \eqref{S2-T-pol}.


\section{Details of in-in calculations for the power spectra}\label{app-PS}
\setcounter{equation}{0}
\renewcommand{\theequation}{C\arabic{equation}}

Here we present the details of the in-in calculations for the power spectra of scalar and tensor modes.

\subsection{Scalar modes}

Having the total quadratic Lagrangian for the scalar modes ${ L}^{\rm SS}$  Eq. \eqref{S2-sigma-s-pm} at hand we can obtain the corresponding quadratic Hamiltonian through the Legendre transformation ${ H}^{\rm SS} = \Pi_J {\cal Q}'^J - { L}^{\rm SS}$ where ${\cal Q}^J \equiv \{ \delta \sigma, \delta s_{+}, \delta s_{-} \}$. Doing so, and separating the quadratic free Hamiltonian ${ H}_0^{\rm SS}$ which is obtained by  the Legendre transformed of the free quadratic Lagrangian ${ L}_0^{\rm SS}$, the interaction Hamiltonian is given by $\delta{ H}^{\rm SS} \equiv { H}^{\rm SS} - { H}_0^{\rm SS}$. Working in the interaction picture, the corresponding interaction Hamiltonian $\delta{ H}_{ I}^{\rm SS}$  can be classified as $\delta{ H}_{ I}^{\rm SS} \equiv \sum_i \delta{ H}^{\rm SS}_{{ I},i}$ with $i=1,..,10$ as follows
\begin{eqnarray}\label{H2-int}
&&\delta{ H}_{{I},1(2)}^{\rm SS} = \mp \frac{4\sqrt{2}}{\tau^2} 
\sqrt{I}\, \overline{\delta\sigma} \delta s_\pm, \hspace{0.5cm} 
\delta{ H}_{{I}, 3(4)}^{\rm SS}= \pm\frac{2\sqrt{2}}{\tau}\sqrt{I}\, 
\overline{\delta\sigma} \delta s'_\pm, \hspace{.5cm} 
\delta{ H}_{{I}, 5}^{\rm SS}=\frac{12}{\tau^2} I \overline{\delta \sigma}^2, \nonumber \\ 
&&\hspace{.25cm} \delta{ H}_{{I}, 6}^{\rm SS} 
= \frac{6 }{\tau^2} I \delta s_+ \delta s_- \,, \hspace{1.3cm}
\delta{ H}_{{I}, 7(8)}^{\rm SS} = - \frac{3}{\tau^2} I \delta s_\pm^2 \,, \hspace{1.15cm}
\delta{ H}_{{I}, 9(10)}^{\rm SS} = \pm \frac{2}{\tau} I k\theta \delta s_\pm^2 \,,
\end{eqnarray}
where for the sake of simple presentation, we have dropped the integrals over the momenta $\int d^3k$.

\begin{figure}
	\begin{center}
		\begin{tikzpicture}[line width=1.5 pt, scale=2]
		\draw[scalarnoarrow,black] (0:1)--(0,0);
		\begin{scope}[shift={(1,0)}]
		\draw[fermionnoarrow] (0:1)--(0,0);
		\end{scope}
		\node at (0.5,0.2){$\delta s_{\pm}$};
		\node at (1.5,0.2){$\overline{\delta \sigma}$};
		\node at (1,0.15){\scriptsize$\sqrt{ I }$};
		\begin{scope}[shift={(3,0)}]
		\draw[scalarnoarrow,black] (0:1)--(0,0);
		\begin{scope}[shift={(1,0)}]
		\draw[scalarnoarrow,black] (0:1)--(0,0);
		\end{scope}
		\node at (0.5,0.2){$\delta s_{\mp}$};
		\node at (1.5,0.2){$\delta s_{\pm}$};
		\node at (1,0.15){\scriptsize $  I$};
		\end{scope}
		\end{tikzpicture}
		\caption{Feynman diagrams for the transfer vertex between adiabatic mode and entropy modes (left panel) and between the entropy modes themselves (right panel).} \label{fig2}
	\end{center}
\end{figure}
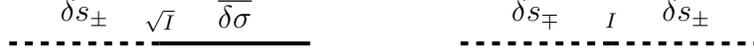

Comparing the above results with Eq. (\ref{L2-int}), we see that $\delta{H}_{{I}}^{\rm SS} \neq - \delta{L}^{\rm SS}$ which is due to the kinetic coupling of the form $\overline{\delta\sigma} \delta{s}'_{\pm}$. The above quadratic interactions $\delta{ H}_{{I},i}^{\rm SS}$ correspond to the exchange of vertices. For interactions $i=1,..,4$, the amplitude of the exchange vertices between the adiabatic mode and the entropy modes $\delta{s}_{\pm}$ are at the order of $\sqrt{I}$. The corresponding Feynman diagram is illustrated in left panel of Fig. \ref{fig2}. On the other hand, the amplitude of the exchange vertices between the entropy modes $\delta{s}_{\pm}$ are at the order of $I$ through the interaction $\delta{\cal H}_{{I}, 6}^{\rm SS}$ which is shown in right panel of Fig. \ref{fig2}.

Having obtained the free and the interaction Hamiltonians, we calculate the two-point correlation functions between all scalar modes which include power spectra and cross-correlations.

\subsubsection{$\langle\overline{\delta\sigma}\,\overline{\delta\sigma}\rangle$}
We first calculate the corrections to the power spectrum of the adiabatic mode $\overline{\delta\sigma}$. In what follows, we use the notation that $\Delta^{(1)}\langle\overline{\delta\sigma}^2\rangle_{i}$ stands for the case where a single Hamiltonian $\delta{H}^{\rm SS}_{{I},i}$ from interactions defined in (\ref{H2-int}) contributes to the two-point correlation function given by the first integral in Eq. (\ref{IN-IN}). On the other hand,  $\Delta^{(2)} \langle{\overline{\delta\sigma}^2}\rangle_{i,j}$ represents  the case of nested integrals in second line of Eq. (\ref{IN-IN}) containing two Hamiltonians where the indices $i, j$ correspond to $\delta{H}^{\rm SS}_{{I},i}(\tau_1)$ and $\delta{H}^{\rm SS}_{{I},j}(\tau_2)$ respectively. Adding all contributions, the total correction to the correlation function $\Delta \langle \overline{\delta\sigma}^2\rangle$ coming from the interaction Hamiltonians in Eq. (\ref{H2-int}) is given by
\begin{eqnarray}\label{correction-dsigma}
&&\Delta\langle {\overline{\delta\sigma}}(\tau_e, {\bf k}) 
{\overline{\delta\sigma}}(\tau_e, {\bf k}') \rangle = \Big(
\Delta^{(1)}\langle\overline{\delta \sigma}^2\rangle_{5} +
\Delta^{(2)} \langle \overline{\delta \sigma}^2\rangle_{1,1}
+\Delta^{(2)}\langle\overline{\delta \sigma}^2\rangle_{2,2}
+\Delta^{(2)}\langle\overline{\delta \sigma}^2\rangle_{1,3}
+\Delta^{(2)}\langle\overline{\delta \sigma}^2\rangle_{3,1}
\nonumber \\
&&
 +\Delta^{(2)}\langle\overline{\delta \sigma}^2 \rangle_{3,3}
+\Delta^{(2)}\langle\overline{\delta \sigma}^2 \rangle_{2,4}
+\Delta^{(2)}\langle\overline{\delta \sigma}^2 \rangle_{4,2} 
+\Delta^{(2)}\langle\overline{\delta \sigma}^2 \rangle_{4,4}
+\Delta^{(2)}\langle\overline{\delta \sigma}^2 \rangle_{5,5} \Big) (2\pi)^3\delta^{(3)}({\bf k}-{\bf k}') .
\end{eqnarray}

We need to calculate all of the above corrections using the in-in formula Eq. (\ref{IN-IN}). This is straightforward but cumbersome and we only present details of two cases as examples: 
\begin{eqnarray}
\label{correction-dsigma-4}\nonumber
\Delta^{(1)}\langle\overline{\delta \sigma}^2\rangle_5
= i \big \langle 0 \Big |\int_{\tau_{0}}^{\tau_{e}} d \tau_{1}\Big[\delta{H}_{{I},5}^{\rm SS}(\tau_{1}), \overline{\delta \sigma}^{2}(\tau_e,{\bf k})\Big] \Big | 0 \big \rangle  = -48 I\, {\rm Re} \Big[ i  \int_{\tau_{0}}^{\tau_{e}} \frac{d\tau_1}{\tau_1^2} 
\big( \overline{\delta\sigma}(\tau_1) \overline{\delta\sigma}^*(\tau_e) \big)^2 \Big]
= \frac{8I N_k}{k^3 \tau_e^2} ,
\end{eqnarray}
and
\begin{eqnarray}\label{correction-dsigma-11}
\Delta^{(2)} \langle\overline{\delta\sigma}^2\rangle_{1,1}
=\Big \langle0 \Big | \int_{\tau_{0}}^{\tau_{e}} d\tau_{1} \int_{\tau_{0}}^{\tau_{1}} d\tau_{2} 
\Big[ \delta{H}_{{I},1}^{\rm SS}(\tau_2), \Big[ \delta{H}_{{I},1}^{\rm SS}(\tau_1), 
\overline{\delta\sigma}^2(\tau_e,{\bf k}) \Big] \Big] \Big | 0 \Big \rangle \hspace{4.5cm} \\
\nonumber 
= 256 I \int_{\tau_0}^{\tau_e} \frac{d\tau_1}{\tau_1^2} \int_{\tau_0}^{\tau_1} 
\frac{d\tau_2}{\tau_2^2} {\rm Im}\Big[\overline{\delta\sigma}(\tau_{1}) 
\overline{\delta\sigma}^*(\tau_e)\Big] 
{\rm Im} \Big[\overline{\delta\sigma}(\tau_{2}) \overline{\delta\sigma}^*(\tau_e) 
\delta{s}_{+}(\tau_{2}) \delta{s}_{+}^*(\tau_1) \Big] 
=\frac{4 I N_k^2 \left(e^{8 \pi  \theta }-1\right)}{9 k^3\tau_e^2 \pi  
	\left(16 \theta^3+\theta \right)} ,
\end{eqnarray}
where $N_k = -\ln(-k \tau_e)$ is the number of e-folds when the mode of interest $k$ leaves the horizon till end of inflation. Since $N_k \sim 60$ to solve the flatness and the horizon problems we can ignore the first order corrections containing $I N_k$ in comparison to $I N_k^2$ in (\ref{correction-dsigma-11}). In other words, the dominant contributions to the power spectrum comes from the first two transfer vertices that are illustrated in the left panel of Fig. \ref{fig3}. 

Calculating all corrections in Eq. (\ref{correction-dsigma}), the power spectrum for the total curvature perturbation is obtained as in Eq.  \eqref{PS-R}.

\subsubsection{$\langle\overline{\delta\sigma}\delta{s}_{\pm}\rangle$}
In a way  similar to the previous subsection, using the in-in formula Eq. (\ref{IN-IN}) together with the interaction Hamiltonians Eq. (\ref{H2-int}), the cross-correlations between the adiabatic mode and the entropy modes are obtained to be  
\begin{equation}\label{CF-sigma-s-pm}
\Delta^{(1)} \langle \overline{\delta \sigma}(\tau_e, {\bf k}) {\delta s_{\pm}}(\tau_e, {\bf k}') \rangle 
= \mp \sqrt{2} \, e^{\pm 4 \pi  \theta } \Theta_1(\theta)
\frac{\sqrt{I} N_{k}}{k^3 \tau_{e}^2} (2\pi)^3\delta^{(3)}({\bf k}-{\bf k}')\,.
\end{equation}

From Eq. (\ref{entropy-perturbation}), we define the normalized entropy perturbations ${\cal S}_{\pm}$ corresponding to the entropy modes $\delta{s}_{\pm}$ as
\begin{eqnarray}\label{entropy-perturbation-N}
\mathcal{S}_{\pm} \equiv -\frac{{H}}{\dot \phi} \cos\vartheta \, \frac{\delta s_{\pm}}{a} \,.
\end{eqnarray}
Now, the cross-correlation between the curvature perturbations and the normalized entropy perturbations Eq. (\ref{entropy-perturbation-N}) is defined by $\langle {\cal R} {\cal S}_{\pm} \rangle \equiv \frac{2\pi^2}{k^3} {\mathcal{C}_{ \mathcal{R}\mathcal{S}}}_{\pm} (2\pi)^3\delta^{(3)}({\bf k}-{\bf k}')$, which using the result (\ref{CF-sigma-s-pm}), yields
\begin{eqnarray}\label{PS-SiR}
{\mathcal{C}_{ \mathcal{R}\mathcal{S}}}_{\pm} 
= \mp 2\sqrt{2}\, e^{\pm{4\pi\theta}} \Theta_1(\theta)
\sqrt{I} N_k \, {\cal P}_{\cal R}^{(0)} \, .
\end{eqnarray}

\subsubsection{$\langle\delta{s}_{+}\delta{s}_{-}\rangle$}
The entropy modes $\delta{s}_{\pm}$ are themselves correlated with each other. The corresponding cross-correlation turns out to be
\begin{eqnarray}
\Delta^{(1)} \langle {\delta s_{+}}(\tau_e, {\bf k}) {\delta s_{-}}(\tau_e, {\bf k}') \rangle 
&=& \frac{3}{4} \Delta^{(2)} \langle {\delta s_{+}}(\tau_e, {\bf k}) 
{\delta s_{-}}(\tau_e, {\bf k}') \rangle \\ \nonumber 
&=& 2 \left(1- 48 \theta ^2 \right) \Theta_1(\theta)^2
\frac{I N_{k}}{k^3 \tau_{e}^2} (2\pi)^3\delta^{(3)}({\bf k}-{\bf k}') \,,
\end{eqnarray}
which leads to the following cross-correlation between the associated normalized entropy perturbations 
\begin{eqnarray}\label{PS-S1S2}
{\mathcal{C}_{ \mathcal{S}_+\mathcal{S}}}_- 
= \frac{28}{3} \left(1-48 \theta ^2\right) \Theta_1(\theta)^2
I N_k {\cal P}_{\cal R}^{(0)} \,.
\end{eqnarray}

\subsubsection{$\langle\delta s_{\pm}\delta s_{\pm}\rangle$}

Finally, the two-point correlation function of the entropy modes $\delta{s}_{\pm}$ are given by
\begin{equation}
\Delta^{(1)}\langle \delta s_{\pm}(\tau_e, {\bf k}) \delta s_{\pm}(\tau_e, {\bf k}') \rangle = 
- 2 e^{\pm 8\pi\theta} \left(1+40 \theta^2\right) \Theta_1(\theta)^2
\frac{I N_{k}}{k^3 \tau_{e}^2} (2\pi)^3\delta^{(3)}({\bf k}-{\bf k}') \,,
\end{equation}
\begin{equation}
\Delta^{(2)}\langle \delta s_{\pm}(\tau_e, {\bf k}) \delta s_{\pm}(\tau_e, {\bf k}') \rangle = 
- \frac{8}{3} e^{\pm 8 \pi  \theta} 
\left(1+48 \theta^2\right) \Theta_1(\theta)^2 
\frac{I N_{k}}{k^3 \tau_{e}^2} (2\pi)^3\delta^{(3)}({\bf k}-{\bf k}') \,,
\end{equation}
which lead to the following power spectra for the normalized entropy modes
\begin{eqnarray}\label{PS-Si}
{\cal P}_{{\cal S}_{\pm}} 
= e^{\pm{4\pi\theta}} \Theta_1(\theta) {\cal P}_{\cal R}^{(0)}
\Big( 1 - \frac{4}{3} \left(7+312\theta^2\right) e^{\pm{4\pi\theta}} \Theta_1(\theta) I N_k \Big) \, .
\end{eqnarray}
From the above results we see that for positive $\theta$, the power spectrum ${\cal P}_{{\cal S}_{+}}$ is amplified exponentially which is a manifestation of chirality in 
vector fields perturbations due to parity violating term \cite{Barnaby:2010vf, Barnaby:2011vw, Barnaby:2011qe, Sorbo:2011rz}. In our setup since we have decomposed the perturbations into the adiabatic and entropy modes, the chirality is translated into the enhancement of the power of the entropy mode ${\cal S}_{+}$ compared to ${\cal S}_{-}$.

\subsection{Tensor modes}

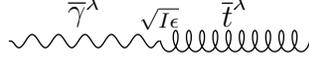
\begin{figure}
	\begin{center}
		\begin{tikzpicture}[line width=0.5 pt, scale=2]
		\draw[vector,black] (0:1)--(0,0);
		\begin{scope}[shift={(1,0)}]
		\draw[gluon,scale=1,line width=0.5 pt,black] (0:1)--(0,0);
		\end{scope}
		\node at (0.5,0.2){$\overline{\gamma}^{\lambda}$};
		\node at (1.5,0.2){$\overline{t}^{\lambda} $};
		\node at (1,0.15){\scriptsize $ \sqrt{I \epsilon }$};
		\end{tikzpicture}
		\caption{Exchange vertex Feynman diagrams between GWs tensor modes ${\overline\gamma}^\lambda$ and vector fields tensor modes ${\overline{t}}^\lambda$ with the same polarizations $\lambda=+,\times$.} \label{fig4}
	\end{center}
\end{figure}

Similar to what we did for the scalar modes, we can find the interaction Hamiltonian from the total quadratic Lagrangian (\ref{S2-T-pol}) which can be classified as $\delta{ H}^{\rm TT}_{I} \equiv \sum_i \delta{ H}^{\rm TT}_{{I},i}$ with
\begin{eqnarray}\label{H2-int-T}
&&\delta{ H}_{{I}, 1 }^{\rm TT} = \frac{8}{\tau^2} \sqrt{I\epsilon} \sum_{\lambda}
\overline{\gamma}^{\lambda} \overline{t}^{\lambda} \,, \hspace{0.5cm} 
\delta{ H}_{{I}, 2 }^{\rm TT} = -\frac{4}{\tau} \sqrt{I\epsilon}
\sum_{\lambda}  \overline{\gamma}^{\lambda } \overline{t'}^{\lambda}  \,, \\ 
&&\delta{ H}_{{I}, 3 }^{\rm TT} = \frac{5}{\tau^2} I\epsilon \sum_{\lambda}
\overline{t}^{\lambda } \overline{t}^{\lambda} \,, \hspace{.9cm}
\delta{ H}_{{I}, 4 }^{\rm TT} = \frac{2}{\tau^2} I\epsilon
\sum_{\lambda} \overline{\gamma}^{\lambda } \overline{\gamma}^{\lambda} \,,\nonumber
\end{eqnarray}
where again the integral over momenta in Fourier space are dropped for the sake of simple presentation. The Feynman diagrams associated with the first two interactions above are presented in Fig \ref{fig4} where the exchange vertex between $\overline{\gamma}^{\lambda }$ and $\overline{t}^{\lambda }$ is shown.

\subsubsection{$\langle\overline{\gamma}\,\overline{\gamma}\rangle$}
In this case, the leading corrections are given by
\begin{eqnarray}\label{CF-h-corrections}
&&\Delta \langle {\overline{\gamma^{\lambda}}} (\tau_e, {\bf k})\, 
\overline{\gamma^{\lambda}}' (\tau_e, {\bf k'})\rangle 
= \Big( \Delta^{(1)} \langle({\overline{\gamma}^{\lambda}})^2\rangle_{4} +
\Delta^{(2)} \langle({\overline{\gamma}^{\lambda}})^2\rangle_{11}
+\Delta^{(2)} \langle({\overline{\gamma}^{\lambda}})^2\rangle_{12} \nonumber \\
&& \hspace{4.9cm} +\Delta^{(2)} \langle({\overline{\gamma}^{\lambda}})^2\rangle_{21}
+ \Delta^{(2)} \langle({\overline{\gamma}^{\lambda}})^2\rangle_{22} 
\Big) \delta_{\lambda \lambda'} (2\pi)^3\delta^{(3)}({\bf k}-{\bf k}') \,.
\end{eqnarray}
Implementing the in-in formula Eq. (\ref{IN-IN}) and using the relevant interaction Hamiltonians from Eq.  (\ref{H2-int-T}), we find
\begin{eqnarray}
\Delta^{(1)} \langle({\overline{\gamma}^{\lambda}})^2\rangle_{4}
= 4 I \epsilon\, {\rm Re}\Big[ i \int_{\tau_{0}}^{\tau_{e}} \frac{d \tau_{1}}{\tau_{1}^2} \big( \overline{\gamma}_{k}(\tau_1) \overline{\gamma}^*_{k}(\tau_e) \big)^2 \Big] =
\frac{4 I \epsilon N_k}{3 k^3 \tau_e^2} , 
\,\,\,\, 
\end{eqnarray}
and

\begin{eqnarray}\label{nested-T}
\Delta^{(2)} \langle ({\overline{\gamma^{\lambda}}})^2 \rangle_{11}
&=& + 512 I\epsilon \int_{\tau_{0}}^{\tau_{e}} \frac{d \tau_{1}}{\tau_1^2} 
\int_{\tau_{0}}^{\tau_{1}} \frac{d\tau_{2}}{\tau_2^2} {\rm Im} \big[ \overline{\gamma}_{k}(\tau_1) 
\overline{\gamma}^*_{k}(\tau_e) \big] {\rm Im} \Big[\overline{\gamma}_{k}(\tau_2) \overline{\gamma}^*_{k}(\tau_e) 
\overline{t}_{k}(\tau_2) \overline{t}^*_{k}(\tau_1) \Big] \nonumber \\
&=& + \frac{64}{9} \Theta_{1}(\theta) e^{4 \lambda \pi \theta}
\big( 24 \theta^2-1+N_{k}\big) \frac{I \epsilon N_k}{\tau_e^2 k^3}
\,, \nonumber \\
\Delta^{(2)} \langle ({\overline{\gamma^{\lambda}}})^2\rangle_{12}
&=& -256 I\epsilon \int_{\tau_{0}}^{\tau_{e}} \frac{d \tau_{1}}{\tau_1^2} \int_{\tau_{0}}^{\tau_{1}} 
\frac{d \tau_{2}}{\tau_{2}}  {\rm Im} \big[ \overline{\gamma}_{k}(\tau_1) \overline{\gamma}^*_{k}(\tau_e) \big] 
{\rm Im} \Big[\overline{\gamma}_{k}(\tau_2) \overline{\gamma}^*_{k}(\tau_e) \overline{t'}_{k}(\tau_2)
\overline{t}^*_{k}(\tau_1) \Big] \nonumber \\ 
&=& -\frac{16}{9}  \Theta_{1}(\theta) e^{4 \lambda \pi \theta} 
\big( 96 \theta^2 + N_{k} \big) \frac{I \epsilon N_k}{\tau_e^2 k^3}
\,, \nonumber \\
\Delta^{(2)} \langle ({\overline{\gamma^{\lambda}}})^2\rangle_{21}
&=& -256 I\epsilon \int_{\tau_{0}}^{\tau_{e}} \frac{d\tau_{1}}{\tau_1} \int_{\tau_{0}}^{\tau_{1}} 
\frac{d\tau_{2}}{\tau_{2}^2}  {\rm Im} \big[ \overline{\gamma}_{k}(\tau_1) \overline{\gamma}^*_{k}(\tau_e) \big] 
{\rm Im} \Big[\overline{\gamma}_{k}(\tau_2) \overline{\gamma}^*_{k}(\tau_e) \overline{t}_{k}(\tau_2) 
\overline{t'}^*_{k}(\tau_1) \Big] \nonumber \\
&=& + \frac{32}{9} \Theta_{1}(\theta) e^{4 \lambda \pi \theta}
\big(24 \theta^2 +N_{k} \big) \frac{I \epsilon N_k}{\tau_e^2 k^3} 
\,, \nonumber \\
\Delta^{(2)} \langle ({\overline{\gamma^{\lambda}}})^2\rangle_{22}
&=& + 128 I\epsilon \int_{\tau_{0}}^{\tau_{e}} \frac{d\tau_{1}}{\tau_1} \int_{\tau_{0}}^{\tau_{1}} 
\frac{d\tau_{2}}{\tau_2}  {\rm Im} \big[ \overline{\gamma}_{k}(\tau_1) \overline{\gamma}^*_{k}(\tau_e) \big] 
{\rm Im} \Big[\overline{\gamma}_{k}(\tau_2) \overline{\gamma}^*_{k}(\tau_e) \overline{t'}_{k}(\tau_2) 
\overline{t'}^*_{k}(\tau_1) \Big] \nonumber \\ 
&=& - \frac{8}{9} \Theta_{1}(\theta) e^{4 \lambda \pi \theta} \big(-2+N_{k} \big) 
\frac{I \epsilon N_k}{\tau_e^2 k^3} 
\,.
\end{eqnarray}

Substituting the above results into Eq. (\ref{CF-h-corrections}) the total corrections to the 
power spectrum of ${\overline \gamma}^\lambda$ is obtained to be
\begin{equation}\label{CF-h-corrections-compute}
\Delta \langle {\overline{\gamma^{\lambda}}} (\tau_e, {\bf k})\, \overline{\gamma^{\lambda}}' (\tau_e, {\bf k'})\rangle
=\frac{8}{3} \Big( 1+ \big( 32 \theta^2 - 2 + 3 N_k \big) \Theta_{1}(\theta) e^{4 \lambda \pi \theta} \Big) 
\frac{I \epsilon N_k}{\tau_e^2 k^3} \delta_{\lambda \lambda'} (2\pi)^3 \delta^{(3)}({\bf k}-{\bf k}')\,.
\end{equation}
Since $N_k \sim 60$, the second term above is the leading correction which comes from the interaction Hamiltonians $\delta{ H}_{I,1}^{\rm TT} $ and $\delta{ H}_{I,2}^{\rm TT}$ in the nested integral in Eq. (\ref{nested-T}). These dominant corrections are corresponding to the Feynman diagrams Fig. \ref{fig5}. From the result Eq. \eqref{CF-h-corrections-compute} we find the power spectrum for the different polarizations of the gravitational tensor modes given in Eq. \eqref{PS-hh-pol}.

\subsubsection{$\langle\overline{t}\,\overline{t}\rangle$}

The dominant corrections to the power spectrum of the vector field tensor modes are given by
\begin{eqnarray}\label{CF-t-corrections}
\Delta \langle {\overline{t^{\lambda}}}(\tau_e, {\bf k}) \, \, \overline{t^{\lambda}}'
(\tau_e, {\bf k'})\rangle &=&\Big[
\Delta^{(1)} \langle ({\overline{t^{\lambda}}})^2\rangle_{3}
+ \Delta^{(2)} \langle({\overline{t^{\lambda}}})^2\rangle_{11}
+\Delta^{(2)} \langle({\overline{t^{\lambda}}})^2\rangle_{12} +\Delta^{(2)} \langle({\overline{t^{\lambda}}})^2\rangle_{21} \nonumber  \\ \nonumber 
&+& \Delta^{(2)} \langle({\overline{t^{\lambda}}})^2\rangle_{22} 
+ \Delta^{(2)} \langle({\overline{t^{\lambda}}})^2\rangle_{33}+\Delta^{(2)} 
\langle({\overline{t^{\lambda}}})^2\rangle_{31}+\Delta^{(2)} 
\langle({\overline{t^{\lambda}}})^2\rangle_{13}  \\ 
&+&\Delta^{(2)} \langle({\overline{t^{\lambda}}})^2\rangle_{23} 
+\Delta^{(2)} 
\langle({\overline{t^{\lambda}}})^2\rangle_{32}    \Big] \, 
\delta_{\lambda \lambda'} (2\pi)^3\delta^{(3)}({\bf k}-{\bf k}') \,.
\end{eqnarray}
Performing the corresponding in-in integral as in the case of GWs, we find 
\begin{eqnarray}
\Delta^{(1)} \langle ({\overline{t^{\lambda}}})^2\rangle_{3} &=&
+\frac{10}{3} \Theta_{1}(\theta)^{2} e^{8 \lambda \pi \theta}
\big(1+48 \theta^2\big) \frac{I \epsilon N_k}{k^3\tau_e^2} 
\,, \nonumber \\
\Delta^{(2)} \langle ({\overline{t^{\lambda}}})^2\rangle_{11} &=&
+\frac{64 }{9} \Theta_{1}(\theta)^2 e^{8 \lambda \pi \theta}
\Big[\Big(40 \theta^2+16 \lambda \frac{k \theta}{\tau_{e}}-1\Big)+N_{k}\Big] 
\frac{I \epsilon N_k}{k^3\tau_e^2}
\,, \nonumber \\
\Delta^{(2)} \langle ({\overline{t^{\lambda}}})^2\rangle_{12}&=&
-\frac{16}{9} \Theta_{1}(\theta)^2 e^{8 \lambda \pi \theta}
\Big[ \Big( 208 \theta^2+16 \lambda \frac{k \theta}{\tau_{e}} \Big) + N_k \Big] 
\frac{I \epsilon N_k}{k^3\tau_e^2}  \,, \nonumber \\
\Delta^{(2)} 
\langle ({\overline{t^{\lambda}}})^2\rangle_{21} &=&
-\frac{64}{9} \Theta_{1}(\theta)^2 e^{8\lambda \pi \theta}
\Big[\Big(52 \theta^2+16\lambda \frac{k \theta}{\tau_{e}}\Big)+N_k\Big]
\frac{I \epsilon N_k}{k^3\tau_e^2}   \,, \nonumber \\ 
\Delta^{(2)} \langle ({\overline{t^{\lambda}}})^2\rangle_{22} &=&
+\frac{16}{9} \Theta_{1}(\theta)^2 e^{8 \lambda\pi \theta}
\Big[\Big(112 \theta^2+16 \lambda \frac{k\theta}{\tau_{e}}+1\Big)+N_{k}\Big] \frac{I \epsilon N_k}{k^3\tau_e^2}  \,, \nonumber
\end{eqnarray}
with $\Delta^{(2)} \langle({\overline{t^{\lambda}}})^2\rangle_{33} \approx \mathcal{O}(I^2 \epsilon^{2}) $ and $\Delta^{(2)} \langle({\overline{t^{\lambda}}})^2\rangle_{13} =\Delta^{(2)} \langle({\overline{t^{\lambda}}})^2\rangle_{31} =\Delta^{(2)} \langle({\overline{t^{\lambda}}})^2\rangle_{23}=\Delta^{(2)} \langle({\overline{t^{\lambda}}})^2\rangle_{32}=0$. 

Summing the above corrections, the power spectrum of the tensor perturbations associated with the vector fields perturbations to leading order becomes Eq. \eqref{PS-t}.

\subsubsection{$\langle\overline{\gamma}\,\overline{t}\rangle$}

The cross-correlation between gravitational tensor modes and vector field tensor modes is non-vanishing in our setup as
\begin{eqnarray}\label{PS-ht-pol}
\Delta^{(1)} \langle \overline{\gamma^{\lambda}}(\tau_e, {\bf k})\,  
\overline{t^{\lambda'}}(\tau_e, {\bf k}') \rangle 
&=& 8 \sqrt{I \epsilon}\, {\rm Re} \Big[ i  \int_{\tau_{0}}^{\tau_{e}} \frac{d\tau_1}{\tau_1^2} (\tau_1-2) \overline{\gamma}_{k}(\tau_1) \overline{\gamma}^*_{k}(\tau_e)
\overline{\gamma}_{k}(\tau_1) \overline{\gamma}^*_{k}(\tau_e) \Big] \nonumber \\
&=& 2 \Theta_{1}(\theta) e^{4 \lambda  \pi \theta } \frac{\sqrt{I \epsilon}N_{k}}{k^3\tau_e^2} 
\delta_{\lambda\lambda'}(2\pi)^3\delta^{(3)}({\bf k}-{\bf k}') \, ,
\end{eqnarray}
from which we find Eq. \eqref{PS-ht-f}. 

\section{Cubic Interactions}
\setcounter{equation}{0}
\renewcommand{\theequation}{D\arabic{equation}}

In order to calculate the NGs we need the cubic interactions. In this appendix we first present the cubic Lagrangians and then obtain the corresponding cubic interaction Hamiltonians.

\subsection{Cubic Lagrangians}\label{app-A}

Expanding the action (\ref{action}) around background configuration (\ref{bac-A}) and (\ref{FRW}) up to third order for scalar and tensor perturbations defined in Eq. (\ref{perturbations}) we find the cubic Lagrangian $\delta{L}^{(3)}$ which we classify it as follows
\begin{equation}\label{cubic-L}
	\delta{L}^{(3)} = \delta{L}^{\rm SSS} + \delta{L}^{\rm SST} + \delta{L}^{\rm STT} + \delta{L}^{\rm TTT} \,,
\end{equation}
where, similar to the quadratic case, the superscripts ${\rm S}$ and ${\rm T}$ denote scalar and tensor modes respectively so that $\delta{L}^{\rm SST}$ represents cubic Lagrangian that includes interactions with two scalar modes and one tensor mode and so on.

\subsubsection{Scalar-scalar-scalar}

After tedious calculations and making some integration by parts, the cubic action of the form scalar-scalar-scalar, which includes only scalar modes, simplifies to
\begin{equation}\label{cubic-LSSS}
	\delta{L}^{\rm SSS} = \frac{H}{\sqrt{2\epsilon}} \int d^3x 
	\left( { \delta\mathcal{L}}_1^{\rm SSS} + \sqrt{I} \, 
	{ \delta\mathcal{L}}_2^{\rm SSS} + I \, { \delta\mathcal{L}}_3^{\rm SSS} \right) \,,
\end{equation}
with
\begin{eqnarray}
	\label{cubic-LSSS1}
	&&{ \delta\mathcal{L}}_1^{\rm SSS} = - \tau \overline{\delta\sigma} 
	\Big\{ \tau^4 \Big[\big( \tau^{-2}\overline{\delta{s}}\big)' \Big]^2 
	+ 2 \tau^4 \Big[\big( \tau^{-2}\partial\tilde{U} \big)' \Big]^2 
	+ (\partial\partial \overline{Y})^2 \Big\} \nonumber \\
	&& \hspace{1.5cm} + \tau \overline{\delta\sigma} 
	\Big[ 2 (\partial\overline{\delta s})^2 + (\partial^2 \tilde{U})^2 + (\partial\partial \tilde{U})^2 \Big] \nonumber \\
	&&\hspace{1.5cm} - 4 \theta \tau \Big[ \Big( 2 \tau^{-4} (\tau^4\overline{\delta\sigma})' \partial\overline{\delta s} 
	- \tau^2 \big( \tau^{-2}\overline{\delta{s}} \big)' \partial \overline{\delta\sigma} 
	+ \overline{\delta\sigma} \partial\overline{\delta s}' \Big) 
	\partial \tilde{U} + \overline{\delta\sigma} \partial\partial \tilde{U} \partial\partial \overline{Y} \Big] \,, \\ 
	\label{cubic-LSSS2}
	&&{ \delta\mathcal{L}}_2^{\rm SSS} 
	= -\tau \overline{\delta s} \Big[ \tau^4 \Big[\big( \tau^{-2}\overline{\delta{s}}\big)' \Big]^2 
	+ 2 \tau^4 \Big[\big( \tau^{-2}\partial\tilde{U} \big)' \Big]^2 
	+ (\partial\partial \overline{Y})^2 \Big] \nonumber \\
	&&\hspace{1.5cm}+ \tau \overline{\delta s} \Big[ 2 (\partial\overline{\delta s})^2 
	+ (\partial^2 \tilde{U})^2 + (\partial\partial \tilde{U})^2 \Big] \nonumber \\
	&&\hspace{1.5cm}+ \frac{8}{\tau} \overline{\delta s}\,  \overline{\delta\sigma}^2 - 2 \overline{\delta\sigma}^2\,  \overline{\delta s}' + 2\tau \overline{\delta\sigma} \, 
	\overline{\delta\sigma}' \,  \overline{\delta s}' - 4\tau \overline{\delta\sigma} \, \partial\overline{\delta\sigma}\,  \partial \overline{\delta s} \nonumber \\ 
	&&\hspace{1.5cm} - 4\theta \tau \Big[ ( \overline{\delta\sigma} \partial\overline{\delta\sigma} 
	- \overline{\delta s} \partial \overline{\delta s} ) \big( \partial \tilde{U}' 
	- \frac{9}{\tau} \partial \tilde{U} \big) 
	+ \overline{\delta s} \partial\partial \tilde{U} \partial\partial \tilde{Y} \Big] \,, \\
	\label{cubic-LSSS3}
	&&{ \delta\mathcal{L}}_3^{\rm SSS} = 
	\frac{4}{\tau} \overline{\delta \sigma}^3 + \frac{12}{\tau} \overline{\delta\sigma} \, \overline{\delta s}^2 
	+ \tau \overline{\delta\sigma}\,  \overline{\delta s}'^2 + 2\tau \overline{\delta s} \, \overline{\delta s}'  \, \overline{\delta\sigma}' - \tau \overline{\delta\sigma} \, \overline{\delta\sigma}'^2 
	\nonumber \\
	&&\hspace{1.5cm}- 2\tau \overline{\delta\sigma} (\partial\overline{\delta s})^2 - 4\tau \overline{\delta s} \partial\overline{\delta s} \partial\overline{\delta\sigma} 
	+ 2\tau \overline{\delta\sigma} (\partial\overline{\delta\sigma})^2
	\nonumber \\
	&&\hspace{1.5cm} 
+ 2 \theta \big( 12 \overline{\delta\sigma} \partial\overline{\delta s} 
+ \tau \overline{\delta\sigma} \partial \overline{\delta s}' 
	+ 18 \overline{\delta s} \partial\overline{\delta\sigma} 
	+ 3\tau \overline{\delta s}' \partial \overline{\delta\sigma} + 2\tau \overline{\delta s} 
	\partial\overline{\delta\sigma}' \big) \partial \tilde{U} ,
\end{eqnarray}
where $\overline{\delta\sigma}$ and $\overline{\delta{s}}$ are the adiabatic and entropy modes that are defined in Eqs. (\ref{adiabtic}) and (\ref{entropic}) and we have dropped spatial indices for the sake of simplicity in notation. In addition,  similar to the other vector fields scalar modes in (\ref{canonical-S0}), we have defined the canonical field $\overline{Y}$ associated to the non-dynamical scalar mode $Y$ as 
\begin{equation}\label{Y-canonical}
	\overline{Y} \equiv \sqrt{2}f Y \,.
\end{equation}

We therefore need to substitute $\overline{Y}$ to the linear order in the above interaction Lagrangian. Substituting from Eqs. (\ref{adiabtic}) and (\ref{entropic}) in (\ref{Y}) and then using Eq. (\ref{Y-canonical}), we find the solution at the linear order as
\begin{equation}\label{y-can}
	\partial^2\overline{Y} = \frac{\sqrt{2}}{2} \tau^2 \Big[ \Big(\frac{\overline{\delta s}}{\tau^2}\Big)' 
	- \sqrt{I} \Big(\frac{\overline{\delta \sigma}}{\tau^2}\Big)' \Big] \,,
\end{equation}
which expresses the spatial Laplacian of $\overline{Y}$ in terms of the adiabatic and entropy modes. We also need $\partial\partial{\overline Y}$ up to linear order to substitute in the cubic action. Solving Eq. (\ref{Y0}) to first order of perturbation, we find
\begin{eqnarray}\label{y-can-ij}
	\partial_i \partial_j {\overline Y} - \frac{1}{3} \partial^2 \overline{Y} \delta_{ij} = M_{ij} \,,
\end{eqnarray} 
where $\partial^2 \overline{Y}$ is given by Eq. (\ref{y-can}) and $M_{ij}$ is a symmetric spatial rank two tensor satisfying
\begin{eqnarray}\label{Mij}
M^i{}_i = 0 \,, \hspace{1cm} \partial^i\partial^j M_{ij} = 0 \,.
\end{eqnarray}

To find the explicit form of $M_{ij}$, we first note that since our model (\ref{action}) is isotropic, following the SVT theorem, the scalar, vector and tensor modes do not couple to each other at the linear order of perturbations. Second, we note that $\delta_{ij}$ and the Levi-Civita tensor $\epsilon_{ijm}$ are the only invariant tensors on the spatial manifold. In this regard, $M_{ij}$ would have the following general form
\begin{eqnarray}\label{M-generalform}
M_{ij} = \partial_i \partial_j S_1 - \frac{1}{3} \partial^2S_1 \delta_{ij} \,,
\end{eqnarray}
where $S_1$ is a first order scalar perturbation that can be generally constructed from the linear combinations of dynamical scalar modes $(\overline{\delta\sigma},\overline{\delta{s}},\tilde{U})$. 
The traceless condition is satisfied trivially by the above solution while the transverse condition implies that $\partial^2\partial^2S_1=0$. The scalar modes $(\overline{\delta\sigma},\overline{\delta{s}},\tilde{U})$ are independent and therefore we conclude $S_1=0$ which results in
\begin{eqnarray}\label{y-can-ij-2}
	\partial_i \partial_j {\overline Y} = \frac{1}{3} \partial^2 \overline{Y} \delta_{ij} \,,
\end{eqnarray} 
with the explicit form of $\partial^2 \overline{Y}$  given by Eq. (\ref{y-can}). 

Moreover, we compute bispectra at the super-horizon limit $k\tau\to0$ so we discard 
the cubic interactions that are suppressed in this limit. We therefore compare the cubic interactions looking at their spatial derivatives. Taking this fact into account, we are left  only with the leading interactions and the cubic actions then take simple forms. 

The canonical scalar and entropy modes $\overline{\delta\sigma}$ and $\overline{\delta s}$ are free of any spatial derivative and we take them to be of the order ${\cal O}(k^0)$ in the amplitude of wave vector. From Eq. (\ref{Y-canonical}), we conclude that $\overline{Y}$ is of the order ${\cal O}(k^{-2})$ which shows that $\partial\partial\overline{Y}$ is of the same order as $\overline{\delta\sigma}$ and $\overline{\delta{s}}$. In the same manner, we conclude that $\tilde{U}$ is of the order ${\cal O}(k^{-1})$ and then $\partial\tilde{U}$ is of the order ${\cal O}(k^{0})$. In this regard, all the interactions in the first line of the Lagrangian (\ref{cubic-LSSS1}) are of the order ${\cal O}(k^0)$, the interactions in the second line are of the order ${\cal O}({k^2})$, and the interactions in third line are of the order ${\cal O}(k\theta)$. We can neglect the interactions in the second line in comparison with the first line. Similarly, we ignore all other terms that are suppressed in the super-horizon limit $k\tau\to0$ in Eqs. (\ref{cubic-LSSS1})-(\ref{cubic-LSSS3}). 

Substituting from Eq. (\ref{y-can-ij-2}), the leading interactions for the cubic Lagrangian with three scalar modes which is defined in Eq. (\ref{cubic-LSSS}) are given by
\begin{eqnarray}\label{cubic-LSSS-red}
	&&\delta{L}^{\rm SSS} \approx 
	- \frac{H}{\sqrt{2\epsilon}} \int d^3x
	\bigg\{ 
	\overline{\delta\sigma} \tau^5 \Big[ \big( \tau^{-2}\overline{\delta{s}}\big)' \Big]^2
	+ \tau (\overline{\delta\sigma} + \sqrt{I} \overline{\delta{s}}) \Big( 
	2 \tau^4 \Big[ \big( \tau^{-2}\partial\tilde{U}\big)' \Big]^2 
	+ \frac{1}{3}(\partial^2 \overline{Y})^2 \Big) \nonumber \\
	&& \hspace{1.5cm}+ 4 \theta \tau \Big[ \Big( 2 \tau^{-4} (\tau^4\overline{\delta\sigma})' \partial\overline{\delta s} 
	- \tau^2 \big( \tau^{-2}\overline{\delta{s}} \big)' \partial \overline{\delta\sigma} 
	+ \overline{\delta\sigma} \partial\overline{\delta s}' \Big) 
	\partial \tilde{U} + \frac{1}{3}\overline{\delta\sigma} \partial^2 \overline{Y} \partial^2 \tilde{U} \Big] \nonumber \\ 
	&& \hspace{1.5cm}+ \sqrt{I} \Big[
	\tau^5 \overline{\delta s} \Big[ \big( \tau^{-2}\overline{\delta{s}}\big)' \Big]^2
	- \frac{8}{\tau}\,  \overline{\delta s}\,  \overline{\delta\sigma}^2 + 2 \overline{\delta\sigma}^2 \overline{\delta{s}}' 
	- 2\tau \overline{\delta\sigma} \, \overline{\delta\sigma}' \, \overline{\delta{s}}'  \\ \nonumber
	\nonumber 
	&& \hspace{1.5cm}+ 4\theta \tau \Big[ ( \overline{\delta\sigma} \partial\overline{\delta\sigma} 
	- \overline{\delta s} \partial \overline{\delta s} ) 
	\big( \partial \tilde{U}' - \frac{9}{\tau} \partial \tilde{U} \big) 
	+ \frac{1}{3} \overline{\delta s} \partial^2 \overline{Y} \partial^2 \tilde{U} \Big]\Big] \, \\ \nonumber
	&& \hspace{1.5cm}- I \, \Big(
	\frac{4}{\tau} \overline{\delta\sigma}^3 + \frac{12}{\tau} \overline{\delta\sigma}\,  \overline{\delta s}^2 + \tau \overline{\delta\sigma}\,  \overline{\delta{s}}'^2 
	+ 2\tau \overline{\delta s} \, \overline{\delta{s}}' \, \overline{\delta{\sigma}}'-\tau \overline{\delta\sigma} \partial\tilde{U}'^2
	\\ \nonumber
	&& \hspace{1.5cm}- 2 \theta \big( 12 \overline{\delta\sigma} \partial\overline{\delta s} 
	+ \tau \overline{\delta\sigma} \partial \overline{\delta s}' 
	+ 18 \overline{\delta s} \partial\overline{\delta\sigma} + 3\tau \overline{\delta s}' \partial \overline{\delta\sigma} 
	+ 2\tau \overline{\delta s} \partial\overline{\delta\sigma}' \big) \partial \tilde{U} \Big)
	\bigg\} \,,
\end{eqnarray}
where the explicit form of $\partial^2 \overline{Y}$ is given by Eq. (\ref{y-can}). The first two lines do not include factor $I$ and they are larger than the other interactions.

\subsubsection{Scalar-scalar-tensor}

The next term in Eq. (\ref{cubic-L}) is the cubic action of the form scalar-scalar-tensor which, after direct calculation, turns out to be
\begin{equation}\label{cubic-LSST}
	\delta{L}^{\rm SST} = \frac{H}{\sqrt{\epsilon}} \int d^3x
	\left( { \delta\mathcal{L}}_1^{\rm SST} + \sqrt{I} \, { \delta\mathcal{L}}_2^{\rm SST} 
	+ I \, { \delta\mathcal{L}}_3^{\rm SST} \right) \,,
\end{equation}
with
\begin{eqnarray}
	\label{cubic-LSST1}
	{ \delta\mathcal{L}}_1^{\rm SST} &=& 
	2\tau \overline{\delta\sigma} \Big[ \Big( \tau^2 \big(\tau^{-2}{\overline t}_{ij}\big)' 
	- 2 \theta \epsilon_j^{mn} \partial_n {\overline t}_{im} \Big) \partial^i\partial^j \overline{Y} + \Big( 2\theta \tau^2 
	\big( \tau^{-2}{\overline t}_{ij} \big)' 
	+ \epsilon_j^{mn} \partial_n {\overline t}_{im} \Big) \partial^i\partial^j \tilde{U} \Big] , \nonumber \\
	\label{cubic-LSST2}
	{ \delta\mathcal{L}}_2^{\rm SST} &=& 2\tau\overline{\delta s} \Big[ \tau^2 
	\big( \tau^{-2}{\overline t}_{ij} \big)' 
	\partial^i\partial^j (\overline{Y} + 2\theta \tilde{U}) + \epsilon_i^{mn} \partial_n {\overline t}_{jm} \partial^i\partial^j (\tilde{U} 
	- 2\theta \overline{Y}) \Big] \nonumber \\
	\label{cubic-LSST3}
	{ \delta\mathcal{L}_3^{\rm SST}} &=& \frac{\sqrt{\epsilon}}{2} \tau 
	\Big[ \big( \frac{6}{\tau}\overline{\delta s} 
	+ \overline{\delta{s}}' \big) \partial^i\partial^j \overline{Y} + (\partial^i\overline{\delta\sigma}\partial^j\overline{\delta\sigma} 
	- \partial^i\overline{\delta s} \partial^j\overline{\delta s}) + \epsilon^{jnm} \partial_n \overline{\delta s} \partial_m \partial^i \tilde{U}
	\Big] {\overline \gamma}_{ij} \,. \nonumber
\end{eqnarray}

Substituting Eq. (\ref{y-can-ij-2}), we find $\partial_{i}\partial_j Y$ does not contribute. Moreover, $\gamma_{ij}$ and $t_{ij}$ are of the order of ${\cal O}(k^0)$, and we have ignored suppressed terms in the limit $k\tau\to0$, so Eq. (\ref{cubic-LSST}) simplifies to
\begin{eqnarray}\label{cubic-LSST-red-f}
	\delta{L}^{\rm SST} \approx 4\theta \tau^3 \frac{H}{\sqrt{\epsilon}} \int d^3x
\	(\overline{\delta\sigma}
	+\sqrt{I} \, \overline{\delta s}) \Big(\frac{{\overline t}_{ij}}{\tau^2}\Big)' \partial^i\partial^j \tilde{U} \,.
\end{eqnarray}

\subsubsection{Scalar-tensor-tensor}

For the cubic action of the form scalar-tensor-tensor, we find 
\begin{eqnarray}\label{cubic-LSTT}
&&	\delta{L}^{\rm STT} = - \sqrt{\frac{2}{\epsilon}} H \tau \int d^3x\bigg\{
	(\overline{\delta\sigma}+\sqrt{I}\,\overline{\delta{s}}) 
	\bigg( \tau^4 \Big[ \big(\tau^{-2}{\overline t}_{ij}\big)' \Big]^2 
	+ (\partial_j {\overline t}_{in}-\partial_n {\overline t}_{ij} ) \partial^n {\overline t}^{ij} 
	\\ \nonumber && \hspace{1.5cm}
	-4 \theta \tau^2 \overline{\delta\sigma} \big( \tau^{-2}{\overline t}_{ij} \big)' 
	\epsilon_j^{mn} \partial_n {\overline t}_{im} \bigg)  
	+ \frac{\sqrt{\epsilon}}{2} I \Big( (6 \tau \overline{\delta s} + \tau^2 \overline{\delta{s}}') 
	\Big(\frac{{\overline t}^{ij}}{\tau^2}\Big)' - \partial_n \overline{\delta s} (\partial^n {\overline t}^{ij} - \partial^i {\overline t}^{nj})
	\Big){\overline \gamma}_{ij}  \bigg\} . \nonumber
\end{eqnarray}
The leading terms in the super-horizon limit are obtained to be,
\begin{eqnarray}\label{cubic-LSTT-red}
\delta{L}^{\rm STT} &\approx& - \sqrt{\frac{2}{\epsilon}} \tau H \int d^3x \Big[ 
(\overline{\delta\sigma}+\sqrt{I}\, 
\overline{\delta s}) \tau^4  \Big(\frac{{\overline t}_{ij}}{\tau^2}\Big)'
- 4 \tau^2 \theta \overline{\delta\sigma} \epsilon_j^{mn} \partial_n \overline{t}_{im} 
\Big] \Big(\frac{{\overline t}^{ij}}{\tau^2}\Big)' \,.
\end{eqnarray}

\subsubsection{Tensor-tensor-tensor}

The cubic action for pure tensor modes turns out to be
\begin{eqnarray}\label{cubic-LTTT}
	\delta{L}^{\rm TTT} &=& 
	\tau H \int d^3 x\Big\{ \tau^4 \Big[ \big(\tau^{-2}{\overline t}_{ij}\big)' \Big]^2 
	- \big( \partial^i {\overline t}^{mn} \partial^j {\overline t}_{mn} - 2 \partial_n {\overline t}^{im} \partial^j {\overline t}_m^n 
	+ \partial_n {\overline t}^{im} \partial^n {\overline t}^j_m \big) \nonumber \\
	&+& 4 \theta \tau \epsilon^{jml} \Big[ \Big(\frac{{\overline t}^{in}}{\tau^2}\Big)'  \partial_l {\overline t}_{mn} 
	+ \Big(\frac{{\overline t}^{n}_l}{\tau^2}\Big)' (\partial_m {\overline t}^{i}_n - \partial^i {\overline t}_{mn}) \Big] 
	\Big\} {\overline \gamma}_{ij} + \delta{L}^{{\overline \gamma}\, {\overline \gamma}\, {\overline \gamma}} \,,
\end{eqnarray}
where $\delta{L}^{{\overline \gamma}\, {\overline \gamma}\, {\overline \gamma}}$ is the cubic Lagrangian for the gravitational tensor modes. We know that NG induced by  $\delta{L}^{{\overline \gamma}\, {\overline \gamma}\, {\overline \gamma}}$ is small \cite{Maldacena:2002vr} and therefore we do not consider it here.

Taking the super-horizon limit, the leading interactions in the above cubic Lagrangian are given by
\begin{eqnarray}\label{cubic-LTTT-red}
\delta{L}^{\rm TTT} \approx \tau H \int d^3x
\Big[ \tau^4 \Big[ \big(\tau^{-2}{\overline t}_{ij}\big)' \Big]^2  + 4 \theta \tau \epsilon^{jml} 
\Big[ \Big(\frac{{\overline t}^{in}}{\tau^2}\Big)'  \partial_l {\overline t}_{mn} + 
\Big(\frac{{\overline t}^{n}_l}{\tau^2}\Big)' (\partial_m {\overline t}^{i}_n - \partial^i {\overline t}_{mn}) \Big] \Big] {\overline \gamma}_{ij} \,.
\end{eqnarray}

Finally we comment on the contributions that can potentially come from the quadratic action. We have already obtained the linear equation of motion for non-dynamical mode $Y$ in Eq. (\ref{Y0}) at the first order of perturbations and also we have solved it in Eq. (\ref{y-can-ij}). We, however, note that there are some other second order corrections to the equation of motion of $Y$ which can be obtained if we take into account the effects of the cubic Lagrangians $\delta{L}^{\rm SSS}$ and $\delta{L}^{\rm SST}$ defined in Eqs. (\ref{cubic-LSSS}) and (\ref{cubic-LSST}) respectively. To see this fact, we note that in quadratic Lagrangian (\ref{S2-S0}), the non-dynamical mode appeared as $\delta{L}^{\rm SS} \supset \partial^2 Y ( \partial^2 Y - 2S_2)$ where $S_2\equiv\delta Q'+2 A' f^{-1}f_{,\phi} \delta \phi $ is the solution at the first order $\partial^2 Y^{(1)} = S_2$ as can be seen from (\ref{Y}). The cubic contribution that may come from the second order corrections to the non-dynamical mode then would take the form $\delta{L}^{(3)} \supset \partial^2 Y^{(2)} (\partial^2 Y^{(1)}-S_2) = 0$ where we have used the linear equation of motion in the parenthesis. Therefore, there is not any cubic contribution coming from the second order corrections to the equation of motion of non-dynamical field $Y$.

\subsection{Cubic interaction Hamiltonians}\label{app-B}

Here we calculate the cubic interaction Hamiltonians from the cubic Lagrangians that we computed in the previous subsection. We have already obtained the quadratic interaction Hamiltonians for the scalar and tensor modes in Eqs. (\ref{H2-int}) and (\ref{H2-int-T}) respectively. Similar to the case of quadratic interaction Hamiltonians, we have $\delta{ H}_{I}^{(3)} \neq - \delta{ L}^{(3)}$ due to the existence of the terms with time derivatives. Moreover, we note that quadratic Lagrangian gives some cubic contributions to the cubic Hamiltonian when we work in interaction picture \cite{Chen:2009zp}. Therefore, we work with the total Lagrangian up to the cubic order that is the sum of quadratic Lagrangians (\ref{LSS-sigma-s}) and (\ref{LTT}), and cubic Lagrangians Eqs. (\ref{cubic-LSSS-red}), (\ref{cubic-LSST-red-f}), (\ref{cubic-LSTT-red}), and (\ref{cubic-LTTT-red}) as 
\begin{equation}\label{L-tot}
{L}^{\rm tot} = {L}^{\rm SS} + {L}^{\rm TT} + \delta{L}^{\rm SSS}
+\delta{L}^{\rm SST} + \delta{L}^{\rm STT} + \delta{L}^{\rm TTT} \,.
\end{equation}

We should perform Legendre transformation on this total Lagrangian to find the total Hamiltonian. In order to do this, we use the compact notation of $\mathcal{Q}^J \equiv \{ \overline{\delta \sigma},\overline{\delta s}, \partial_{i} \tilde{U}, \overline{\gamma}_{ij}, \overline{t}_{ij} \}$. The associated conjugate momenta $\Pi_J \equiv \{\Pi^{\delta \sigma},\Pi^{\delta s},\Pi^{U}_{i},\Pi^{\gamma}_{ij}, \Pi^{t}_{ij}\}$ are given by the Legendre transformation as follows
\begin{equation}
\Pi^{\delta \sigma} = \frac{\partial { L}^{\rm tot}}{\partial \overline{\delta \sigma}'} , \hspace{0.5cm} 
\Pi^{\delta s}=\frac{\partial { L}^{\rm tot}}{\partial \overline{\delta s}'} , \hspace{0.5cm} 
\Pi^{U}_{i}=\frac{\partial { L}^{\rm tot}}{\partial(\partial^{i} \tilde{U}')} , \hspace{0.5cm} 
\Pi^{\gamma}_{ij}=\frac{\partial { L}^{\rm tot}}{\partial \overline{\gamma}'^{ij}} , \hspace{0.5cm} 
\Pi^{t}_{ij}=\frac{\partial { L}^{\rm tot}}{\partial \overline{t}'^{ij}} \,.
\end{equation}

The total Hamiltonian is given by ${ H}^{\rm tot} = \Pi_J {\cal Q}'^J - { L}^{\rm tot}$ as
\begin{equation}
{H}^{\rm tot} = \Pi^{\delta \sigma} \overline{\delta \sigma}'+\Pi^{\delta s} \overline{\delta s}'
+\Pi^{U}_{i} \partial^{i} \tilde{U}' +\Pi^{\gamma}_{ij}  \overline{\gamma}'^{ij}+\Pi^{t}_{ij}  \overline{t}'^{ij} - {L}^{\rm tot} \,.
\end{equation}

Substituting from Eqs. (\ref{cubic-LSSS-red}), (\ref{cubic-LSST-red-f}), (\ref{cubic-LSTT-red}), and (\ref{cubic-LTTT-red}) in (\ref{L-tot}) and then using the result in the above relation, we obtain the explicit expression for the total Hamiltonian. Working with the interaction picture fields and then expressing the results in terms of the time derivative of interaction picture fields, it is cumbersome but straightforward to find the total Hamiltonian  in the interaction picture.

In the case of three scalar modes, the leading terms in the cubic interaction Hamiltonian in interaction picture are given by
\begin{eqnarray}\label{H-int-SSS}
&&\delta{H}_{I}^{\rm SSS} \approx \frac{H}{\sqrt{2\epsilon}} \int d^3 x \Bigg\{ 
I \bigg[ \frac{2}{\tau} \overline{\delta \sigma}^3 + \frac{5}{2} \tau \overline{\delta \sigma} \, \overline{ \delta \sigma}'^2
+ 55 \overline{\delta s}^2 \, \overline{ \delta \sigma}' 
- \tau^6\Big( ( \tau^{-2}\overline{\delta{s}} )' \Big)^2 \overline{\delta \sigma} 
- 5 \tau \overline{\delta s}\,  \overline{\delta s}' \, \overline{ \delta \sigma}' 
\bigg] \nonumber \\
\nonumber 
&& + \tau^6 ( \overline{\delta \sigma} + \sqrt{I}\, \overline{\delta{s}} ) 
\bigg[ \frac{5}{2} \Big( ( \tau^{-2}\overline{\delta{s}} )' \Big)^2 
+ 2 \Big( ( \tau^{-2}\partial\tilde{U} )' \Big)^2
- 20 \frac{\sqrt{I}}{\tau^6} \Big( \overline{\delta s} \, \overline{ \delta \sigma} 
+ \frac{5}{2} \tau \overline{\delta s} \, \overline{ \delta \sigma}' 
+ \frac{\tau^2}{4} \overline{ \delta s}' \,  \overline{ \delta \sigma}' \Big) \bigg] \nonumber \\ 
&& + 2 \theta 
\Bigg[ 4 \tau^3 \overline{\delta \sigma}\partial \overline{\delta s} \Big(\frac{\partial\tilde{U}}{\tau^2}\Big)'
-(2+\sqrt{2})\overline{\delta \sigma} \partial^2 \tilde{U} 
\Big(\frac{\overline{\delta s}}{\tau^2}\Big)' + \sqrt{I}\bigg( 
\partial^2 \tilde{U} \tau^3\Big(\overline{\delta \sigma}
\Big(\frac{\overline{\delta\sigma}}{\tau^2}\Big)'-\overline{\delta s}
\Big(\frac{\overline{\delta s}}{\tau^2}\Big)'\Big) \\ \nonumber 
&&\hspace{1.5cm} +4 \tau^3 \big( \overline{\delta s}\partial \overline{\delta s}
-\overline{\delta \sigma}\partial \overline{\delta \sigma} \big)
\Big(\frac{\partial\tilde{U}}{\tau^2}\Big)'+2 \sqrt{2} 
(\overline{\delta \sigma}^2-\overline{\delta s}^2) \partial^2 \tilde{U}
-\sqrt{2} \tau (\overline{\delta \sigma} \,  \overline{\delta \sigma}'-\overline{\delta s}\,  
\overline{\delta s}') \partial^2 \tilde{U}\bigg) \\
\nonumber 
&&- I \tau^3 \bigg( \overline{\delta s} \Big( \partial \overline{\delta \sigma} 
\Big(\frac{\partial \tilde{U}}{\tau^2}\Big)'
-\sqrt{2} \partial^2 \tilde{U}\Big(\frac{\partial \overline{\delta \sigma}}{\tau^2}\Big)' \Big)
- \overline{\delta \sigma} \partial \overline{\delta s} 
\Big(\frac{\partial \tilde{U}}{\tau^2}\Big)' +\Big(2 \overline{\delta s} 
\Big(\frac{\overline{\delta \sigma}}{\tau^2}\Big)'+\overline{\delta \sigma} 
\Big(\frac{\overline{\delta s}}{\tau^2}\Big)'\Big) \partial^2 \tilde{U} \bigg) \Bigg] \Bigg\},
\end{eqnarray}
where the subscript $I$ in the left hand side denotes that the fields are in the interaction picture while we do not use a new notation for the fields for the sake of simplicity. Comparing the cubic interaction Hamiltonian (\ref{H-int-SSS}) with the corresponding cubic Lagrangian (\ref{cubic-LSSS-red}), we see that $\delta{H}_{I}^{\rm SSS} \neq - \delta{L}^{\rm SSS}$ and there are some other contributions which are originated from the cross terms in the quadratic Lagrangians that include time derivative.

For the case of two scalar modes and one tensor, the interaction Hamiltonian in the interaction picture is given by
\begin{equation}
\label{H-int-SST}
\delta{H}_{I}^{\rm SST} \approx -\frac{4\theta}{\sqrt{\epsilon}} \tau^3 H \int d^3x
(\overline{\delta\sigma}
+\sqrt{I}\overline{\delta s}) \Big(\frac{{\overline t}_{ij}}{\tau^2}\Big)' \partial^i\partial^j \tilde{U} \,.
\end{equation}

In the case of one scalar mode and two tensor modes we find
\begin{eqnarray}\label{H-int-STT}
\delta{H}_{I}^{\rm STT} \approx \sqrt{\frac{2}{\epsilon}} \tau H \int d^3x \Big[ 
(\overline{\delta\sigma}+\sqrt{I}\, 
\overline{\delta s}) \tau^4  \Big(\frac{{\overline t}_{ij}}{\tau^2}\Big)'
- 4 \tau^2 \theta \overline{\delta\sigma} \epsilon_j^{mn} \partial_n \overline{t}_{im} 
\Big] \Big(\frac{{\overline t}^{ij}}{\tau^2}\Big)' \, ,
\end{eqnarray}
where only the leading terms are kept.

For the  case of three tensor modes, the leading terms to the cubic interaction Hamiltonian are given by
\begin{equation}\label{H-int-TTT}
\delta{H}_{I}^{\rm TTT}\approx - \tau^3 H \int d^3x \bigg[
\tau^2 \big( (\tau^{-2}{\overline t}_{ij})' \big)^2  + 4\theta \epsilon^{jml} 
\Big[ \Big(\frac{{\overline t}^{in}}{\tau^2}\Big)'  \partial_l {\overline t}_{mn} + \Big(\frac{{\overline t}^{n}_l}{\tau^2}\Big)' 
(\partial_m {\overline t}^{i}_n - \partial^i {\overline t}_{mn}) \Big]  \bigg] {\overline \gamma}_{ij} \,.
\end{equation} 

Comparing the above cubic interaction Hamiltonians with the corresponding cubic Lagrangians Eqs. (\ref{cubic-LSST-red-f}), (\ref{cubic-LSTT-red}), and (\ref{cubic-LTTT-red}), we find that $\delta{H}_{I}^{\rm SST} = - \delta{L}^{\rm SST}$, $\delta{H}_{I}^{\rm STT} = - \delta{L}^{\rm STT}$, and $\delta{H}_{I}^{\rm TTT} = - \delta{L}^{\rm TTT}$, respectively.

Having obtained all quadratic interaction Hamiltonians in Eqs. (\ref{H2-int}) and (\ref{H2-int-T}), and all cubic interaction Hamiltonians in Eqs. (\ref{H-int-SSS}), (\ref{H-int-SST}), (\ref{H-int-STT}), and (\ref{H-int-TTT}), we can compute any three-point function in our model by means of the in-in formalism (\ref{IN-IN-def}).

\section{Calculation of $ \langle {\cal RRR} \rangle$}\label{app-RRR}
\setcounter{equation}{0}
\renewcommand{\theequation}{E\arabic{equation}}

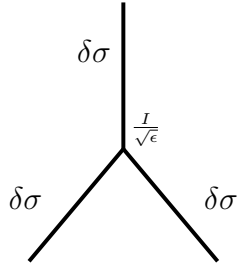
\begin{figure}
	\begin{center}
		\begin{tikzpicture}[line width=1.5 pt, scale=1.3]
		\begin{scope}[rotate=90]
		\draw[fermionnoarrow] (-140:1.5)--(0,0);
		\draw[fermionnoarrow] (140:1.5)--(0,0);
		\draw[fermionnoarrow] (0:1.5)--(0,0);
		\node at (-0.5,-1) {$\delta \sigma$};
		\node at (-0.5,1) {$\delta \sigma$};
		\node at (1,0.3)  {$\delta \sigma$};	
		\node at (0.2,-0.23)  {{\scriptsize  $\frac{I}{\sqrt{\epsilon}}$}};	
		\end{scope}
		\end{tikzpicture}
		\caption{Tree-level diagram for one-vertex contribution to the NG of the curvature perturbations.} \label{fig7}
	\end{center}
\end{figure}

In this appendix, we present the details of in-in calculations for the three-point function of the curvature perturbations. In order to do so, from Eq. (\ref{curvature-perturbation}), we see that we need to calculate the bispectra of $\overline{\delta\sigma}$.  The interaction Hamiltonian required to calculate the  bispectra of scalar modes is given in Eq. (\ref{H-int-SSS}). Naively, the one-vertex tree level Feynman diagram which is shown in Fig \ref{fig7} seems to give the dominant contribution and, therefore, we start by this diagram. Only the first two terms in the total interaction Hamiltonian (\ref{H-int-SSS})
\begin{eqnarray}\label{H-int-SSS-1}
\delta{H}_{{I},1}^{{\rm SSS}} = I \sqrt{\frac{2}{\epsilon}} \frac{H}{\tau} \int d^3 x 
\Big( \overline{\delta \sigma}^3 + \frac{5}{4} \tau^2 \overline{\delta \sigma} \, \overline{ \delta \sigma}'^2
\Big) \,,
\end{eqnarray}
contribute to this diagram. Performing the Fourier transformation for the above interaction Hamiltonian and substituting the result into the in-in formula (\ref{IN-IN-def}) and expanding to the first order, yields
\begin{eqnarray}
\nonumber && \big \langle \delta\sigma({\bf k}_1) \delta\sigma({\bf k}_2)
\delta\sigma({\bf k}_3)  \big \rangle|_{\rm Fig. \ref{fig7}} 
= i \int_{\tau_{0}} ^{\tau_{e} }d\tau_1 \langle \big[ \delta{H}_{{I},1}^{\rm SSS} (\tau_1), 
\delta\sigma(\tau_{e},{\bf k}_1) \delta\sigma(\tau_{e},{\bf k}_2) 
\delta\sigma(\tau_{e},{\bf k}_3) \big] \rangle \\ \nonumber 
&& = \frac{2IH}{\sqrt{2\epsilon}} \int_{\tau_{0}} ^{\tau_{e} } \frac{d\tau _1}{\tau_{1}} 
{\rm Im} \Big[\delta\sigma_{k_{1}}^{*}(\tau_{e}) \delta\sigma_{k_{2}}^{*}(\tau_{e}) \delta\sigma_{k_{3}}^{*}(\tau_{e})
\Big( 12 \delta \sigma_{k_{1}}(\tau_{1})\delta\sigma_{k_{2}}(\tau_{1})\delta\sigma_{k_{3}}(\tau_{1})
\\ \nonumber 
&& - 20 \tau_{1} \big( \delta \sigma_{k_{1}}'(\tau_{e})\delta\sigma_{k_{2}}(\tau_{e})\delta\sigma_{k_{3}}(\tau_{e})+2 {\rm perm} \big) + 5 \tau_{1}^{2} \big(\delta \sigma_{k_{1}}'(\tau_{e})\delta\sigma_{k_{2}}'(\tau_{e})\delta\sigma_{k_{3}}(\tau_{e})+2 {\rm perm} \big) \Big)\Big]
\\
&& = \sqrt{\frac{2}{\epsilon}} \frac{H}{8\tau_{e}^3} \frac{\sum_i k_{i}^3}{\Pi_ik_{i}^3} \, I N_K \,
(2\pi)^3 \delta^3(\textbf{k}_{1}+\textbf{k}_{2}+\textbf{k}_{3}) \,,
\end{eqnarray}
where $K \equiv \frac{1}{3}(k_1+k_2+k_3)$ is a reference momentum and $N_K=-\ln(-K\tau_e)$ is the number of e-folds associated to $K$ till the end of inflation.

Now, using the definition (\ref{curvature-perturbation}), it is straightforward to show that the contribution from the Feynman diagram Fig \ref{fig7} to the three-point function of the curvature perturbations in the super horizon limit is given by
\begin{equation}\label{RRR-1V}
\langle \mathcal{R}(\textbf{k}_{1})\mathcal{R}(\textbf{k}_{2})\mathcal{R}(\textbf{k}_{3})\rangle|_{\rm Fig. \ref{fig7}} = - \frac{H^4}{16  \epsilon^2} I N_K \frac{\sum_i k_{i}^3}{\Pi_ik_{i}^3} (2 \pi)^3 \delta^{3}(\textbf{k}_{1}+\textbf{k}_{2}+\textbf{k}_{3}) \,.
\end{equation}

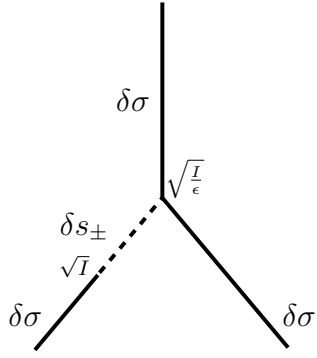
\begin{figure}
	\begin{center}
		\begin{tikzpicture}[line width=1.5 pt, scale=1.3]
		\begin{scope}[rotate=90]
		\draw[fermionnoarrow] (-140:2)--(0,0);
		\draw[scalarnoarrow,black] (140:1)--(0,0);
		\draw[fermionnoarrow] (0:2)--(0,0);
		\node at (-1.2,1.4) {$\delta \sigma$};
		\node at (-1.2,-1.4) {$\delta \sigma$};
		\node at (-0.3,0.8) {$\delta s_{\pm}$};
		\node at (1,.3)  {$\delta \sigma$};	
		\node at (-0.7,0.9) {\scriptsize $ \sqrt{I}$};
		\node at (0.2,-0.23) {\scriptsize $\sqrt{\frac{ I}{ \epsilon} }$};
		\begin{scope}[shift={(-0.79,0.66)}]
		\draw[fermionnoarrow] (140:1)--(0,0);
		\end{scope}
		\end{scope}
		\end{tikzpicture}
		\caption{ Tree-level diagrams for two-vertex contributions to the NG of the curvature perturbations.
		} \label{fig8}
	\end{center}
\end{figure}

In a similar way, we can compute the two-vertex contributions to the three-point function of the curvature perturbations. For the sake of simplicity in computation, we work with $\delta{s}_{\pm}$ rather than $\overline{\delta{s}}$ and $\tilde{U}$ and all the results can be easily translated in terms of these physical variables through the linear transformations (\ref{trans-deltas}). The relevant Feynman diagrams are shown in Fig \ref{fig8} which has one three-leg vertex and one two-leg vertex. The two-leg vertex is determined by $\delta{ H}_{I,i}^{\rm SS}$ with $i=1,2,3,4$ in (\ref{H2-int}) while the relevant three-leg vertex interaction Hamiltonian in (\ref{H-int-SSS}) is given by
\begin{eqnarray}\label{H-int-SSS-2}
&&\delta{H}_{{I},2}^{\rm SSS} \approx - 5 \sqrt{\frac{I}{2\epsilon}} H \int d^3 x 
\Big( 4\overline{\delta s}\,\,   \overline{ \delta \sigma}^2
+ 10 \tau \overline{\delta s}\, \,  \overline{\delta \sigma}\,\,  \overline{ \delta \sigma}' 
+ \tau^2 \overline{ \delta s}' \, \, \overline{\delta \sigma}\,\,   \overline{ \delta \sigma}' \Big) \,.
\end{eqnarray}

As an example, we present some details of calculations for $\delta{s}_+$ in Fig. \ref{fig8}. Substituting from Eq. (\ref{trans-deltas}) in the above interaction Hamiltonian and going to the Fourier space, we can find the contributions of $\delta{s}_+$ to the cubic interaction (\ref{H-int-SSS-2}). Substituting this result together with the quadratic interaction Hamiltonians $\delta{ H}_{I,1,3}^{\rm SS}$ defined in Eq. (\ref{H2-int}) into the in-in formula Eq. (\ref{IN-IN-def}), and expanding up to the first order in parameter $I$, we find
\begin{eqnarray}
&&\big\langle \delta\sigma({\bf k}_1) \delta\sigma({\bf k}_2)\delta\sigma({\bf k}_3) \big\rangle
|_{\rm Fig. \ref{fig8}} = \nonumber \\  
&& - \int_{\tau_{0}}^{\tau_{e} }d\tau _1 \int_{\tau_{0}}^{\tau _1}d\tau_2 \langle 
\big[ \delta{H}_{{I},2}^{\rm SSS}(\tau_2), \big[ \delta{H}_{{I},1}^{\rm SS} (\tau_1) 
+ \delta{H}_{{I},3}^{\rm SS} (\tau_1), 
\delta\sigma(\tau_{e},{\bf k}_1) \delta\sigma(\tau_{e},{\bf k}_2)
\delta\sigma(\tau_{e},{\bf k}_3) \big] \big] \rangle 
\nonumber \\
&& - \int_{\tau_{0}}^{\tau_{e} }d\tau_1 \int_{\tau_{0}}^{\tau_1} d\tau_2 
\langle \big[ \delta{H}_{{I},1}^{\rm SS}(\tau_2 ) + \delta{H}_{{I},3}^{\rm SS} (\tau_2 ) , 
\big[ \delta{H}_{{I},2}^{\rm SSS}(\tau _1 ) , \delta\sigma(\tau_{e},{\bf k}_1)
\delta\sigma(\tau_{e},{\bf k}_2) \delta\sigma(\tau_{e},{\bf k}_3) \big] \big] \rangle
\nonumber \\
&& = - \sqrt{\frac{2}{\epsilon}} \frac{85 H}{3\tau_{e}^3} \cosh(4 \pi \theta) \Theta_1(\theta)\, I N_K^2 \, 
\frac{\sum_i k_{i}^3}{\Pi_ik_{i}^3}
(2\pi)^3 \delta^3(\textbf{k}_{1}+\textbf{k}_{2}+\textbf{k}_{3}) \,.
\end{eqnarray}
The contribution coming from $\delta{s}_-$ can also be obtained in a similar way. 

After summing all contributions and using the definition  (\ref{curvature-perturbation}), we obtain the following result at the super horizon limit
\begin{equation}\label{RRR-2V}
\langle \mathcal{R}(\textbf{k}_{1}) \mathcal{R}(\textbf{k}_{2}) 
\mathcal{R}(\textbf{k}_{3})\rangle|_{\rm Fig. \ref{fig8}}
= \frac{85 H^4 }{ 6 \epsilon^2}  \cosh(4 \pi \theta) \Theta_1(\theta)\, I N_K^2 \,
\frac{\sum_i k_{i}^3}{\Pi_ik_{i}^3}  
(2 \pi)^3 \delta^{3}(\textbf{k}_{1}+\textbf{k}_{2}+\textbf{k}_{3}) \,.
\end{equation}

We see that the ratio of the amplitude of the two-vertex Feynman diagram \ref{fig8} to the one-vertex diagram \ref{fig7} is proportional to $N_K$ which shows that the two-vertex contributions are larger than the one-vertex one. In other words, as mediator particles, the entropy modes $\delta{s}_{\pm}$ enhance the amplitude of the three-point functions. Therefore, we expect that the three-vertex contributions would be even larger than the two-vertex contribution. The Feynman diagrams for the three-vertex contribution are shown in Fig. \ref{fig88}. Therefore, similar to the two-vertex case, we need the second order Hamiltonians $\delta{H}_{I,i}^{\rm SS}$ with $i=1,2,3,4$ which are defined in (\ref{H2-int}). Looking at the transformation (\ref{trans-deltas}), we see that the cubic interaction Hamiltonian which contributes to the Feynman diagram \ref{fig88} is given by the first two terms in the second line of the total interaction Hamiltonian (\ref{H-int-SSS})
\begin{eqnarray}\label{H-int-SSS-3}
\delta{H}_{{I},3}^{\rm SSS} = \frac{\tau^6}{2} \frac{H}{\sqrt{2\epsilon}} \int d^3 x 
\bigg[ 5 \big( ( \tau^{-2}\overline{\delta{s}} )' \big)^2 
+ 4 \big( ( \tau^{-2}\partial\tilde{U} )' \big)^2 \bigg] \overline{\delta \sigma} \,.
\end{eqnarray}
Had we worked with the original variables $\overline{\delta{s}}$ and ${\tilde U}$, from the quadratic action (\ref{LSS-sigma-s}) and cubic interactions (\ref{H-int-SSS}), 
only the first term in the above interaction Hamiltonian would have contributed to the three-point function since the results are independent of the fields that we use. 

Going to the Fourier space and expanding the in-in formula (\ref{IN-IN-def}) up to the third order in the perturbations, for diagram (a) in Fig. (\ref{fig88}) and only for $\delta{s}_+$ as the mediator,  we find
\begin{eqnarray}
&&\langle \delta\sigma({\bf k}_1) \delta\sigma({\bf k}_2) \delta\sigma({\bf k}_3) 
\rangle|_{\rm Fig. \ref{fig88} (a)} =
- i \int_{\tau_{0}} ^{\tau_{e} }d\tau_1 \int_{\tau_{0}}^{\tau_1 }
d\tau_2 \int_{\tau_{0}} ^{\tau_2}d\tau _3  \nonumber \\ \nonumber
&&
\Bigg \{ \big\langle \big[ \delta{H}_{{I},3}^{\rm SSS} (\tau_3 ) , \big[ \delta{H}_{{I},1}^{\rm SS}(\tau _2 ) 
+ \delta{H}_{{I},2}^{\rm SS} (\tau_2 ) , \big[ \delta{H}_{{I},1}^{\rm SS}(\tau_1 )
+ \delta{H}_{{I},3}^{\rm SS} (\tau_1 ) , \delta\sigma(\tau_{e},{\bf k}_1)
\delta\sigma(\tau_{e},{\bf k}_2) \delta\sigma(\tau_{e},{\bf k}_3) \big] \big] \big] \big\rangle \\
\nonumber
&&+ \big\langle \big[ \delta{H}_{{I},1}^{\rm SS}(\tau_3 ) 
+ \delta{H}_{{I},2}^{\rm SS} (\tau_3 ) , 
\big[ \delta{H}_{{I},3}^{\rm SSS} (\tau_2 ) , 
\big[ \delta{H}_{{I},1}^{\rm SS}(\tau_1 )
+ \delta{H}_{{I},3}^{\rm SS} (\tau_1 ) , \delta\sigma(\tau_{e},{\bf k}_1)
\delta\sigma(\tau_{e},{\bf k}_2) \delta\sigma(\tau_{e},{\bf k}_3) \big] \big] \big] \big\rangle \\
\nonumber
&& + \big\langle \big[ \delta{H}_{{I},1}^{\rm SS}(\tau _3) 
+ \delta{H}_{{I},2}^{\rm SS} (\tau_3 ) , 
\big[ \delta{H}_{{I},1}^{\rm SS}(\tau _2 )
+ \delta{H}_{{I},3}^{\rm SS} (\tau _2 ) , \big[ \delta{H}_{{I},3}^{\rm SSS} (\tau _1 ) , 
\delta\sigma(\tau_{e},{\bf k}_1)
\delta\sigma(\tau_{e},{\bf k}_2) \delta\sigma(\tau_{e},{\bf k}_3) \big] \big] \big]  \big\rangle \Bigg\} \\ 
&& = - \sqrt{\frac{2}{\epsilon}} \frac{27 H}{4\tau_{e}^3} (e^{8 \pi \theta}-1) 
\Theta_1(\theta)^{2}\,  I N_K^3 
\frac{\sum_i k_{i}^3}{\Pi_ik_{i}^3}
(2\pi)^3 \delta^3 (\textbf{k}_{1}+\textbf{k}_{2}+\textbf{k}_{3}) \,.
\end{eqnarray}
For panel (b) of Fig. (\ref{fig88}),  and again for $\delta{s}_+$ as the mediator, 
we also find
\begin{eqnarray}
\langle \delta \sigma({\bf k}_1)\delta\sigma({\bf k}_2)\delta\sigma({\bf k}_3)
\rangle|_{\rm Fig. \ref{fig88} (b)}=-\sqrt{\frac{2}{\epsilon}}  \frac{3 H}{  \tau_{e}^3} 
\Theta_1(\theta)^2\, I N_K^3  
\frac{\sum_i k_{i}^3}{\Pi_ik_{i}^3}
(2\pi)^3 \delta^3(\textbf{k}_{1}+\textbf{k}_{2}+\textbf{k}_{3}) .
\end{eqnarray}
In a similar manner we obtain the contributions coming from the other entropy mode $\delta{s}_-$ of the diagrams shown in Fig. \ref{fig88}. 

Similar to the  two vertices case, the sum of all contributions associated to linearized $\theta$ terms in Lagrangian (\ref{H-int-SSS-3}) vanishes in above Hamiltonian permutations. Adding the results for both $\delta{s}_+$ and $\delta{s}_-$ mediators 
we find the following result for the three-point function of curvature perturbations
associated to Fig. \ref{fig88}
\begin{equation}\label{RRR-3V-app}
\langle \mathcal{R}(\textbf{k}_{1})\mathcal{R}(\textbf{k}_{2})\mathcal{R}(\textbf{k}_{3})\rangle|_{\rm Fig. \ref{fig88}}
= \frac{ 3 H^4 }{2 \epsilon^2} (1+9 \cosh(8 \pi \theta)) \Theta_1(\theta)^{2} \, I N_K^3 \, 
\frac{\sum_i k_{i}^3}{\Pi_ik_{i}^3}
(2\pi)^3 \delta^{3}(\textbf{k}_{1}+\textbf{k}_{2}+\textbf{k}_{3}) \,.
\end{equation}

The three-point function for the curvature perturbations is the sum of all contributions coming from the diagrams shown in Figs. \ref{fig7}, \ref{fig8}, and \ref{fig88} which are calculated in (\ref{RRR-1V}), (\ref{RRR-2V}), and (\ref{RRR-3V-app}) respectively. The three-vertex contribution (\ref{RRR-3V-app}) is proportional to $N_K^3$ which is larger than the one-vertex and two-vertex contributions. Therefore, the dominant 
contribution to the bi-spectrum of the curvature perturbations is given by (\ref{RRR-3V-app}).

{}

\end{document}